\documentclass[letterpaper,journal]{IEEEtran}
\usepackage[latin9]{inputenc}
\usepackage{array}
\usepackage{float}
\usepackage{mathtools}
\usepackage{multirow}
\usepackage{amsmath}
\usepackage{amsthm}
\usepackage{amssymb}
\usepackage{graphicx}
\PassOptionsToPackage{normalem}{ulem}
\usepackage{ulem}
\usepackage{color}
\usepackage{cite}
\makeatletter

\pdfpageheight\paperheight
\pdfpagewidth\paperwidth

\providecommand{\tabularnewline}{\\}
\floatstyle{ruled}
\newfloat{algorithm}{tbp}{loa}
\providecommand{\algorithmname}{Algorithm}
\floatname{algorithm}{\protect\algorithmname}

  \theoremstyle{definition}
  \newtheorem{defn}{\protect\definitionname}
  \theoremstyle{plain}
  \newtheorem{lem}{\protect\lemmaname}
  \theoremstyle{remark}
  \newtheorem{rem}{\protect\remarkname}
  \theoremstyle{plain}
  \newtheorem{thm}{\protect\theoremname}

\usepackage{float}
\usepackage{graphicx}
\usepackage{tabularx}
\usepackage{amsmath}
\usepackage{amsthm}
\usepackage{algorithm,algpseudocode}
\usepackage{lipsum}
\usepackage{url} 
\usepackage{epstopdf}
\usepackage{amssymb}
\usepackage[font=footnotesize]{caption}
\@ifundefined{showcaptionsetup}{}{%
 \PassOptionsToPackage{caption=false}{subfig}}
\usepackage{subfig}
\makeatother

\providecommand{\definitionname}{Definition}
\providecommand{\lemmaname}{Lemma}
\providecommand{\remarkname}{Remark}
\providecommand{\theoremname}{Theorem}

\begin{document}

\title{{\huge{}iSTRICT: An Interdependent Strategic Trust Mechanism for
the Cloud-Enabled Internet of Controlled Things}}

\author{Jeffrey Pawlick, Juntao Chen, and Quanyan Zhu
\thanks{The authors are with the Department of Electrical and Computer Engineering, Tandon School of Engineering, New York University, Brooklyn, NY, 11201 USA. E-mail: \{jpawlick,jc6412,qz494\}@nyu.edu.}
\thanks{This work is partially supported by an NSF IGERT grant through the Center for Interdisciplinary Studies 
in Security and Privacy (CRISSP) 
at New York University, by the grants CNS-1544782, 
EFRI-1441140, and SES-1541164 from National Science 
Foundation (NSF) and DE-NE0008571 from the 
Department of Energy. }}
\maketitle

\begin{abstract}
The cloud-enabled Internet of controlled things (IoCT) envisions a
network of sensors, controllers, and actuators connected through a
local cloud in order to intelligently control physical devices. Because
cloud services are vulnerable to advanced persistent threats (APTs),
each device in the IoCT must strategically decide whether to trust
cloud services that may be compromised. In this paper, we present
iSTRICT, an \uline{i}nterdependent \uline{s}trategic \uline{tr}ust
mechanism for the cloud-enabled \uline{I}o\uline{CT}. iSTRICT
is composed of three interdependent layers. In the cloud layer, iSTRICT
uses \texttt{FlipIt} games to conceptualize APTs. In the communication
layer, it captures the interaction between devices and the cloud using
signaling games. In the physical layer, iSTRICT uses optimal control
to quantify the utilities in the higher level games. Best response
dynamics link the three layers in an overall ``game-of-games,''
for which the outcome is captured by a concept called Gestalt Nash
equilibrium (GNE). We prove the existence of a GNE under a set of
natural assumptions and develop an adaptive algorithm to iteratively
compute the equilibrium. Finally, we apply iSTRICT to trust management
for autonomous vehicles that rely on measurements from remote sources.
 We show that strategic trust in the communication layer achieves
a worst-case probability of compromise for any attack and defense
costs in the cyber layer.\end{abstract}


\section{Introduction\label{sec:Introduction}}

The Internet of Things (IoT) will impact a diverse set of consumer,
public sector, and industrial systems. Smart homes and buildings,
autonomous vehicles and transportation \cite{swan_sensor_2012}, and
the interaction between wearable fitness devices and social networks
\cite{_internet_2015} provide a few examples of application areas
which will be particularly impacted by the IoT. One definition of
the IoT is a ``dynamic global network infrastructure
with self-configuring capabilities based on standard and interoperable
communication protocols where physical and virtual `things' have identities,
physical attributes, and virtual personalities'' \cite{_CERP_2015}.
This definition envisions a decentralized, heterogeneous network with
plug-and-play capabilities. The related concept of cyber-physical
systems (CPS) refers to ``smart networked systems with embedded sensors,
processors, and actuators'' \cite{_CPS_2015}. \textcolor{black}{\cite{liu2017review} provides a detailed introduction to CPS and reports on its development status.} The term CPS emphasizes
the ``systems'' nature of these networks. In both IoT and CPS, ``the
joint behavior of the `cyber' and physical elements of the system
is critical---computing, control, sensing, and networking can be
integrated into every component'' \cite{_CPS_2015}. The importance
of sensing, actuation, and control to devices in the IoT has given
rise to the term ``Internet of \emph{controlled} things,'' or IoCT.
Hereafter, we refer to the IoCT as a way to address challenges of
both CPS and IoT.

\begin{figure}
\begin{centering}
\includegraphics[width=0.7\columnwidth]{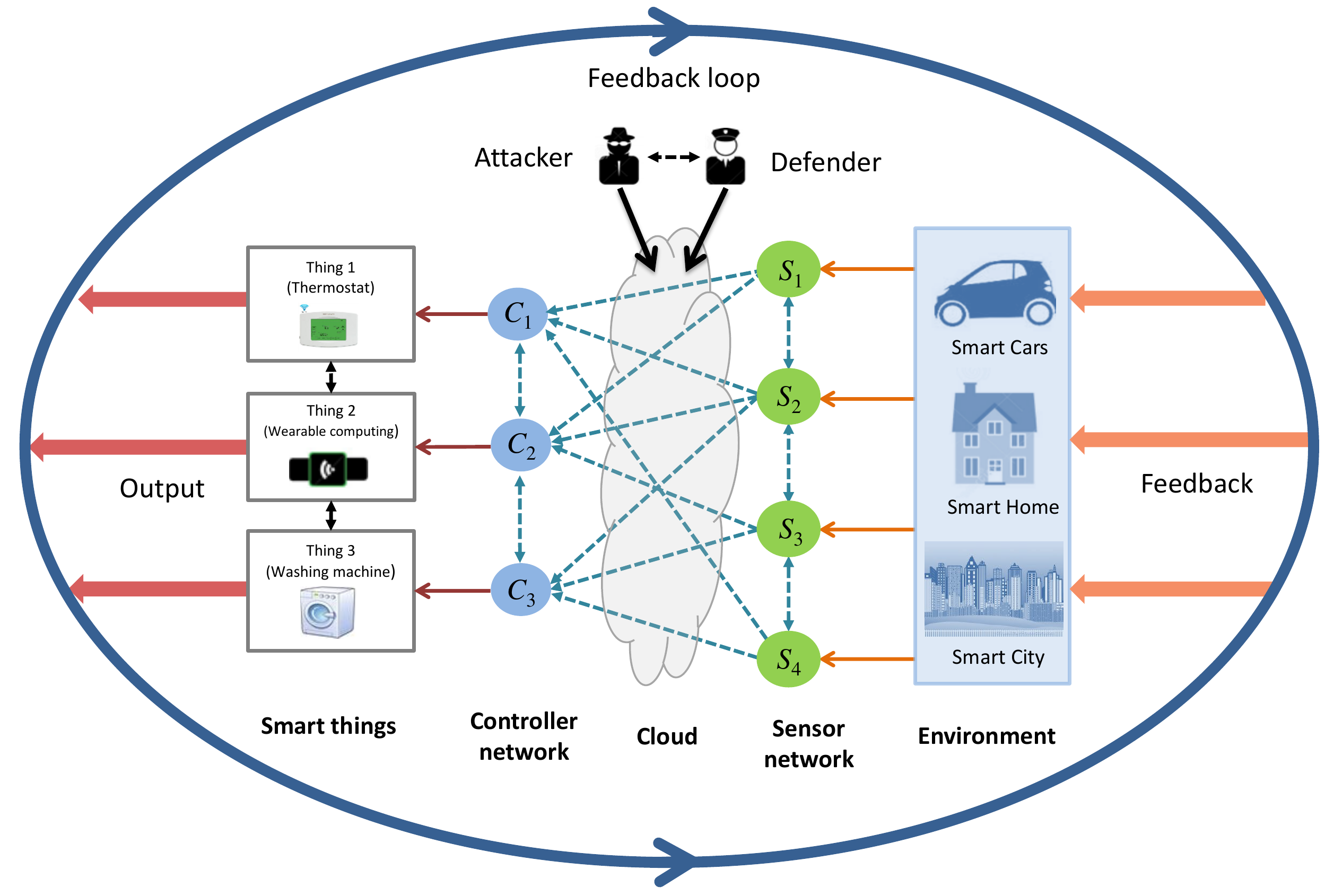} 
\par\end{centering}

\caption{\label{fig:network}iSTRICT addresses security and trust issues for a cloud-enabled
IoCT. The cloud-enabled IoCT consists of connected sensors and devices,
with a cloud as the interface. Adversaries are capable of compromising
cloud services and modifying the control signals that they transmit
to the devices. \textcolor{black}{The trust issue lies between the cloud (sender) and IoCT (receiver). Each IoCT device should determine which signals to trust from cloud services strategically.}}
 
\end{figure}

The IoCT requires an interface between heterogeneous components. Local
clouds (or \emph{fogs} or \emph{cloudlets}) offer promising solutions.
In these networks, a cloud provides services for data aggregation,
data storage, and computation. In addition, the cloud provides a market
for the services of software developers and computational intelligence
experts \cite{jin2014information}. Figure \ref{fig:network} depicts
a cloud-enabled IoCT. In this network, sensors push environment data
to the cloud, where it is aggregated and sent to devices (or ``things''),
which use the data for feedback control. These devices modify the
environment, and the cycle continues. \textcolor{black}{Note that the control design of the IoCT is distributed, since each device can determine which cloud services to use for feedback control.}

\subsection{Advanced Persistent Threats in the Cloud-Enabled IoCT}

Unfortunately, cyberattacks on the cloud are increasing as more businesses
utilize cloud services \cite{alert2015}. \textcolor{black}{To provide reliable support for IoCT applications, sensitive data provided by the cloud services needs to be well protected \cite{fernandes16flowfence}.} In this paper we focus on
the attack model of advanced persistent threats (APTs): ``cyber attacks
executed by sophisticated and well-resourced adversaries targeting
specific information in high-profile companies and governments, usually
in a long term campaign involving different steps'' \cite{chen2014study}.
In the initial stage of an APT, an attacker penetrates the network through
techniques such as social engineering, malicious hardware injection,
theft of cryptographic keys, or zero-day exploits \cite{bowers2012defending}.
For example, the \emph{Naikon APT}, which targeted governments around
the South China Sea in 2010-2015, used a bait document that appeared
to be a Microsoft Word file but which was actually a malicious executable
that installed spware \cite{securelist2015naikon}. The cloud is particularly
vulnerable to initial penetration through application-layer attacks,
because many applications are required for developers and clients
to interface with the cloud. Our iSTRICT can be applied to many cyberattack scenarios. For example, cross-site scripting (XSS) and SQL injection
are two types of application-layer attacks. In SQL injection, attackers
insert malicious SQL code into fields which do not properly process
string literal escape characters. The malicious code targets the server,
where it could be used to modify data or bypass authentication systems.
By contrast, XSS targets the execution of code in the browser on the
client side. All of these attacks give attackers an initial entry
point into a system, from which they can begin to gain more complete,
insider control. This control of the cloud can be used to transmit
malicious signals to CPS and cause physical damage.

\subsection{Strategic Trust}
\textcolor{black}{
Given the threat of insider attacks on the cloud, each IoCT device must decide which signals to trust from cloud services.} Trust refers
to positive beliefs about the perceived reliability of, dependability
of, and confidence in another entity \cite{fogg1999elements}. These
entities may be agents in an IoCT with misaligned incentives. Many
specific processes in the IoCT require trust, such as data collection,
aggregation and processing, privacy protection, and user-device trust
in human-in-the-loop interactions \cite{yan2014trust}. While many
factors influence trust, including subjective beliefs, we focus on
objective properties of trust. These include 1) reputation, 2) promises,
and 3) interaction context. Many trust management systems are based
on tracking reputation over multiple interactions. Unfortunately,
agents in the IoCT may interact only once, making reputation difficult
to accrue \cite{minhas_multifaceted_2011}. This property of IoCT
also limits the effectiveness of promises such as contracts or policies.
Promises may not be enforceable for entities that interact only once.
Therefore we focus on strategic trust that is predictive rather than
reactive. We use game-theoretic utility functions to capture the motivations
for entities to be trustworthy. These utility functions change based
on the particular context of the interaction. In this sense, our model
of strategic trust is \emph{incentive-compatible}, \emph{i.e.}, consistent
with each agent acting in its own self-interest.

\subsection{Game-Theoretic iSTRICT Model}

We propose a framework called iSTRICT, which is composed of three
interacting layers: a cloud layer, a communication layer, and a physical
layer. In the first layer, the cloud-services are threatened by attackers
capable of APTs and defended by network administrators (or ``defenders'').
The interaction at each cloud-service is modeled using the \texttt{FlipIt}
game recently proposed by Bowers et al. \cite{bowers2012defending}
and van Dijk et al. \cite{vanDijk2013Flip}. iSTRICT uses one \texttt{FlipIt}
game per cloud-service. In the communication layer, the cloud-services---which
may be controlled by the attacker or defender according to the outcome
of the \texttt{FlipIt} game---transmit information to a device which
decides whether to trust the cloud-services. This interaction is captured
using a signaling game. At the physical layer, the utility parameters
for the signaling game are determined using optimal control. The cloud,
communication, and physical layers are interdependent. This motivates
an overall equilibrium concept called \emph{Gestalt Nash equilibrium}
(GNE). GNE requires each game to be solved optimally given the results
of the other games. Because this is a similar idea to \emph{best-response}
in Nash equilibrium, we call the multi-game framework a \emph{game-of-games}.

\subsection{Contributions}

In summary, we present the following contributions: 
\begin{enumerate}
\item \textbf{Trust Model:} We develop a multi-layer framework (iSTRICT)
and associated equilibrium concept (GNE) to capture interdependent
strategic trust in the cloud-enabled IoCT. iSTRICT combines analysis
at the cloud, communication, and physical layers. 
\item \textbf{GNE Analysis: }We prove the existence of GNE, and we show
that strategic trust in the communication layer guarantees a worst-case
probability of compromise regardless of attack costs in the cyber
layer.
\item \textbf{Adaptive Algorithm:} We present an adaptive algorithm using
best-response dynamics to compute a GNE.
\item \textbf{Autonomous Vehicle Application:} We simulate the control of
a pair of autonomous vehicles using iSTRICT, and show improvement over the
performance under naive policies.
\end{enumerate}
The rest of the paper proceeds as follows. In Section \ref{sec:broadModel},
we give a broad outline of the iSTRICT model. Section \ref{sec:detailedModel}
presents the details of the \texttt{FlipIt} game, signaling game,
physical layer control system, and equilibrium concept. Then, in Section
\ref{sec:Gestalt}, we study the equilibrium analytically using an
adaptive algorithm. Finally, we apply the framework to the control
of autonomous vehicles in Section \ref{sec:app}.

\subsection{Related Work}
\textcolor{black}{Designing trustworthy cloud service systems has been investigated extensively in the literature. Various methods, including a feedback evaluation component, Bayesian game, and domain partition have been proposed  \cite{siadat2017identifying,zhang2018domain,zhu2015authenticated}. Trust models to predict the cloud trust values (or reputation) can be mainly divided into  objective and subjective classes. The first are based on the quality of service parameters, and the second are based on feedback from cloud service users  \cite{siadat2017identifying,fan2014novel}.} 

\textcolor{black}{In the IoCT, however, agents may not have sufficient number of interactions, which makes reputation challenging
to obtain \cite{minhas_multifaceted_2011}. In addition, trust value-based cloud trust management systems can be compromised by reputation attacks through fake feedback which can severely degrade the system performance \cite{siadat2017identifying,noor2013reputation}. Therefore, in this work, we aim to design a strategic trust mechanism which is predictive rather than reactive through an integrative game-theoretic framework. Rather than using trust value  \cite{haq2010sla,noor2013reputation}, IoCT devices in our iSTRICT model make decisions based on the strategies of players at the cloud layer as well as based on the physical system performance. This multi-layer design provides resilience to reputation attacks.}

\textcolor{black}{Cyber-physical systems security becomes a critical concern due to the prevailing threats from both cyber and physical components in the system \cite{xie2015security,mo2012cyber,chen2017interdependent}. To facilitate a secure system design,
game theory has been widely adopted to model and capture the strategic interactions between the attackers and defenders \cite{pawlick2015flip,manshaei2013game,zhu2015game}. Our iSTRICT framework builds on two existing game models. One is the signaling game which has been used in intrusion
detection systems \cite{alpcan2003game} and network defense \cite{pawlick2015deception}. The other one is the \texttt{FlipIt} game \cite{bowers2012defending,vanDijk2013Flip} which has been applied to security of a single cloud service
\cite{pawlick2015flip,chen2016optimal} as well as AND/OR combinations of cloud services \cite{laszka2014flipthem}. In contrast to previous works, in this paper we propose a three-layer interdependent model to enable devices to decide whether to trust cloud services that may be compromised.} Specifically, trust management decisions are coupled by the
dynamics of cloud-enabled devices, because data provided by the cloud services
is used for feedback control. Devices must balance the need for as
many data sources as possible (in order to increase the quality of
the feedback control) with the imperative to reject data sources
that are compromised by attackers. 

\textcolor{black}{In terms of the technical framework, iSTRICT builds on existing achievements in IoCT architecture design
\cite{jin2014information,chen2017security,sun2016Internet,cenedese2014padova}, which
describe the roles of different layers of the IoCT at which data is
collected, processed, and accessed by devices \cite{sun2016Internet}.}
Each layer of the IoCT consists of different enabling technologies
such as wireless sensor networks and data management systems \cite{cenedese2014padova}.
Our perspective, however, is distinct from this literature because
we emphasize an integrated mathematical framework. iSTRICT leverages
game theory to obtain optimal defense strategies for IoCT components,
and it uses control theory to quantify the performance of devices.

\section{iSTRICT Overview\label{sec:broadModel}}

We consider a cyber-physical attack in which an adversary penetrates
a cloud service in order to transmit malicious signals to a physical
device and cause improper operation. This type of cross-layer attack
is increasingly relevant in IoCT settings. Perhaps the most famous
cross-layer attack is the Stuxnet attack that damaged Iran's nuclear
program. But even more recently, an attacker allegedly penetrated
the supervisory control and data acquisition (SCADA) system that controls
the Bowman Dam, located less than $20$ miles north of Manhattan.
The attacker gained control of a sluice gate which manages water level\footnote{The sluice gate happened to be disconnected for manual repair at the
time, however, so the attacker could not actually change water levels.} \cite{usdoj2013rye}. Cyber-physical systems ranging from the smart
grid to public transportation need to be protected from similar attacks. 

The iSTRICT framework offers a defense-in-depth approach to IoCT security.
In this section, we introduce each of the three layers of iSTRICT
very briefly, in order to focus on the interaction between the layers.
We describe an equilibrium concept for the simultaneous steady-state
of all three layers. Later, Section \ref{sec:detailedModel} describes
each layer in detail. \textcolor{black}{Table \ref{tab:notationPhyTrust} lists the notation for the paper.}

\begin{table}
\caption{\label{tab:notationPhyTrust} \textcolor{black}{Nomenclature}}

\centering{}%
\begin{tabular}{|c|c|}
\hline 
Notation & Meaning\tabularnewline
\hline 
\hline 
$\mathbb{S}=\{1,2,\ldots,N\}$ & Cloud services (CSs)\tabularnewline
\hline 
$\mathcal{A}^{i},$ $\mathcal{D}^{i},$ $i\in\mathbb{S},$ $\mathcal{R}$  & Attackers, defenders, device\tabularnewline
\hline 
$v_{\mathcal{A}}=[v_{\mathcal{A}}^{i}]_{i\in\mathbb{S}}$ & Values of CSs for $\mathcal{A}$ \tabularnewline
\hline 
 $v_{\mathcal{D}}=[v_{\mathcal{D}}^{i}]_{i\in\mathbb{S}}$ & Values of CSs for $\mathcal{D}$\tabularnewline
\hline 
$p_{\mathcal{A}}=[p_{\mathcal{A}}^{i}]_{i\in\mathbb{S}}$  & Probabilities that $\mathcal{A}$ controls CSs\tabularnewline
\hline 
$p_{\mathcal{D}}=[p_{\mathcal{D}}^{i}]_{i\in\mathbb{S}}$ & Probabilities that $\mathcal{D}$ controls CSs\tabularnewline
\hline 
$p_{\mathcal{A}}^{i*}=T^{F_{i}}(v_{\mathcal{A}}^{i},v_{\mathcal{D}}^{i})$ & \texttt{FlipIt} mapping for CS $i$\tabularnewline
\hline 
$(v_{\mathcal{A}}^{*},v_{\mathcal{D}}^{*})\in T^{S}(p_{\mathcal{A}})$ & Signaling game mapping\tabularnewline
\hline 
$f_{\mathcal{A}}^{i},$ $f_{\mathcal{D}}^{i}$ & Frequencies $\mathcal{A}^{i}$ and $\mathcal{D}^{i}$\tabularnewline
\hline 
$u_{\mathcal{A}}^{F_{i}}(f_{\mathcal{A}}^{i},f_{\mathcal{D}}^{i})$
 &  $\mathcal{A}^{i}$'s utility in \texttt{FlipIt}
game $i$\tabularnewline
\hline 
$u_{\mathcal{D}}^{F_{i}}(f_{\mathcal{A}}^{i},f_{\mathcal{D}}^{i})$ &  $\mathcal{D}^{i}$'s utility in \texttt{FlipIt}
game $i$\tabularnewline
\hline 
$\theta=[\theta^{i}]_{i\in\mathbb{S}}$  & Types of CSs\tabularnewline
\hline 
$\theta^{i}\in\Theta=\{\theta_{\mathcal{A}},\theta_{\mathcal{D}}\}$ & Type spaces of CSs\tabularnewline
\hline 
$m=[m^{i}]_{i\in\mathbb{S}}$  & Messages from CSs\tabularnewline
\hline 
$m^{i}\in M=\{m_{L},m_{H}\}$ & Low or high risk message\tabularnewline
\hline 
$a=[a^{i}]_{i\in\mathbb{S}}$  & Actions for CSs\tabularnewline
\hline 
$a^{i}\in A=\{a_{T},a_{N}\}$ & Trust or not trust action\tabularnewline
\hline 
$u_{\mathcal{A}}^{S_{i}}(m,a),$ $u_{\mathcal{D}}^{S_{i}}(m,a)$  & Signaling game utility of $\mathcal{A}^{i}$ and $\mathcal{D}^{i}$ \tabularnewline
\hline 
$u_{\mathcal{R}}^{S}(\theta,m,a)$ & Signaling game utility for $\mathcal{R}$\tabularnewline
\hline 
$\sigma_{\mathcal{A}}^{i}(m)\in\Sigma_{\mathcal{A}}$ & Signaling game mixed strategies of $\mathcal{A}^{i}$ \tabularnewline
\hline 
 $\sigma_{\mathcal{D}}^{i}(m)\in\Sigma_{\mathcal{D}}$  & Signaling game mixed strategies of $\mathcal{D}^{i}$\tabularnewline
\hline 
$\sigma_{\mathcal{R}}(a\,|\,m)\in\Sigma_{\mathcal{R}}^{N}$ & Signaling game mixed strategy for $\mathcal{R}$\tabularnewline
\hline 
$\mu(\theta\,|\,m)=[\mu^{i}(\theta\,|\,m)]_{i\in\mathbb{S}}$ & Beliefs of $\mathcal{R}$ about CSs \tabularnewline
\hline 
$\bar{u}_{\mathcal{A}}^{S_{i}}(\sigma_{\mathcal{R}};\sigma_{\mathcal{A}}^{i},\sigma_{\mathcal{A}}^{-i};\sigma_{\mathcal{D}}^{-i})$  & Signaling game utility for $\mathcal{A}^{i}$\tabularnewline
\hline 
$\bar{u}_{\mathcal{D}}^{S_{i}}(\sigma_{\mathcal{R}};\sigma_{\mathcal{A}}^{-i};\sigma_{\mathcal{D}}^{i},\sigma_{\mathcal{D}}^{-i})$ & Signaling game utility for $\mathcal{D}^{i}$\tabularnewline
\hline 
$x[k],$ $\hat{x}[k],$ $u[k],$ & State, estimated state, control\tabularnewline
\hline 
$\Delta_{\mathcal{A}}^{i}[k],$ $\Delta_{\mathcal{D}}^{i}[k],$  & bias terms of $\mathcal{A}^{i}$ and $\mathcal{D}^{i}$\tabularnewline
\hline 
$\Xi_{\theta}[k]$ & cloud type matrix\tabularnewline
\hline 
$y[k],$ $\tilde{y}[k],$ & Measurements without and with biases \tabularnewline
\hline 
$\xi,$ $\zeta$ & Covariance matrices of noises\tabularnewline
\hline 
$\nu[k],$ $\epsilon$ & Innovation, innovation thresholds\tabularnewline
\hline 
$D_{\sigma_{\mathcal{R}}}(\nu[k])$ & Innovation gate\tabularnewline
\hline 
$v_{\mathcal{A}\mathcal{D}}^{i}=v_{\mathcal{A}}^{i}/v_{\mathcal{D}}^{i},$
$i\in\mathbb{S}$ & Ratios of CSs' value for $\mathcal{A}^{i}$ and $\mathcal{D}^{i}$\tabularnewline
\hline 
$\mathbb{V}^{i},$ $\mathbb{PR}^{i}$ & Spaces of $v_{\mathcal{A}\mathcal{D}}^{i}$ and $p_{\mathcal{A}}^{i}$
in a GNE\tabularnewline
\hline 
$p_{\mathcal{A}}^{i*}=\tilde{T}^{F_{i}}(v_{\mathcal{A}\mathcal{D}}^{i})$ & Redefined \texttt{FlipIt} mapping for CS $i$\tabularnewline
\hline 
$v_{\mathcal{A}\mathcal{D}}^{*}\in\tilde{T}^{S}(p_{\mathcal{A}})$ & Redefined signaling game mapping\tabularnewline
\hline 
$v_{\mathcal{A}\mathcal{D}}^{*}\in\tilde{T}^{S\circ F}(v_{\mathcal{A}\mathcal{D}})$ & Composition of $\tilde{T}^{F_{i}},$ $i\in\mathbb{S}$ and $\tilde{T}^{S}$ \tabularnewline
\hline 
$v_{\mathcal{A}\mathcal{D}}^{\dagger}\in\tilde{T}^{S\circ F}(v_{\mathcal{A}\mathcal{D}}^{\dagger})$ & Fixed-point requirement for a GNE\tabularnewline
\hline 
\end{tabular}
\end{table}

\subsection{Cloud Layer}

Consider a cloud-enabled IoCT composed of sensors that push data to
a cloud, which aggregates the data and sends it to devices. For example,
in a cloud-enabled smart home, sensors could include lighting sensors,
temperature sensors, and blood pressure or heart rate sensors that
may be placed on the skin or embedded within the body. Data from these
sensors is processed by a set of cloud services $\mathbb{S}=\left\{ 1,\ldots,N\right\} ,$
which make data available for control.

For each cloud service $i\in\mathbb{S},$ let $\mathcal{A}^{i}$ denote
an attacker who attempts to penetrate the service using zero-day exploits,
social engineering, or other techniques described in Section \ref{sec:Introduction}.
Similarly, let $\mathcal{D}^{i}$ denote a defender or network administrator
attempting to maintain the security of the cloud service. $\mathcal{A}^{i}$
and $\mathcal{D}^{i}$ attempt to claim or reclaim control of the
each cloud service at periodic intervals. We model the interactions
at all of the services using \texttt{FlipIt} games, one for each
of the $N$ services.

In the \texttt{FlipIt} game \cite{bowers2012defending,vanDijk2013Flip},
an attacker and a defender gain utility proportional to the amount
of time that they control a resource (here a cloud service), and pay
attack costs proportional to the number of times that they attempt
to claim or reclaim the resource. We consider a version of the game
in which the attacker and defender are restricted to attacking at
fixed frequencies. The equilibrium of the game is a Nash equilibrium. 

Let $v_{\mathcal{A}}^{i}\in\mathbb{R}$ and $v_{\mathcal{D}}^{i}\in\mathbb{R}$
denote the values of each cloud service $i\in\mathbb{S}$ to $\mathcal{A}^{i}$
and $\mathcal{D}^{i},$ respectively. These quantities represent the
inputs of the \texttt{FlipIt} game. The outputs of the \texttt{FlipIt}
game are the proportions of time for which $\mathcal{A}^{i}$ and
$\mathcal{D}^{i}$ control the cloud service. Denote these proportions
by $p_{\mathcal{A}}^{i}\in\left[0,1\right]$ and $p_{\mathcal{D}}^{i}=1-p_{\mathcal{A}}^{i},$
respectively. To summarize each of the \texttt{FlipIt} games, define
a set of mappings $T^{F_{i}}:\,\mathbb{R}\times\mathbb{R}\to\left[0,1\right],$
$i\in\mathbb{S},$ such that 
\begin{equation}
p_{\mathcal{A}}^{i*}=T^{F_{i}}\left(v_{\mathcal{A}}^{i},v_{\mathcal{D}}^{i}\right)\label{eq:TfIntro}
\end{equation}
maps the values of cloud service $i$ for $\mathcal{A}^{i}$ and $\mathcal{D}^{i}$
to the proportion of time $p_{\mathcal{A}}^{i*}$ for which the service
will be compromised in equilibrium. We will study this mapping further
in Section \ref{sec:cyber}.

\subsection{Communication Layer}

In the \emph{communication layer}, the cloud services $i\in\mathbb{S},$
which each may be controlled by $\mathcal{A}^{i}$ or $\mathcal{D}^{i},$
send data to a device $\mathcal{R},$ which decides whether to trust
the signals. This interaction is modeled by a signaling game. The
signaling game \emph{sender} is the cloud service. The two \emph{types}
of the sender are attacker or defender. The signaling game \emph{receiver}
is the device $\mathcal{R}.$ While we used $N$ \texttt{FlipIt}
games to describe the cloud layer, we use only one signaling game
to describe the communication layer, because $\mathcal{R}$ must decide
which services to trust all at once.

The prior probabilities in the communication layer are the equilibrium
proportions $p_{\mathcal{A}}^{i}$ and $p_{\mathcal{D}}^{i}=1-p_{\mathcal{A}}^{i},$
$i\in\mathbb{S}$ from the equilibrium of the cloud layer. Denote
the vectors of the prior probabilities for each sensor by $p_{\mathcal{A}}=\left[p_{\mathcal{A}}^{i}\right]_{i\in\mathbb{S}},$
$p_{\mathcal{D}}=\left[p_{\mathcal{D}}^{i}\right]_{i\in\mathbb{S}}.$
These prior probabilities are the inputs of the signaling game.

The outputs of the signaling game are the equilibrium utilities received
by the senders. Denote these utilities by $v_{\mathcal{A}}^{i}$ and
$v_{\mathcal{D}}^{i},$ $i\in\mathbb{S}.$ Importantly, these are
the same quantities that describe the incentives of $\mathcal{A}$
and $\mathcal{D}$ to control each cloud service in the \texttt{FlipIt}
game, because the party which controls each service is awarded the
opportunity to be the sender in the signaling game. Define vectors
to represent each of these utilities by $v_{\mathcal{A}}=\left[v_{\mathcal{A}}^{i}\right]_{i\in\mathbb{S}},$
$v_{\mathcal{D}}=\left[v_{\mathcal{D}}^{i}\right]_{i\in\mathbb{S}}.$ 

Finally, let $T^{S}:\,\left[0,1\right]^{N}\to\mathcal{P}(\mathbb{R}^{2N})$
be a mapping that summarizes the signaling game, where $\mathcal{P}(\mathbb{X})$
is the power set of $\mathbb{X}.$ According to this mapping, the
set of vectors of signaling game equilibrium utility ratios $v_{\mathcal{A}}^{*}$
and $v_{\mathcal{D}}^{*}$ that result from the vector of prior probabilities
$p_{\mathcal{A}}$ is given by 
\begin{equation}
\left(v_{\mathcal{A}}^{*},v_{\mathcal{D}}^{*}\right)=T^{S}\left(p_{\mathcal{A}}\right).\label{eq:TsIntro}
\end{equation}
This mapping summarizes the signaling game. We study the mapping in
detail in Section \ref{sec:comm}.

\subsection{Physical Layer}

Many IoCT devices such as pacemakers, cleaning robots, appliances,
and electric vehicles are dynamic systems that operate using feedback
mechanisms. The physical-layer control of these devices requires remote
sensing of the environment and the data stored or processed in the
cloud. The security at the cloud and the communication layers of the
system are intertwined with the performance of the controlled devices
at the physical layer. Therefore the trustworthiness of the data has
a direct impact on the control performance of the devices. This control
performance determines the utility of the device $\mathcal{R}$ as
well as the utility of each of the attackers $\mathcal{A}^{i}$ and
defenders $\mathcal{D}^{i}.$ The control performance is quantified
using a cost criterion for observer-based optimal feedback control.
The observer uses data from the cloud services that $\mathcal{R}$
elects to trust, and ignores the cloud services that $\mathcal{R}$
decides not to trust. We study the physical layer control in Section
\ref{sub:phy}.

\subsection{Coupling of the Cloud and Communication Layers}

\begin{figure}
\begin{centering}
\includegraphics[width=0.55\columnwidth]{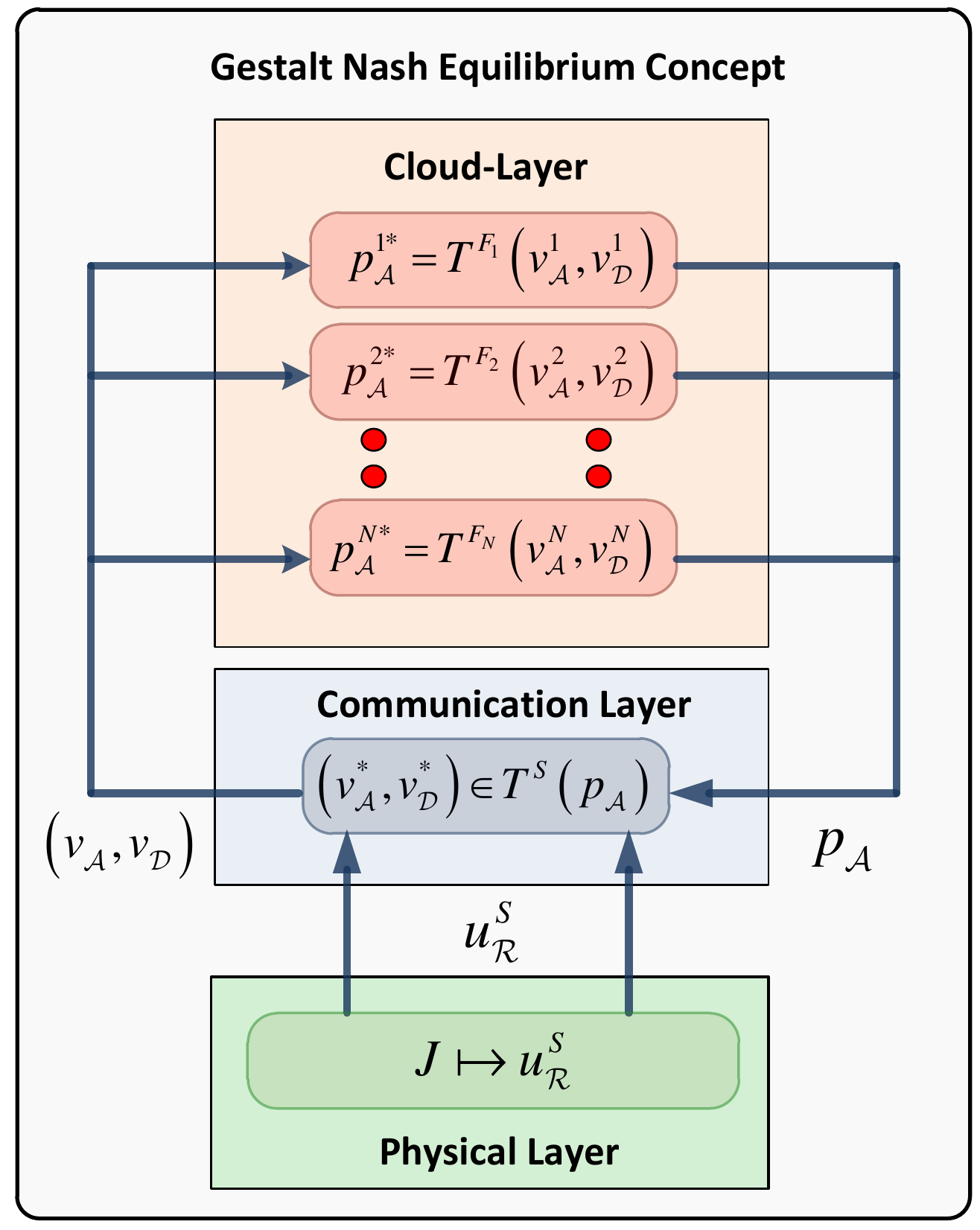} 
\par\end{centering}

\caption{\label{fig:layers}In iSTRICT, \texttt{FlipIt} games model attacks
on the set of cloud services. $T^{F_{i}},$ $i\in\mathbb{S},$ map
the value of each service to the proportion of time that it will be
compromised in equilibrium. The communication layer is modeled by
a signaling game. $T^{S}$ maps probabilities of compromise to the
value of each cloud service. The cloud layer and communication layer
are interdependent. The physical layer performance quantifies the
utilities for the signaling game.}
 
\end{figure}

Clearly, the cloud and communication layers are coupled through Eq.
(\ref{eq:TfIntro}) and Eq. (\ref{eq:TsIntro}). \textcolor{black}{The cloud layer security serves as an input to the communication layer. The resulting  utilities of the signaling game at the communication layer further becomes an input to the FlipIt game at the cloud layer. In addition, the physical layer performance quantifies the utilities for the signaling games.} Fig. \ref{fig:layers} depicts this concept. In order to predict
the behavior of the whole cloud-enabled IoCT, iSTRICT considers an
equilibrium concept which we call \textit{Gestalt Nash equilibrium}
(GNE). Informally, a triple $\left(p_{\mathcal{A}}^{\dagger},v_{\mathcal{A}}^{\dagger},v_{\mathcal{D}}^{\dagger}\right)$
is a GNE if it simultaneously satisfies Eq. \eqref{eq:TfIntro} and
Eq. \eqref{eq:TsIntro}.  

GNE is useful for three reasons. First, cloud-enabled IoCT networks
are dynamic. The modular structure of GNE requires the \texttt{FlipIt}
games and the signaling game to be at equilibrium given the parameters
that they receive from the other type of game. This imposes the requirement
of \emph{perfection}, in the sense that each game must be optimal
given the other game. In GNE, perfection applies in both directions,
because there is no clear chronological order or directional flow
of information between the two games. Actions in each sub-game must
be chosen by prior-commitment relative to the results of the other
sub-game.

Second, GNE draws upon established results from \texttt{FlipIt} games
and signaling games instead of attempting to analyze one large game.
IoCT networks promise plug-and-play capabilities, in which devices
and users are easily able to enter and leave the network. This also
motivates plug-and-play availability of solution concepts. The solution
to one sub-game should not need to be totally recomputed if an actor
enters or leaves another subgame. GNE follows this approach.

Finally, GNE serves as an example of a solution approach which could
be called \emph{game-of-games}. The equilibrium solutions to the \texttt{FlipIt}
games and signaling game must be rational ``best responses'' to
the solution of the other type of game.

\section{Detailed iSTRICT Model\label{sec:detailedModel}}

In this section, we define more precisely the three layers of the
iSTRICT framework.

\subsection{Cloud Layer: \texttt{FlipIt} Game\label{sec:cyber}}
We use a \texttt{FlipIt} game to model the interactions between the attacker and the defender over each cloud service. 

\subsubsection{\texttt{FlipIt} Actions}

For each service, $\mathcal{A}^{i}$ and $\mathcal{D}^{i}$ choose
$f_{\mathcal{A}}^{i}$ and $f_{\mathcal{D}}^{i},$ the frequencies
with which they claim or reclaim control of the service. These frequencies
are chosen by prior commitment. Neither player knows the other player's
action when she makes her choice. Figure \ref{fig:FlipIt} depicts
the \texttt{FlipIt} game. The green boxes above the horizontal axis
represent control of the service by $\mathcal{D}^{i}$ and the red
boxes below the axis represent control of the service by $\mathcal{A}^{i}.$

From $f_{\mathcal{A}}^{i}$ and $f_{\mathcal{D}}^{i},$ it is easy
to compute the expected proportions of the time that $\mathcal{A}$
and $\mathcal{D}$ control service $i$ \cite{bowers2012defending,vanDijk2013Flip}. 
Let $\mathbb{R}_{+}$ denote the set of non-negative real numbers.
Define the function $\rho:\,\mathbb{R}_{+}\times\mathbb{R}_{+}\to\left[0,1\right],$
such that $p_{\mathcal{A}}^{i}=\rho\left(f_{\mathcal{A}}^{i},f_{\mathcal{D}}^{i}\right)$
gives the proportion of the time that $\mathcal{A}^{i}$ will control
the cloud service if he attacks with frequency $f_{\mathcal{A}}^{i}$
and $\mathcal{D}^{i}$ renews control of the service (through changing
cryptographic keys or passwords, or through installing new hardware)
with frequency $f_{\mathcal{D}}^{i}.$  
We have 
\begin{equation}
\rho\left(f_{\mathcal{A}}^{i},f_{\mathcal{D}}^{i}\right)=\begin{cases}
0, & \text{ if }f_{\mathcal{A}}^{i}=0,\\
\frac{f_{\mathcal{A}}^{i}}{2f_{\mathcal{D}}^{i}}, & \text{ if }f_{\mathcal{D}}^{i}\geq f_{\mathcal{A}}^{i}>0,\\
1-\frac{f_{\mathcal{D}}^{i}}{2f_{\mathcal{A}}^{i}}, & \text{ if }f_{\mathcal{A}}^{i}>f_{\mathcal{D}}^{i}\geq0.
\end{cases}\label{eq:pA}
\end{equation}

\textcolor{black}{ Notice that when $f_{\mathcal{A}}^{i}>f_{\mathcal{D}}^{i}\geq 0$, i.e., the attacking frequency of $\mathcal{A}^i$ is greater than the renewal frequency of $\mathcal{D}^i$, the proportion of time that service $i$ is insecure is $\rho\left(f_{\mathcal{A}}^{i},f_{\mathcal{D}}^{i}\right)>\frac{1}{2}$, and when $f_{\mathcal{D}}^{i}\geq f_{\mathcal{A}}^{i}>0$, we obtain $\rho\left(f_{\mathcal{A}}^{i},f_{\mathcal{D}}^{i}\right)\leq \frac{1}{2}$.}

%

\subsubsection{\texttt{FlipIt} Utility Functions}

\begin{figure}
\begin{centering}
\includegraphics[width=0.65\columnwidth]{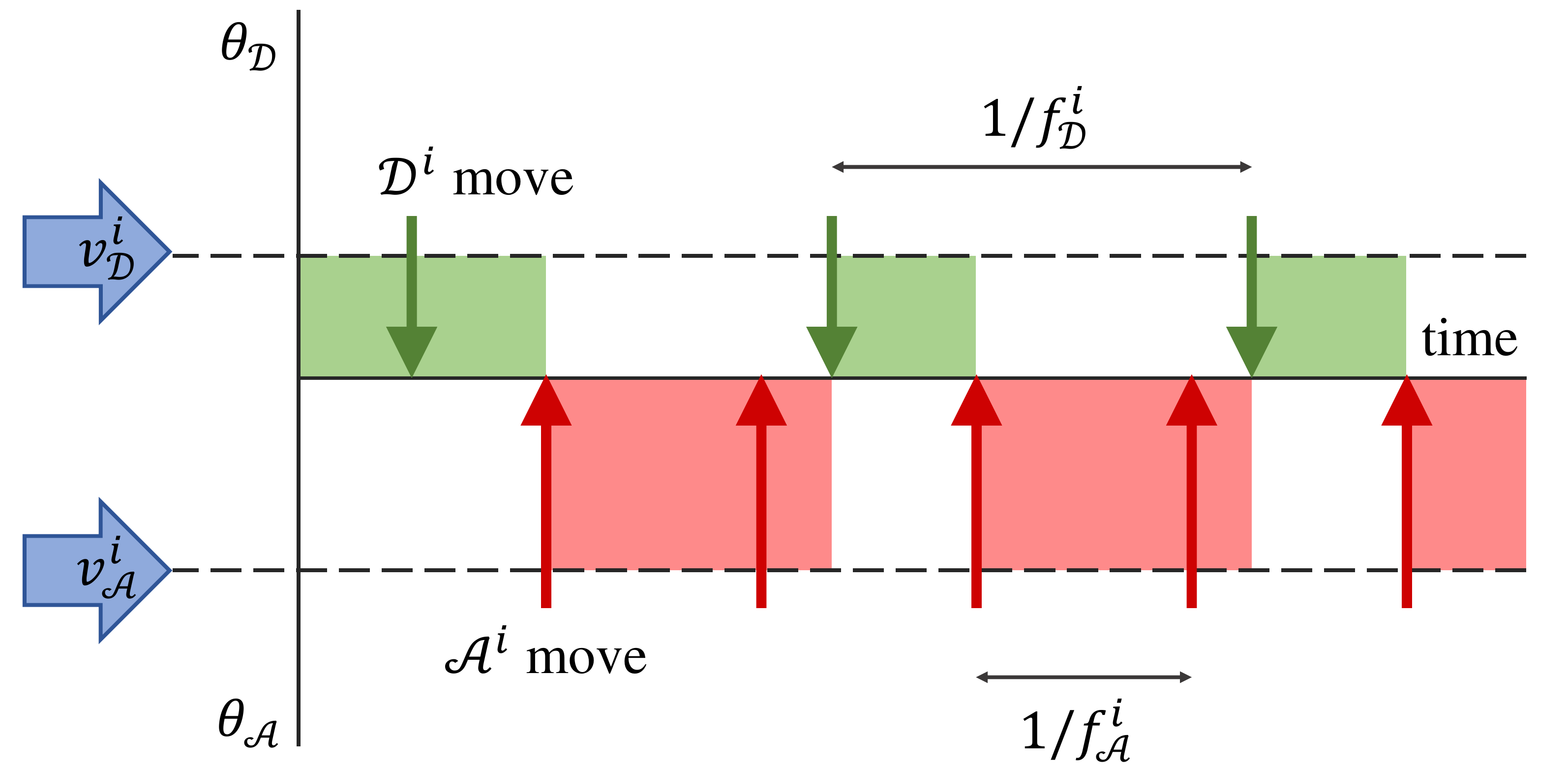} 
\par\end{centering}

\caption{\label{fig:FlipIt}In each \texttt{FlipIt} game, $\mathcal{A}^{i}$
and $\mathcal{D}^{i}$ periodically claim control of cloud service
$i.$ The values of the service for each player are given by $v_{\mathcal{A}}^{i}$
and $v_{\mathcal{D}}^{i},$ which depend on the equilibrium of the
signaling game.}
 
\end{figure}

Recall that $v_{\mathcal{A}}^{i}$ and $v_{\mathcal{D}}^{i}$ denote
the value of controlling service $i\in\mathbb{S}$ for $\mathcal{A}^{i}$
and $\mathcal{D}^{i},$ respectively. These quantities define the
heights of the red and green boxes in Fig. \ref{fig:FlipIt}. Denote
the costs of renewing control of the cloud service for the two players
by $\alpha_{\mathcal{A}}^{i}$ and $\alpha_{\mathcal{D}}^{i}.$ Finally,
let $\bar{u}_{\mathcal{A}}^{F_{i}}:\,\mathbb{R}_{+}\times\mathbb{R}_{+}\to\mathbb{R}$
and $\bar{u}_{\mathcal{D}}^{F_{i}}:\,\mathbb{R}_{+}\times\mathbb{R}_{+}\to\mathbb{R}$
be expected utility functions for each \texttt{FlipIt} game. The
utilities of each player are given in Eq. \eqref{eq:expUtilDefFlipIt}
and Eq. \eqref{eq:expUtilAtkFlipIt} by the values $v_{\mathcal{D}}^{i}$
and $v_{\mathcal{A}}^{i}$ of controlling the service multiplied by
the proportions $p_{\mathcal{D}}^{i}$ and $p_{\mathcal{A}}^{i}$
with which the service is controlled, minus the costs $\alpha_{\mathcal{D}}^{i}$
and $\alpha_{\mathcal{A}}^{i}$ of attempting to renew control of
the service.
\begin{equation}
\bar{u}_{\mathcal{D}}^{F_{i}}\left(f_{\mathcal{A}}^{i},f_{\mathcal{D}}^{i}\right)=v_{\mathcal{D}}^{i}\left(1-\rho\left(f_{\mathcal{A}}^{i},f_{\mathcal{D}}^{i}\right)\right)-\alpha_{\mathcal{D}}^{i}f_{\mathcal{D}}^{i}.\label{eq:expUtilDefFlipIt}
\end{equation}
\begin{equation}
\bar{u}_{\mathcal{A}}^{F_{i}}\left(f_{\mathcal{A}}^{i},f_{\mathcal{D}}^{i}\right)=v_{\mathcal{A}}^{i}\rho\left(f_{\mathcal{A}}^{i},f_{\mathcal{D}}^{i}\right)-\alpha_{\mathcal{A}}^{i}f_{\mathcal{A}}^{i}.\label{eq:expUtilAtkFlipIt}
\end{equation}
\textcolor{black}{Therefore, based on the attacker's action $f_{\mathcal{A}}^{i}$, the defender determines $f_{\mathcal{D}}^{i}$ strategically to maximize the proportional time of controlling the cloud service $i$, $1-\rho(f_{\mathcal{A}}^{i},f_{\mathcal{D}}^{i})$, and minimize the cost of choosing $f_{\mathcal{D}}^{i}$. }

\textcolor{black}{Note that in the game, the attacker knows $v_{\mathcal{A}}^{i}$ and $\alpha_{\mathcal{A}}^{i}$, and the defender knows $v_{\mathcal{D}}^{i}$ and $\alpha_{\mathcal{D}}^{i}$. Furthermore, $\rho(f_{\mathcal{A}}^{i},f_{\mathcal{D}}^{i})$ is public information, and hence both players know the frequencies of control of the cloud through \eqref{eq:pA}.  Therefore, the communication between two players at the cloud layer is not necessary when determining their strategies.}

\subsubsection{\texttt{FlipIt} Equilibrium Concept}

The equilibrium concept for the \texttt{FlipIt} game is Nash equilibrium,
since it is a complete information game in which strategies are chosen
by prior commitment. 
\begin{defn}
\label{def:NEflip}(\emph{Nash Equilibrium}) A Nash equilibrium of
the \texttt{FlipIt} game played for control of service $i\in\left\{ 1,\ldots,N\right\} $
is a strategy profile $\left(f_{\mathcal{A}}^{i*},f_{\mathcal{D}}^{i*}\right)$
such that 
\begin{equation}
f_{\mathcal{D}}^{i*}\in\underset{f_{\mathcal{D}}^{i}\in\mathbb{R}_{+}}{\arg\max}\:\bar{u}_{\mathcal{D}}^{F_{i}}\left(f_{\mathcal{A}}^{i*},f_{\mathcal{D}}^{i*}\right),\label{defender_p}
\end{equation}
\begin{equation}
f_{\mathcal{A}}^{i*}\in\underset{f_{\mathcal{A}}^{i}\in\mathbb{R}_{+}}{\arg\max}\:\bar{u}_{\mathcal{A}}^{F_{i}}\left(f_{\mathcal{A}}^{i*},f_{\mathcal{D}}^{i*}\right),\label{attacker_p}
\end{equation}
where $\bar{u}_{\mathcal{D}}^{F_{i}}$ and $\bar{u}_{\mathcal{A}}^{F_{i}}$
are computed by Eq. \eqref{eq:expUtilDefFlipIt} and Eq. \eqref{eq:expUtilAtkFlipIt}. 
\end{defn}
From the equilibrium frequencies $f_{\mathcal{D}}^{i*}$ and $f_{\mathcal{A}}^{i*},$
let the equilibrium proportion of time that $\mathcal{A}^{i}$ controls
cloud service $i$ be given by $p_{\mathcal{A}}^{i*}$ according to
Eq. \eqref{eq:pA}. The Nash equilibrium solution can then be used
to determine the mapping in Eq. \eqref{eq:TfIntro} from the cloud
service values $v_{\mathcal{A}}^{i}$ and $v_{\mathcal{D}}^{i}$ to
the equilibrium attacker control proportion $p_{\mathcal{A}}^{i*}$,
where $T^{F_{i}}:\,\mathbb{R}\times\mathbb{R}\to\left[0,1\right]$.
The $T^{F_{i}}$ mappings, $i\in\mathbb{S},$ constitute the top layer
of Fig. \ref{fig:layers}.

\subsection{Communication Layer: Signaling Game\label{sec:comm}}

\begin{figure}
\begin{centering}
\includegraphics[width=0.65\columnwidth]{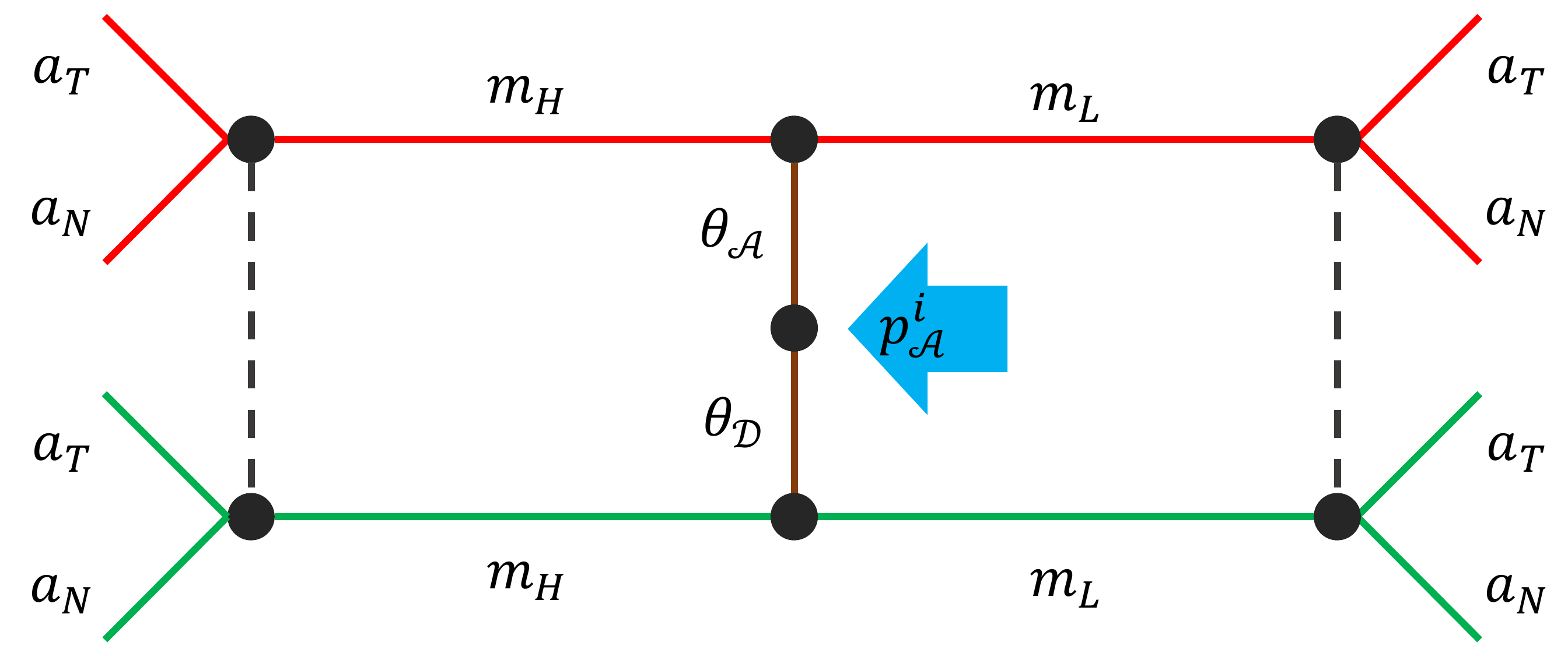} 
\par\end{centering}

\caption{\label{fig:signaling}The vector of types $\theta\in\Theta^{N}$ defines
whether each cloud service $i\in\mathbb{S}$ is controlled by an attacker
or defender. Each prior probability $p_{\mathcal{A}}^{i}$ comes from
the corresponding \texttt{FlipIt} game. The player who controls each
service chooses $m^{i}.$ $\mathcal{R}$ observes all $m^{i}$ and
chooses $a^{i},$ $i\in\mathbb{S}$ simultaneously. Here, we show
one service, although all of the services are coupled. }
 
\end{figure}

Because the cloud services are vulnerable, devices which depend on
data from the services should rationally decide whether to trust them.
This is captured using a signaling game. In this model, the device
$\mathcal{R}$ updates a belief about the state of each cloud service
and decides whether to trust it. Figure \ref{fig:signaling} depicts the actions that correspond to
one service of the signaling game. \textcolor{black}{Compared to the trust value-based cloud trust management system where the reputation attack can significantly influence the trust decision \cite{siadat2017identifying,noor2013reputation}, in iSTRICT, $\mathcal{R}$'s decision is based on the strategies of each $\mathcal{A}^{i}$ and $\mathcal{D}^{i}$
at the cloud layer as well as the physical layer performance, and hence it does not depend on the feedback of cloud services from users which could be malicious due to attacks.} We next present the detailed model of signaling game.

\subsubsection{Signaling Game Types}

The \emph{types} of each cloud service $i\in\mathbb{S}$ are $\theta^{i}\in\Theta=\left\{ \theta_{\mathcal{A}},\theta_{\mathcal{D}}\right\} ,$
where $\theta^{i}=\theta_{\mathcal{A}}$ indicates that the service
is compromised by $\mathcal{A}^{i}$, and $\theta=\theta_{\mathcal{D}}$
indicates that the service is controlled by $\mathcal{D}^{i}$. Denote
the vector of all the service types by $\theta=\left[\theta^{i}\right]_{i\in\mathbb{S}}\triangleq\left[\begin{array}{cccc}
\theta^{1} & \theta^{2} & \ldots & \theta^{m}\end{array}\right]^{T}.$

\subsubsection{Signaling Game Messages}

Denote the risk level of the data from each service $i$ by $m^{i}\in M=\left\{ m_{L},m_{H}\right\} ,$
where $m_{L}$ and $m_{H}$ indicate low-risk and high-risk messages,
respectively. (We define this risk level in Section \ref{sub:phy}.)
Further, define the vector of all of the risk levels by $m=\left[m^{i}\right]_{i\in\mathbb{S}}.$ 

Next, define mixed strategies for $\mathcal{A}^{i}$ and $\mathcal{D}^{i}.$
Let $\sigma_{\mathcal{A}}^{i}:\,M\to\left[0,1\right]$ and $\sigma_{\mathcal{D}}^{i}:\,M\to\left[0,1\right]$
be functions such that $\sigma_{\mathcal{A}}^{i}\left(m_{\mathcal{A}}^{i}\right)\in\Sigma_{\mathcal{A}}$
and $\sigma_{\mathcal{D}}^{i}\left(m_{\mathcal{D}}^{i}\right)\in\Sigma_{\mathcal{D}}$
give the proportions with which $\mathcal{A}^{i}$ and $\mathcal{D}^{i}$
send messages with risk levels $m_{\mathcal{A}}^{i}$ and $m_{\mathcal{D}}^{i},$
respectively, from each cloud service $i$ that they control. Note
that $\mathcal{R}$ only observes $m_{\mathcal{A}}^{i}$ or $m_{\mathcal{D}}^{i},$
depending on who controls the service $i.$ Let 
\[
m^{i}=\begin{cases}
m_{\mathcal{A}}^{i},\quad\text{if} & \theta^{i}=\theta_{\mathcal{A}}\\
m_{\mathcal{D}}^{i},\quad\text{if} & \theta^{i}=\theta_{\mathcal{D}}
\end{cases},
\]
denote risk level of the message that $\mathcal{R}$ actually observes.
Finally, define the vector of observed risk levels by $m=\left[m^{i}\right]_{i\in\mathbb{S}}.$

\subsubsection{Signaling Game Beliefs and Actions}

Based on the risk levels $m$ that $\mathcal{R}$ observes, it updates
its vector of prior beliefs $p_{\mathcal{A}}.$ Define $\mu^{i}:\,\Theta\to\left[0,1\right],$
such that $\mu^{i}\left(\theta\,|\,m^{i}\right)$ gives the belief
of $\mathcal{R}$ that service $i\in\mathbb{S}$ is of type $\theta$
given that $\mathcal{R}$ observes risk level $m^{i}.$ Also write
the vector of beliefs as $\mu\left(\theta\,|\,m\right)=\left[\mu^{i}\left(\theta^{i}\,|\,m^{i}\right)\right]_{i\in\mathbb{S}}.$ \textcolor{black}{As a direction for future work, we note that evidence-based signaling game approaches could be used to update belief in a manner robust to reputation attacks \cite{pawlick2015deception,pawlick2017imperfect,pawlick2018leakyWEIS}.}

Based on these beliefs, $\mathcal{R}$ chooses which cloud services
to trust. For each service $i,$ $\mathcal{R}$ chooses $a^{i}\in A=\left\{ a_{T},a_{N}\right\} $
where $a_{T}$ denotes trusting the service (\emph{i.e.}, using it
for observer-based optimal feedback control) and $a_{N}$ denotes
not trusting the service. Assume that $\mathcal{R},$ aware of the
system dynamics, chooses actions for each service simultaneously,
\emph{i.e.}, $a=\left[a^{i}\right]_{i\in\mathbb{S}}.$ 

Next, define $\sigma_{\mathcal{R}}:\,A^{N}\to\left[0,1\right]$ such
that $\sigma_{\mathcal{R}}\left(a\,|\,m\right)\in\Sigma_{\mathcal{R}}^{N}$
gives the mixed strategy probability with which $\mathcal{R}$ plays
the vector of actions $a$ given the vector of risk levels $m.$

\subsubsection{Signaling Game Utility Functions}

Let $\mathcal{R}$'s utility function be denoted by $u_{\mathcal{R}}^{S}:\,\Theta^{N}\times M^{N}\times A^{N}\to\mathbb{R},$
such that $u_{\mathcal{R}}^{S}\left(\theta,m,a\right)$ gives the
utility that $\mathcal{R}$ receives when $\theta$ is the vector
of cloud service types, $m$ is the vector of risk levels, and $\mathcal{R}$
chooses the vector of actions $a.$

For $i\in\mathbb{S},$ define the functions $u_{\mathcal{A}}^{S_{i}}:\,M^{N}\times A^{N}\to\mathbb{R}$
and $u_{\mathcal{D}}^{S_{i}}:\,M^{N}\times A^{N}\to\mathbb{R},$ such
that $u_{\mathcal{A}}^{S_{i}}\left(m,a\right)$ and $u_{\mathcal{D}}^{S_{i}}\left(m,a\right)$
give the utility that $\mathcal{A}^{i}$ and $\mathcal{D}^{i}$ receive
for service $i$ when the risk levels are given by the vector $m,$
and $\mathcal{R}$ plays the vector of actions $a.$ 

Next, consider expected utilities based on the strategies of each
player. Let $\bar{u}_{\mathcal{R}}^{S}:\,\Sigma_{\mathcal{R}}^{N}\to\mathbb{R}$
denote the expected utility function for $\mathcal{R},$ such that
$\bar{u}_{\mathcal{R}}^{S}\left(\sigma_{\mathcal{R}}\,|\,m,\mu\left(\bullet\,|\,m\right)\right)$
gives $\mathcal{R}$'s expected utility when he plays mixed strategy
$\sigma_{\mathcal{R}}$ given that he observes risk levels $m$ and
has belief $\mu.$ We have 
\begin{equation}
\bar{u}_{\mathcal{R}}^{S}\left(\sigma_{\mathcal{R}}\,|\,m,\mu\right)=\underset{\theta\in\Theta^{m}}{\sum}\underset{a\in A^{m}}{\sum}u_{\mathcal{R}}^{S}\left(\theta,m,a\right)\mu\left(\theta\,|\,m\right)\sigma_{\mathcal{R}}\left(a\,|\,m\right).\label{eq:expUtilR}
\end{equation}

In order to compute the expected utility functions for $\mathcal{A}^{i}$
and $\mathcal{D}^{i},$ define $\sigma_{\mathcal{A}}^{-i}=\left\{ \sigma_{\mathcal{A}}^{j}\,|\,j\in\mathbb{S}\backslash\{i\}\right\} $
and $\sigma_{\mathcal{D}}^{-i}=\left\{ \sigma_{\mathcal{D}}^{j}\,|\,j\in\mathbb{S}\backslash\{i\}\right\} ,$
the sets of the strategies of all of the senders except the sender
on cloud service $i.$ Then define $\bar{u}_{\mathcal{A}}^{S_{i}}:\,\Sigma_{\mathcal{R}}^{N}\times\Sigma_{\mathcal{A}}^{N}\times\Sigma_{\mathcal{D}}^{N-1}\to\mathbb{R}$
such that $\bar{u}_{\mathcal{A}}^{S_{i}}\left(\sigma_{\mathcal{R}};\sigma_{\mathcal{A}}^{i},\sigma_{\mathcal{A}}^{-i};\sigma_{\mathcal{D}}^{-i}\right)$
gives the expected utility to $\mathcal{A}^{i}$ when he plays mixed
strategy $\sigma_{\mathcal{A}}^{i},$ and the attackers and defenders
on the other services play $\sigma_{\mathcal{A}}^{-i}$ and $\sigma_{\mathcal{D}}^{-i}.$
Define the expected utility to $\mathcal{D}^{i}$ by $\bar{u}_{\mathcal{D}}^{S_{i}}\left(\sigma_{\mathcal{R}};\sigma_{\mathcal{A}}^{-i};\sigma_{\mathcal{D}}^{i},\sigma_{\mathcal{D}}^{-i}\right)$
in a similar manner.

Let $\mathcal{X}^{i}\in\left\{ \mathcal{A},\mathcal{D}\right\} $
denote the player that controls service $i$ and $\mathcal{X}\in\left\{ \mathcal{A},\mathcal{D}\right\} ^{N}$
denote the set of players that control each service. Then the expected
utilities are computed by

\begin{multline}
\bar{u}_{\mathcal{A}}^{S_{i}}\left(\sigma_{\mathcal{R}};\sigma_{\mathcal{A}}^{i},\sigma_{\mathcal{A}}^{-i};\sigma_{\mathcal{D}}^{-i}\right)=\underset{m\in M^{N}}{\sum}\underset{a\in A^{N}}{\sum}\\
\underset{\mathcal{X}^{1}\in\left\{ \mathcal{A},\mathcal{D}\right\} }{\sum}\ldots\underset{\mathcal{X}^{i-1}\in\left\{ \mathcal{A},\mathcal{D}\right\} }{\sum}\underset{\mathcal{X}^{i+1}\in\left\{ \mathcal{A},\mathcal{D}\right\} }{\sum}\ldots\underset{\mathcal{X}^{N}\in\left\{ \mathcal{A},\mathcal{D}\right\} }{\sum}\\
u_{\mathcal{A}}^{S_{i}}\left(m,a\right)\sigma_{\mathcal{R}}\left(a\,|\,m\right)\sigma_{\mathcal{A}}^{i}\left(m^{i}\right)\underset{j\in\mathbb{S}\backslash\{i\}}{\prod}\sigma_{\mathcal{X}^{j}}^{j}\left(m^{j}\right)p_{\mathcal{X}^{j}}^{j},\label{eq:expUtilSigA}
\end{multline}
\begin{multline}
\bar{u}_{\mathcal{D}}^{S_{i}}\left(\sigma_{\mathcal{R}};\sigma_{\mathcal{A}}^{-i};\sigma_{\mathcal{D}}^{i},\sigma_{\mathcal{D}}^{-i}\right)=\underset{m\in M^{N}}{\sum}\underset{a\in A^{N}}{\sum}\\
\underset{\mathcal{X}^{1}\in\left\{ \mathcal{A},\mathcal{D}\right\} }{\sum}\ldots\underset{\mathcal{X}^{i-1}\in\left\{ \mathcal{A},\mathcal{D}\right\} }{\sum}\underset{\mathcal{X}^{i+1}\in\left\{ \mathcal{A},\mathcal{D}\right\} }{\sum}\ldots\underset{\mathcal{X}^{N}\in\left\{ \mathcal{A},\mathcal{D}\right\} }{\sum}\\
u_{\mathcal{D}}^{S_{i}}\left(m,a\right)\sigma_{\mathcal{R}}\left(a\,|\,m\right)\sigma_{\mathcal{D}}^{i}\left(m^{i}\right)\underset{j\in\mathbb{S}\backslash\{i\}}{\prod}\sigma_{\mathcal{X}^{j}}^{j}\left(m^{j}\right)p_{\mathcal{X}^{j}}^{j}.\label{eq:expUtilSigD}
\end{multline}

\subsubsection{Perfect Bayesian Nash Equilibrium Conditions}

Finally, we can state the requirements for a perfect Bayesian Nash
equilibrium (PBNE) for the signaling game \cite{fudenberg1991game}. 
\begin{defn}
\label{def:PBNEsig}(\emph{PBNE}) For the device, let $\bar{u}_{\mathcal{R}}^{S}\left(\sigma_{\mathcal{R}}\,|\,m,\mu\right)$
be formulated according to Eq. \eqref{eq:expUtilR}. For each service
$i\in\mathbb{S},$ let $\bar{u}_{\mathcal{A}}^{S_{i}}\left(\sigma_{\mathcal{R}};\sigma_{\mathcal{A}}^{i},\sigma_{\mathcal{A}}^{-i};\sigma_{\mathcal{D}}^{-i}\right)$
be given by Eq. \eqref{eq:expUtilSigA} and $\bar{u}_{\mathcal{D}}^{S_{i}}\left(\sigma_{\mathcal{R}};\sigma_{\mathcal{A}}^{-i};\sigma_{\mathcal{D}}^{i},\sigma_{\mathcal{D}}^{-i}\right)$
be given by Eq. \eqref{eq:expUtilSigD}. Finally, let vector $p_{\mathcal{A}}$
give the prior probabilities of each service being compromised. Then,
a \emph{perfect Bayesian Nash equilibrium} of the signaling game is
a strategy profile $\left(\sigma_{\mathcal{R}}^{*};\sigma_{\mathcal{A}}^{1*},\ldots\sigma_{\mathcal{A}}^{N*};\sigma_{\mathcal{D}}^{1*},\ldots\sigma_{\mathcal{D}}^{N*}\right)$
and a vector of beliefs $\mu\left(\theta\,|\,m\right)$ such that
the following hold: 
\begin{equation}
\forall i\in\mathbb{S},\,\sigma_{\mathcal{A}}^{i*}\left(\bullet\right)\in\underset{\sigma_{\mathcal{A}}^{i}\in\Sigma_{\mathcal{A}}}{\arg\max\:}\bar{u}_{\mathcal{A}}^{S_{i}}\left(\sigma_{\mathcal{R}}^{*};\sigma_{\mathcal{A}}^{i},\sigma_{\mathcal{A}}^{-i*};\sigma_{\mathcal{D}}^{-i*}\right),\label{sigma_a}
\end{equation}
\begin{equation}
\forall i\in\mathbb{S},\,\sigma_{\mathcal{D}}^{i*}\left(\bullet\right)\in\underset{\sigma_{\mathcal{D}}^{i}\in\Sigma_{\mathcal{D}}}{\arg\max\:}\bar{u}_{\mathcal{D}}^{S_{i}}\left(\sigma_{\mathcal{R}}^{*};\sigma_{\mathcal{A}}^{-i*};\sigma_{\mathcal{D}}^{i},\sigma_{\mathcal{D}}^{-i*}\right),\label{sigma_d}
\end{equation}
\begin{equation}
\forall m\in M,\:\sigma_{\mathcal{R}}^{*}\in\underset{\sigma_{\mathcal{R}}\in\Sigma_{\mathcal{R}}^{m}}{\arg\max\:}\bar{u}_{\mathcal{R}}^{S}\left(\sigma_{\mathcal{R}}\,|\,m,\mu\left(\bullet\,|\,m\right)\right),\label{sigma_r}
\end{equation}
and $\forall i\in\mathbb{S},$ 
\begin{equation}
\mu^{i}\left(\theta_{\mathcal{A}}\,|\,m^{i}\right)=\frac{\sigma_{\mathcal{A}}^{i*}\left(m^{i}\right)p_{\mathcal{A}}^{i}}{\sigma_{\mathcal{A}}^{i*}\left(m^{i}\right)p_{\mathcal{A}}^{i}+\sigma_{\mathcal{D}}^{i*}\left(m^{i}\right)\left(1-p_{\mathcal{A}}^{i}\right)},\label{belief_a}
\end{equation}
if $\sigma_{\mathcal{A}}^{i*}\left(m^{i}\right)p_{\mathcal{A}}^{i}+\sigma_{\mathcal{D}}^{i*}\left(m^{i}\right)p_{\mathcal{D}}^{i}\neq0,$
and $\mu^{i}\left(\theta_{\mathcal{A}}\,|\,m^{i}\right)\in\left[0,1\right],$
if $\sigma_{\mathcal{A}}^{i*}\left(m^{i}\right)p_{\mathcal{A}}^{i}+\sigma_{\mathcal{D}}^{i*}\left(m^{i}\right)p_{\mathcal{D}}^{i}=0.$
Additionally, $\mu^{i}\left(\theta_{\mathcal{D}}\,|\,m^{i}\right)=1-\mu^{i}\left(\theta_{\mathcal{A}}\,|\,m^{i}\right)$
in both cases. 
\end{defn}
Note that we have denoted the equilibrium utilities for $\mathcal{A}^{i}$
and $\mathcal{D}^{i},$ $i\in\mathbb{S}$ by 
\begin{equation}
v_{\mathcal{A}}^{i}=\bar{u}_{\mathcal{A}}^{S_{i}}\left(\sigma_{\mathcal{R}}^{*};\sigma_{\mathcal{A}}^{i*},\sigma_{\mathcal{A}}^{-i*};\sigma_{\mathcal{D}}^{-i*}\right),\label{va}
\end{equation}
\begin{equation}
v_{\mathcal{D}}^{i}=\bar{u}_{\mathcal{D}}^{S_{i}}\left(\sigma_{\mathcal{R}}^{*};\sigma_{\mathcal{A}}^{-i*};\sigma_{\mathcal{D}}^{i*},\sigma_{\mathcal{D}}^{-i*}\right),\label{vd}
\end{equation}
and the vectors of those values by $v_{\mathcal{A}}=\left[v_{\mathcal{A}}^{i}\right]_{i\in\mathbb{S}},$
$v_{\mathcal{D}}=\left[v_{\mathcal{D}}^{i}\right]_{i\in\mathbb{S}}.$
We now have the complete description of the signaling game mapping
Eq. \eqref{eq:TsIntro}, where $T^{S}:\,\left[0,1\right]^{N}\to\mathcal{P}(\mathbb{R}^{2N}).$
This mapping constitutes the middle layer of Fig. \ref{fig:layers}.

\subsection{Physical Layer: Optimal Control\label{sub:phy}}

\begin{figure*}[!t]
\centering{}\includegraphics[width=0.6\textwidth]{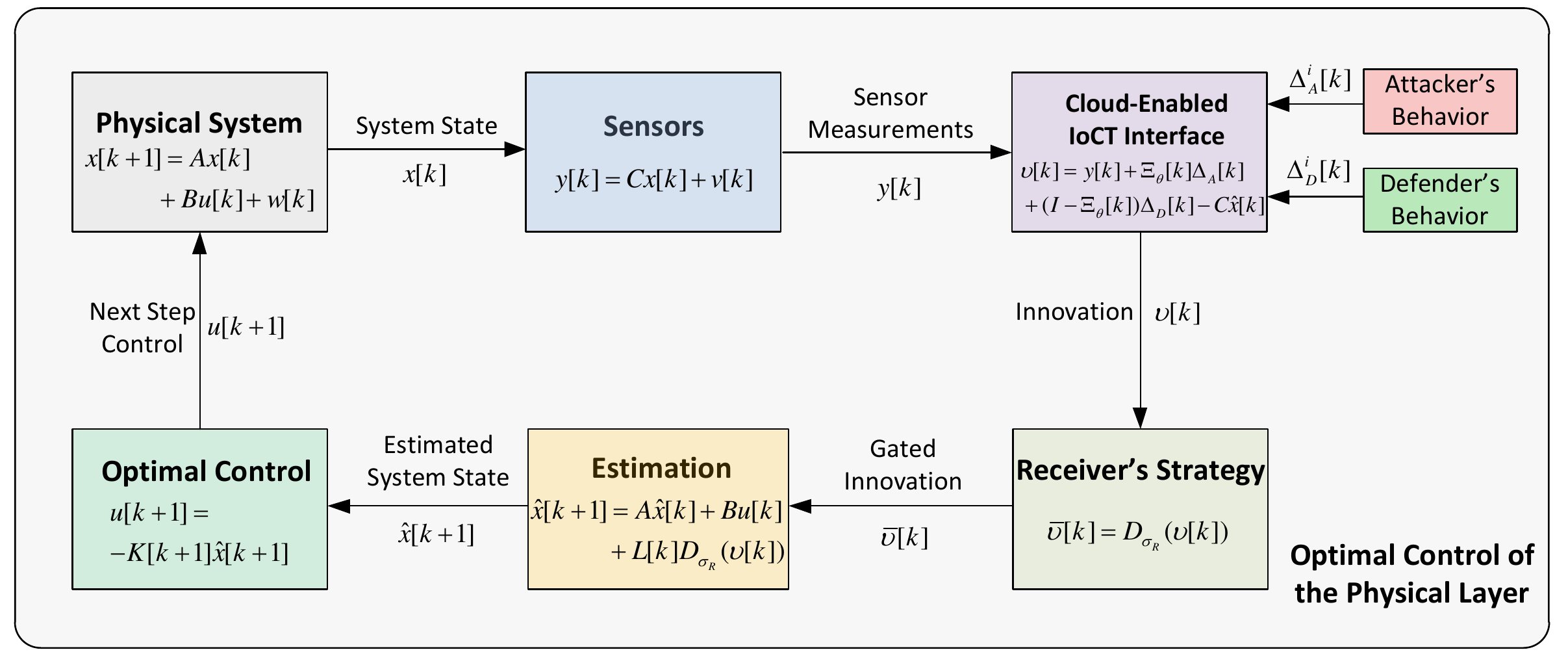}
\caption{A block diagram shows the various components of the control system
in the iSTRICT. \textcolor{black}{The physical system refers to IoCT device whose states are collected by sensors.}
Each $\mathcal{A}^{i}$ and $\mathcal{D}^{i}$ in the cloud layer
can add bias terms to the measured senor data before sending it to
the receiver. $\mathcal{R}$ decides whether to trust or not trust
each of the cloud services, and then designs an optimal control for
the physical system. \textcolor{black}{Since the optimal control is designed over a finite-horizon cost criterion, the loop terminates after $T$ time steps.}}
\label{control_system} 
\end{figure*}

The utility function $u_{\mathcal{R}}^{S}\left(\theta,m,a\right)$
is determined by the performance of the device controller as shown
in Fig. \ref{fig:layers}. A block illustration of the control system
is shown in Fig. \ref{control_system}. \textcolor{black}{Note that the physical system in the diagram refers to the IoCT devices.}

\subsubsection{Device Dynamics}

Each device in the IoCT is governed by dynamics. We can capture the
dynamics of the things by the linear system model
\begin{equation}
x[k+1]=Ax[k]+Bu[k]+w[k],\label{eq:state}
\end{equation}
where $A\in\mathbb{R}^{n\times n}$, $B\in\mathbb{R}^{n\times q}$,
$x[k]\in\mathbb{R}^{n}$ is the system state, $u[k]\in\mathbb{R}^{q}$
is the control input, $w[k]$ denotes the system white noise, and
$x[0]=x_{0}\in\mathbb{R}^{n}$ is given. Let $y[k]\in\mathbb{R}^{N}$
represent data from cloud services which suffers from white, additive
Gaussian sensor noise given by the vector $v[k].$ We have $y[k]=Cx[k]+v[k],$
where $C\in\mathbb{R}^{N\times n}$ is the output matrix. Let the
system and sensor noise processes have known covariance matrices $\mathbb{E}\left\{ w[k]w'[k]\right\} =\xi,\;\mathbb{E}\left\{ v[k]v'[k]\right\} =\zeta,$
where $\xi$ and $\zeta$ are symmetric, positive, semi-definite matrices,
and $w'[k]$ and $v'[k]$ denote the transposes of the noise vectors.

In addition, for each cloud service $i\in\mathbb{S},$ the attacker
$\mathcal{A}^{i}$ and defender $\mathcal{D}^{i}$ in the signaling
game choose whether to add bias terms to the measurement $y^{i}[k].$
Let $\Delta_{\mathcal{A}}^{i}[k],\Delta_{\mathcal{D}}^{i}[k]\in\mathbb{R}$
denote these bias terms. The actual noise levels that $\mathcal{R}$
observes depends on who controls the service in the \texttt{FlipIt}
game. Recall that the vector of types of each service is given by
$\theta=\left[\theta^{i}\right]_{i\in\mathbb{S}}.$ Let $\mathbf{1}_{\left\{ \bullet\right\} }$
represent the indicator function, which takes the value of $1$ if
its argument is true and $0$ otherwise. Then, define the matrix 
\[
\Xi_{\theta}=\text{diag}\left\{ \left[\begin{array}{ccc}
\mathbf{1}_{\left\{ \theta^{1}=\theta_{\mathcal{A}}\right\} } & \ldots & \mathbf{1}_{\left\{ \theta^{N}=\theta_{\mathcal{A}}\right\} }\end{array}\right]\right\} .
\]
Including the bias term, the measurements are given by
\begin{equation}
    \tilde{y}[k]=Cx[k]+v[k]+\Xi_{\theta}[k]\Delta_{\mathcal{A}}[k]+\left(I-\Xi_{\theta}[k]\right)\Delta_{\mathcal{D}}[k],
\end{equation}
where $I$ is the $N$-dimensional identity matrix.

\subsubsection{Observer-Based Optimal Feedback Control}

Let $F,$ $Q,$ and $R$ be positive-definite matrices of dimensions
$n\times n,$ $n\times n,$ and $q\times q,$ respectively. The device
chooses the control $u$ that minimizes the operational cost given
by 
\begin{equation}
J=\mathbb{E}\left\{ x'[T]Fx[T]+\sum_{k=0}^{T-1}x'[k]Qx[k]+u'[k]Ru[k]\right\} ,\label{eq:criterion}
\end{equation}
subject to the dynamics of Eq. \eqref{eq:state}.

To attempt to minimize Eq. \eqref{eq:criterion}, the device uses
\emph{observer-based optimal feedback control}. Define $P[k]$ by
the forward Riccati difference equation
\begin{multline*}
P[k+1]=A\Bigl(P[k]-P[k]C'\\
\left(CP[k]C'+\xi\right)^{-1}CP[k]\Bigr)A'+\zeta,
\end{multline*}
with $P[0]=\mathbb{E}\{(x[0]-\hat{x}[0])(x[0]-\hat{x}[0])'\},$ and
let $L[k]=P[k]C'(CP[k]C'+\xi)^{-1}.$ Then the observer is a Kalman
filter given by \cite{franklin1998digital}
\[
\hat{x}[k+1]=A\hat{x}[k]+Bu[k]+L[k]\left(\tilde{y}[k]-C\hat{x}[k]\right).
\]

\subsubsection{Innovation}

In this context, the term $\tilde{y}[k]-C\hat{x}[k]$ is the \emph{innovation}.
Label the innovation by $\nu[k]=\tilde{y}[k]-C\hat{x}[k].$ This term
is used to update the estimate $\hat{x}[k]$ of the state. We consider
the components of the innovation as the signaling-game \emph{messages}
that the device decides whether to trust. Let us label each component
of the innovation as low-risk or high-risk. For each $i\in\mathbb{S},$
we classify the innovation as 
\[
m^{i}=\begin{cases}
m_{L}, & \text{if }\left|\nu^{i}[k]\right|\leq\epsilon^{i}\\
m_{H}, & \text{if }\left|\nu^{i}[k]\right|>\epsilon^{i}
\end{cases},
\]
where $\epsilon\in\mathbb{R}_{++}^{N}$ is a vector of thresholds.
Since $\mathcal{R}$ is strategic, it chooses whether to incorporate
the innovations using the signaling game strategy $\sigma_{\mathcal{R}}(a\,|\,m),$
given the vector of messages $m.$ 

Define a \emph{strategic innovation filter} by $D_{\sigma_{\mathcal{R}}}:\,\mathbb{R}^{N}\to\mathbb{R}^{N}$
such that, given innovation $\nu,$ the components of gated innovation
$\bar{\nu}=D_{\sigma_{\mathcal{R}}}(\nu)$ are given by 
\[
\bar{\nu}^{i}=\begin{cases}
\nu^{i}, & \text{if }a^i=a_T\\
0, & \text{otherwise}
\end{cases},
\]
for $i\in\mathbb{S}.$ Now we incorporate the function $D_{\sigma_{\mathcal{R}}}$
into the estimator by
\[
\hat{x}[k+1]=A\hat{x}[k]+Bu[k]+L[k]D_{\sigma_{\mathcal{R}}}\left(\nu[k]\right).
\]

\subsubsection{Feedback Controller}

The optimal controller is given by the feedback law $u[k]=-K[k]\hat{x}[k],$
with gain
\[
K[k]=\left(B'[k]S[K+1]B+R\right)^{-1}B'S[k+1]A,
\]
where $S[k]$ is obtained by the backward Riccati difference equation
\begin{multline*}
S[k]=A'\Bigl(S[k+1]-S[k+1]B\\
\left(B'S[k+1]B+R\right)^{-1}B'S[k+1]\Bigr)A+Q,
\end{multline*}
with $S[T]=F.$

\subsubsection{Control Criterion to Utility Mapping\label{sub:ctrlCriterion}}

The control cost $J$ determines the signaling game utility of the
device $\mathcal{R}.$ This utility should be monotonically decreasing
in $J.$ We consider a mapping $J\mapsto u_{\mathcal{R}}^{S}$ defined
by $u_{\mathcal{R}}^{S}\left(\theta,m,a\right)=(\bar{v}_{\mathcal{R}}-\underline{v}_{\mathcal{R}})e^{-\beta_{\mathcal{R}}J}+\bar{v}_{\mathcal{R}},$
where $\bar{v}_{\mathcal{R}}$ and $\underline{v}_{\mathcal{R}}$
denote maximum and minimum values of the utility, and $\beta_{\mathcal{R}}$
represents the sensitivity of the utility to the control cost.

\subsection{Definition of Gestalt Nash Equilibrium}

We now define the equilibrium concept for the overall game, which
is called \emph{Gestalt Nash equilibrium }(\emph{GNE}). \textcolor{black}{To differentiate with the equilibria in \texttt{FlipIt} game and signaling game, we use notations with a superscript $\dagger$ to emphasize the solution at GNE. }

\begin{defn}
(Gestalt Nash equilibrium)\label{def:GNE} The triple $\left(p_{\mathcal{A}}^{\dagger},v_{\mathcal{A}}^{\dagger},v_{\mathcal{D}}^{\dagger}\right),$
where $p_{\mathcal{A}}^{\dagger}$ represents the probability of compromise
of each of the cloud services, and $v_{\mathcal{A}}^{\dagger}$ and
$v_{\mathcal{D}}^{\dagger}$ represent the vectors of equilibrium
utilities for $\mathcal{A}^{i}$ and $\mathcal{D}^{i},$ $i\in\mathbb{S},$
constitutes a Gestalt Nash equilibrium of the overall game if both
Eq. \eqref{eq:gestaltFlip} and Eq. \eqref{eq:gestaltSig} are satisfied:
\begin{equation}
\forall i\in\left\{ 1,\ldots,m\right\} ,\;p_{\mathcal{A}}^{i\dagger}=T^{F_{i}}\left(v_{\mathcal{A}}^{i\dagger},v_{\mathcal{D}}^{i\dagger}\right),\label{eq:gestaltFlip}
\end{equation}
\begin{equation}
\left(\left[\begin{array}{c}
v_{\mathcal{A}}^{1\dagger}\\
v_{\mathcal{A}}^{2\dagger}\\
\vdots\\
v_{\mathcal{A}}^{N\dagger}
\end{array}\right],\left[\begin{array}{c}
v_{\mathcal{D}}^{1\dagger}\\
v_{\mathcal{D}}^{2\dagger}\\
\vdots\\
v_{\mathcal{D}}^{N\dagger}
\end{array}\right]\right)\in T^{S}\left(\left[\begin{array}{c}
p_{\mathcal{A}}^{1\dagger}\\
p_{\mathcal{A}}^{2\dagger}\\
\vdots\\
p_{\mathcal{A}}^{N\dagger}
\end{array}\right]\right).\label{eq:gestaltSig}
\end{equation}

\end{defn}
According to Definition \ref{def:GNE}, the overall game is at equilibrium
when, simultaneously, each of the \texttt{FlipIt} games is at equilibrium
and the one signaling game is at equilibrium.

\section{Equilibrium Analysis\label{sec:Gestalt}}

In this section, we give conditions under which a GNE exists. We start
with a set of natural assumptions. Then we narrow the search for feasible
equilibria. We show that the signaling game only supports pooling
equilibria, and that only low-risk pooling equilibria survive selection
criteria. Finally, we create a mapping that composes the signaling
and \texttt{FlipIt} game models. We show that this mapping has a
closed graph, and we use Kakutani's fixed-point theorem to prove the
existence of a GNE. In order to avoid obstructing the flow of the
paper, we briefly summarize the proofs of each lemma, and we refer
readers to the GNE derivations for a single cloud service in \cite{pawlick2015flip}
and \cite{pawlick2017strategic}.

\subsection{Assumptions}

\begin{table*}
\caption{\label{tab:assumptions}Assumptions}

\centering{}%
\begin{tabular}{|c|c|}
\hline 
\# & Assumption ($\forall i\in\mathbb{S}$)\tabularnewline
\hline 
\hline 
A1 & $0=u_{\mathcal{A}}^{S_{i}}\left(m_{L},a_{N}\right)=u_{\mathcal{A}}^{S_{i}}\left(m_{H},a_{N}\right)=u_{\mathcal{D}}^{S_{i}}\left(m_{L},a_{N}\right)=u_{\mathcal{D}}^{S_{i}}\left(m_{H},a_{N}\right).$\tabularnewline
\hline 
A2 & $0<u_{\mathcal{A}}^{S_{i}}\left(m_{L},a_{T}\right)<u_{\mathcal{D}}^{S_{i}}\left(m_{H},a_{T}\right)<u_{\mathcal{D}}^{S_{i}}\left(m_{L},a_{T}\right)<u_{\mathcal{A}}^{S_{i}}\left(m_{H},a_{T}\right).$\tabularnewline
\hline 
\multirow{1}{*}{A3} & $\forall\theta^{-i},m^{-i},a^{-i},\;\;$ $u_{\mathcal{R}}^{S}\left(\theta,m,\bar{a}\right)>u_{\mathcal{R}}\left(\theta,m,\tilde{a}\right),\;\;$
where $\;\;\theta^{i}=\theta_{\mathcal{A}},$ $m^{i}=m_{H},$ $\bar{a}^{-i}=\tilde{a}^{-i}=a^{-i},$
$\bar{a}^{i}=a_{N},$ and $\tilde{a}^{i}=a_{T}.$\tabularnewline
\hline 
\multirow{1}{*}{A4} & $\forall\theta^{-i},m^{-i},a^{-i},\;\;$ $u_{\mathcal{R}}^{S}\left(\theta,m,\bar{a}\right)<u_{\mathcal{R}}^{S}\left(\theta,m,\tilde{a}\right),\;\;$
where $\;\;\theta^{i}=\theta_{\mathcal{D}},$ $m^{i}=m_{L},$ $\bar{a}^{-i}=\tilde{a}^{-i}=a^{-i},$
$\bar{a}^{i}=a_{N},$ and $\tilde{a}^{i}=a_{T}.$\tabularnewline
\hline 
\multirow{1}{*}{A5} & $\forall\theta,m^{-i},a^{-i},\;\;$ $u_{\mathcal{R}}^{S}\left(\theta,\bar{m},a\right)>u_{\mathcal{R}}^{S}\left(\theta,\tilde{m},a\right),\;\;$
where $\;\;a^{i}=a_{T},$ $\bar{m}^{-i}=\tilde{m}^{-i}=m^{-i},$ $\bar{m}^{i}=m_{L},$
and $\tilde{m}^{i}=m_{H}.$\tabularnewline
\hline 
\end{tabular}
\end{table*}
For simplicity, let the utility functions of each signaling game sender
$i$ be dependent only on the messages and actions on cloud service
$i.$ That is, $\forall i\in\mathbb{S},$ $u_{\mathcal{A}}^{S_{i}}\left(m,a\right)\equiv u_{\mathcal{A}}^{S_{i}}(m^{i},a^{i})$
and $u_{\mathcal{D}}^{S_{i}}\left(m,a\right)\equiv u_{\mathcal{D}}^{S_{i}}(m^{i},a^{i}).$
This can be removed, but it makes analysis more straightforward. Table
\ref{tab:assumptions} gives five additional assumptions. Assumption
A1 assumes that each $\mathcal{A}^{i}$ and $\mathcal{D}^{i},$ $i\in\mathbb{S},$
get zero utility when their messages are not trusted. A2 assumes an
ordering among the utility functions for the senders in the signaling
game. It implies that a) $\mathcal{A}^{i}$ and $\mathcal{D}^{i}$
get positive utility when their messages are trusted; b) for trusted
messages, $\mathcal{A}$ prefers $m_{H}$ to $m_{L};$ and c) for
trusted messages, $\mathcal{D}$ prefers $m_{L}$ to $m_{H}.$ These
assumptions are justified if the goal of the attacker is to cause
damage (with a high-risk message), while the defender is able to operate
under normal conditions (with a low-risk message).

Assumptions A3-A4 give natural requirements on the utility function
of the device. First, the worst case utility for $\mathcal{R}$ is
trusting a high-risk message from an attacker. Assume that, on every
channel $i\in\mathbb{S},$ regardless of the messages and actions
on the other channels, $\mathcal{R}$ prefers to play $a^{i}=a_{N}$
if $m^{i}=m_{H}$ and $\theta^{i}=\theta_{\mathcal{A}}.$ This is
given by A3. Second, the best case utility for $\mathcal{R}$ is trusting
a low-risk message from a defender. Assume that, on every channel
$i\in\mathbb{S},$ regardless of the messages and actions on the other
channels, $\mathcal{R}$ prefers to play $a^{i}=a_{T}$ if $m^{i}=m_{L}$
and $\theta^{i}=\theta_{\mathcal{D}}.$ This is given by A4. Finally,
under normal operating conditions, $\mathcal{R}$ prefers trusted
low-risk messages compared to trusted high risk messages from both
an attacker and a defender. This is given by A5.

\subsection{GNE Existence Proof}

We prove the existence of a GNE using Lemmas \ref{lem:regimes}-\ref{lem:sigGameClosed}
and Theorem \ref{thm:kakutani}.

\subsubsection{Narrowing the Search for GNE}

Lemma \ref{lem:regimes} eliminates some candidates for GNE.
\begin{lem}
\label{lem:regimes}(GNE Existence Regimes \cite{pawlick2017strategic})
Every GNE $(p_{\mathcal{A}}^{\dagger},v_{\mathcal{A}}^{\dagger},v_{\mathcal{D}}^{\dagger})$
satisfies: $\forall i\in\mathbb{S},$ $v_{\mathcal{A}}^{i},v_{\mathcal{D}}^{i}>0.$
\end{lem}
The basic idea behind the proof of Lemma \ref{lem:regimes} is that
$v_{\mathcal{A}}^{i}=0$ or $v_{\mathcal{D}}^{i}=0$ cause either
$\mathcal{A}^{i}$ or $\mathcal{D}^{i}$ to give up on capturing or
recapturing the cloud. The cloud becomes either completely secure
or completely insecure, neither of which can result in a GNE. Lemma
\ref{lem:regimes} has a significant intuitive interpretation given
by Remark \ref{rem:notUseless}.
\begin{rem}
\label{rem:notUseless}In any GNE, for all $i\in\mathbb{S},$ $\mathcal{R}$
plays $a^{i}=a_{T}$ with non-zero probability. In other words, $\mathcal{R}$
never completely ignores any cloud service.
\end{rem}

\subsubsection{Elimination of Separating Equilibria}

In signaling games, equilibria in which different types of senders
transmit the same message are called \emph{pooling equilibria}, while
equilibria in which different types of senders transmit distinct messages
are called \emph{separating equilibria} \cite{fudenberg1991game}.
The distinct messages in separating equilibria completely reveal the
type of the sender to the receiver. Lemma \ref{lem:noSep} is typical
of signaling games between players with opposed incentives. 
\begin{lem}
\label{lem:noSep}(No Separating Equilibria \cite{pawlick2017strategic})
Consider all pure-strategy signaling-game equilibria $\left(\sigma_{\mathcal{R}}^{*};\sigma_{\mathcal{A}}^{1*},\ldots\sigma_{\mathcal{A}}^{N*};\sigma_{\mathcal{D}}^{1*},\ldots\sigma_{\mathcal{D}}^{N*}\right)$
in which each $\mathcal{A}^{i}$ and $\mathcal{D}^{i},$ $i\in\mathbb{S},$
receive positive expected utility. All such equilibria satisfy $\sigma_{\mathcal{A}}^{i*}(m)=\sigma_{\mathcal{D}}^{i*}(m)$
for all $m\in M$ and $i\in\mathbb{S}.$ That is, the senders on each
cloud service $i$ use pooling strategies.
\end{lem}
Lemma \ref{lem:noSep} holds because it is never incentive-compatible
for an attacker $\mathcal{A}^{i}$ to reveal his type, in which case
$\mathcal{R}$ would not trust $\mathcal{A}^{i}.$ Hence, $\mathcal{A}^{i}$
always imitates $\mathcal{D}^{i}$ by pooling.

\subsubsection{Signaling Game Equilibrium Selection Criteria}

Four pooling equilibria are possible in the signaling game: $\mathcal{A}^{i}$
and $\mathcal{D}^{i}$ transmit $m^{i}=m_{L}$ and $\mathcal{R}$
plays $a^{i}=a_{T}$ (which we label \textbf{EQ-L1}), $\mathcal{A}^{i}$
and $\mathcal{D}^{i}$ transmit $m^{i}=m_{L}$ and $\mathcal{R}$
plays $a^{i}=a_{N}$ (\textbf{EQ-L2}), $\mathcal{A}^{i}$ and $\mathcal{D}^{i}$
transmit $m^{i}=m_{H}$ and $\mathcal{R}$ plays $a^{i}=a_{T}$ (\textbf{EQ-H1}),
and $\mathcal{A}^{i}$ and $\mathcal{D}^{i}$ transmit $m^{i}=m_{H}$
and $\mathcal{R}$ plays $a^{i}=a_{N}$ (which we label \textbf{EQ-H2}).
In fact, the signaling game always admits multiple equilibria. Lemma
\ref{lem:lowSelected} performs equilibrium selection. 
\begin{lem}
\label{lem:lowSelected}(Selected Equilibria) The \emph{intuitive
criterion} \cite{cho1987signaling} and the criterion of \emph{first
mover advantage} imply that equilibria \textbf{\emph{E}}\textbf{Q-}\textbf{\emph{L1
}}and \textbf{\emph{E}}\textbf{Q-}\textbf{\emph{L2}} will be selected.\end{lem}
\begin{IEEEproof}
The first mover advantage states that, if both $\mathcal{A}^{i}$
and $\mathcal{D}^{i}$ prefer one equilibrium over the others, they
will choose the preferred equilibrium. Thus, $\mathcal{A}^{i}$ and
$\mathcal{D}^{i}$ will always choose \textbf{EQ-L1} or \textbf{EQ-H1}
if either of those is admitted. When neither is admitted, we select
\textbf{EQ-L2}\footnote{This is without loss of generality, since A1 implies that the sender
utilities are the same for \textbf{EQ-L2 }and \textbf{EQ-H2}.}. When both are admitted, we use the intuitive criterion to select
among them. Assumption A2 states that $\mathcal{A}^{i}$ prefers \textbf{EQ-H1},
while $\mathcal{D}^{i}$ prefers \textbf{EQ-L1}. Thus, if a sender
deviates from \textbf{EQ-H1} to \textbf{EQ-L1}, $\mathcal{R}$ can
infer that the sender is a defender, and trust the message. Therefore,
the intuitive criterion rejects \textbf{EQ-H1} and selects \textbf{EQ-L1}.
Finally, Assumption A5 can be used to show that \textbf{EQ-H1} is
never supported without \textbf{EQ-L1}. Hence, only \textbf{EQ-L1}
and \textbf{EQ-L2} survive the selection criteria.
\end{IEEEproof}
At the boundary between the parameter regime that supports \textbf{EQ-L1}
and the parameter regime that supports \textbf{EQ-L2}, $\mathcal{R}$
can choose any mixed strategy, in which he plays both $a^{i}=a_{T}$
and $a^{i}=a_{N}$ with some probability. Indeed, for any cloud service
$i\in\mathbb{S},$ hold $p_{\mathcal{A}}^{j},$ $j\neq i$ and $j\in\mathbb{S},$
constant, and let $p_{\mathcal{A}}^{i\diamond}$ denote the boundary
between the \textbf{EQ-L1} and \textbf{EQ-L2} regions. Then Remark
\ref{rem:worst-case} gives an important property of $p_{\mathcal{A}}^{i\diamond}.$
\begin{rem}
\label{rem:worst-case}By Lemma \ref{lem:regimes}, all GNE satisfy
$p_{\mathcal{A}}^{i}\leq p_{\mathcal{A}}^{i\diamond}.$ Therefore,
$p_{\mathcal{A}}^{i\diamond}$ is a worst-case probability of compromise.
\end{rem}
Remark \ref{rem:worst-case} is a result of the combination of the
signaling and \texttt{FlipIt} games. Intuitively, it states that
strategic trust in the communication layer is able to limit the probability
of compromise of a cloud service, regardless of the attack and defense
costs in the cyber layer.

\subsubsection{\texttt{FlipIt} Game Properties}

For the \texttt{FlipIt} games on each cloud service $i\in\mathbb{S},$
denote the ratio of attacker and defender expected utilities by $v_{\mathcal{AD}}^{i}=v_{\mathcal{A}}^{i}/v_{\mathcal{D}}^{i}.$
For $i\in\mathbb{S},$ define the set $\mathbb{V}^{i}$ by
\[
\mathbb{V}^{i}=\left\{ v\in\mathbb{R}_{+}\,:\,0\leq v\leq u_{\mathcal{A}}^{S_{i}}(m_{L},a_{T})/u_{\mathcal{D}}^{S_{i}}(m_{L},a_{T})\right\} .
\]
Also define the set $\mathbb{PR}^{i},$ $i\in\mathbb{S},$ by $\mathbb{PR}^{i}=$
\[
\left\{ p\in[0,1]\,:\,0<p<T^{F_{i}}\left(u_{\mathcal{A}}^{S_{i}}(m_{L},a_{T}),u_{\mathcal{D}}^{S_{i}}(m_{L},a_{T})\right)\right\} .
\]
Next, for $i\in\mathbb{S},$ define modified \texttt{FlipIt} game
mappings $\tilde{T}^{F_{i}}:\,\mathbb{V}^{i}\to\mathbb{PR}^{i},$
where 
\begin{equation}
p_{\mathcal{A}}^{i*}=\tilde{T}^{F_{i}}\left(v_{\mathcal{AD}}^{i}\right)\iff p_{\mathcal{A}}^{i*}\in T^{F_{i}}\left(v_{\mathcal{A}}^{i},v_{\mathcal{D}}^{i}\right).\label{eq:redefTf}
\end{equation}
Then Lemma \ref{lem:flipGameContin} holds.
\begin{lem}
\label{lem:flipGameContin}(Continuity of $\tilde{T}^{F_{i}}$ \cite{pawlick2015flip})
For $i\in\mathbb{S},$ $\tilde{T}^{F_{i}}(v_{\mathcal{AD}}^{i})$
is continuous in $v_{\mathcal{AD}}^{i}\in\mathbb{V}^{i}.$ 
\end{lem}
The dashed curve in Fig. \ref{fig:oneDbothMappings} gives an example
of $\tilde{T}^{F_{i}}$ for $i=1.$ The independent variable is on
the vertical axis, and the dependent variable is on the horizontal
axis.

\subsubsection{Signaling Game Properties}

Let $v_{\mathcal{AD}}=[v_{\mathcal{AD}}^{i}]_{i\in\mathbb{S}},$ $\mathbb{V}=\prod_{i\in\mathbb{S}}\mathbb{V}^{i},$
and $\mathbb{PR}=\prod_{i\in\mathbb{S}}\mathbb{PR}^{i}.$ Define a
modified signaling game mapping by $\tilde{T}^{S}:\,\mathbb{PR}\to\mathcal{P}(\mathbb{V})$
such that 
\begin{equation}
v_{\mathcal{AD}}^{*}\in\tilde{T}^{S}\left(p_{\mathcal{A}}\right)\iff\left(v_{\mathcal{A}}^{*},v_{\mathcal{D}}^{*}\right)\in T^{S}\left(p_{\mathcal{A}}\right),\label{eq:redefTs}
\end{equation}
where $T^{S}$ selects the equilibria given by Lemma \ref{lem:lowSelected}.
Then we have Lemma \ref{lem:sigGameClosed}.
\begin{lem}
\label{lem:sigGameClosed}(Properties of $\tilde{T}^{S}$) Construct
a graph
\[
\mathbb{G}=\left\{ \left(p_{\mathcal{A}},v_{\mathcal{AD}}^{*}\right)\in\mathbb{PR}\times\mathbb{V}\,:\,v_{\mathcal{AD}}^{*}\in\tilde{T}^{S}\left(p_{\mathcal{A}}\right)\right\} ,
\]
The graph $\mathbb{G}$ is closed. Additionally, for every $p_{\mathcal{A}}\in\mathbb{PR},$
the set of outputs of $\tilde{T}^{S}(p_{\mathcal{A}})$ is non-empty
and convex.\end{lem}
\begin{IEEEproof}
The graph $\mathbb{G}$ is closed because it contains all of its limit
points. The set of outputs is non-empty because a signaling game equilibrium
exists for all $p_{\mathcal{A}}.$ It is convex because expected utilities
for mixed-strategy equilibria are convex combinations of pure strategy
utilities and because assumption A2 implies that convexity also holds
for the ratio of the utilities.
\end{IEEEproof}
The step functions (plotted with solid lines) in Figure \ref{fig:oneDbothMappings}
plot example mappings from $p_{\mathcal{A}}^{1}$ on the horizontal
axis to $v_{\mathcal{AD}}^{1}$ on the vertical axis for $v_{\mathcal{AD}}\in\tilde{T}^{S}(p_{\mathcal{A}}),$
holding $p_{\mathcal{A}}^{i},$ $i\in\{2,3,\ldots,N\}$ fixed. It
is clear that the graphs are closed.

\subsubsection{Fixed-Point Theorem}

By combining Eq. (\ref{eq:gestaltFlip}-\ref{eq:gestaltSig}) with
Eq. (\ref{eq:redefTf}) and Eq. (\ref{eq:redefTs}), we see that the
vector of equilibrium utility ratios $v_{\mathcal{AD}}^{\dagger}=[v_{\mathcal{AD}}^{i\dagger}]_{i\in\mathbb{S}}$
in any GNE $(p_{\mathcal{A}}^{\dagger},v_{\mathcal{A}}^{\dagger},v_{\mathcal{D}}^{\dagger})$
must satisfy 

\[
\left[\begin{array}{c}
v_{\mathcal{AD}}^{1\dagger}\\
v_{\mathcal{AD}}^{2\dagger}\\
\vdots\\
v_{\mathcal{AD}}^{N\dagger}
\end{array}\right]\in\tilde{T}^{S}\left(\left[\begin{array}{c}
\tilde{T}^{F_{1}}\left(v_{\mathcal{AD}}^{1\dagger}\right)\\
\tilde{T}^{F_{2}}\left(v_{\mathcal{AD}}^{2\dagger}\right)\\
\vdots\\
\tilde{T}^{F_{2}}\left(v_{\mathcal{AD}}^{N\dagger}\right)
\end{array}\right]\right).
\]
Denote this composed mapping by $\tilde{T}^{S\circ F}:\,\mathbb{V}\to\mathcal{P}(\mathbb{V})$
such that the GNE requirement can be written by $v_{\mathcal{AD}}^{\dagger}\in\tilde{T}^{S\circ F}(v_{\mathcal{AD}}^{\dagger}).$
Figure \ref{fig:oneDbothMappings} gives a one-dimensional intuition
behind Theorem \ref{thm:exist}. The example signaling game step functions
$\tilde{T}^{S}$ have closed graphs, and the outputs of the functions
are non-empty and convex. The \texttt{FlipIt} curve $\tilde{T}^{F_{1}}$
is continuous. The two mappings are guaranteed to intersect, and the
intersection is a GNE.

\begin{center}
\begin{figure}
\begin{centering}
\includegraphics[width=0.65\columnwidth]{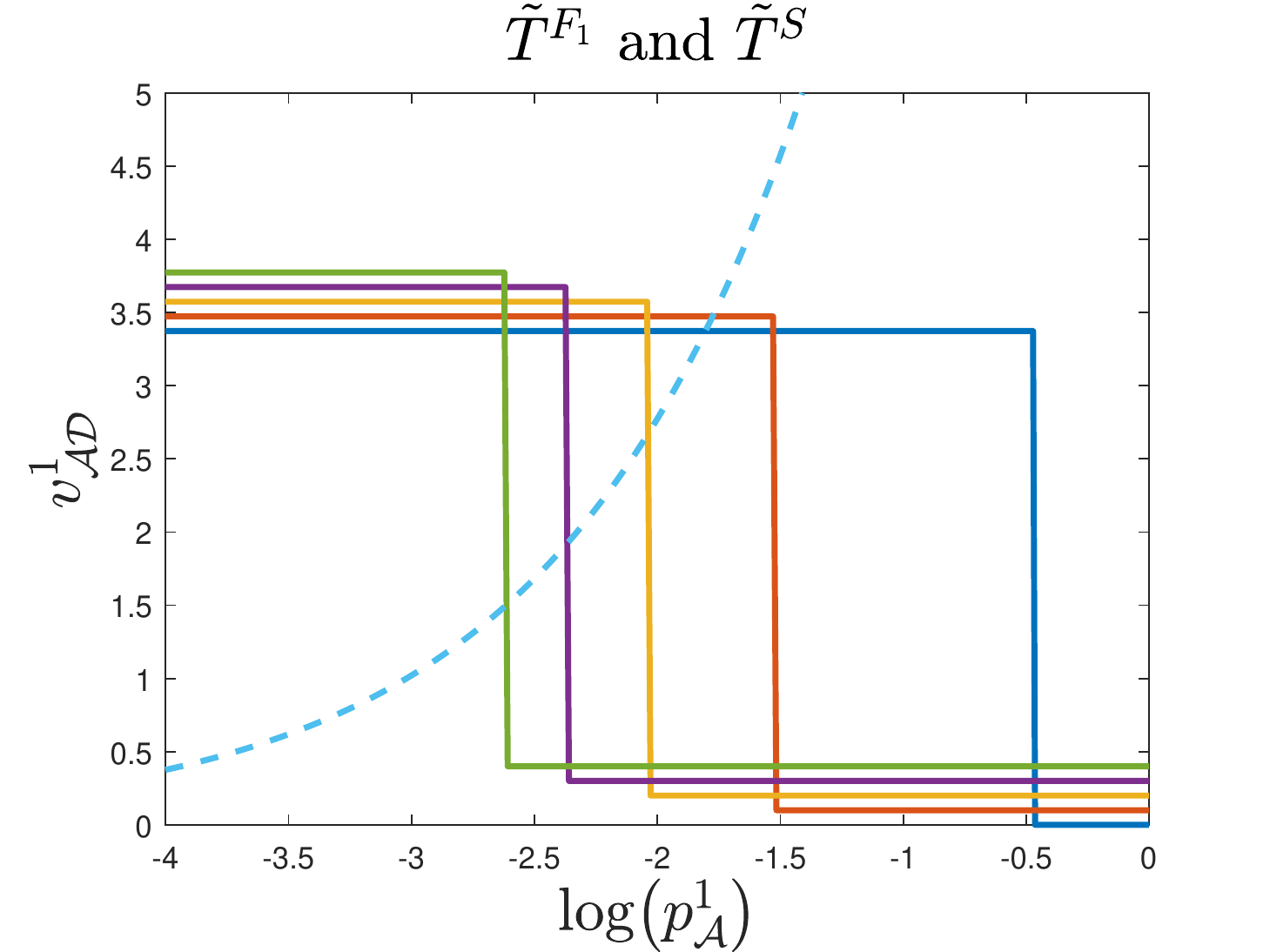}
\par\end{centering}

\caption{\label{fig:oneDbothMappings}The (solid) step-functions depict modified
signaling game mappings $\tilde{T}^{S}$ for five different sets of
parameters. The (dashed) curve depicts a modified \texttt{FlipIt}
game mapping $\tilde{T}^{F_{1}}.$ The intersection is a GNE. The
figure shows only one dimension out of $N$ dimensions. }
\end{figure}

\par\end{center}

According to Lemma \ref{lem:sigGameClosed}, the graph $\mathbb{G}$
of the signaling game mapping is closed, and the set of outputs of
$\tilde{T}^{S}$ is non-empty and convex. Since each modified \texttt{FlipIt}
game mapping $\tilde{T}^{F_{i}},$ $i\in\mathbb{S}$ is a continuous
function, each $\tilde{T}^{F_{i}}$ produces a closed graph and has
non-empty and (trivially) convex outputs. Thus, the graph of the composed
mapping, $\tilde{T}^{S\circ F},$ is also closed, and has non-empty
and convex outputs. Because of this, we can apply Kakutani's fixed-point
theorem, famous for its use in proving Nash equilibrium.
\begin{thm}
\label{thm:kakutani}(Kakutani Fixed-Point Theorem \cite{kakutani1941generalization})
- Let $\Phi$ be a non-empty, compact, and convex subset of some Euclidean
space $\mathbb{R}^{n}.$ Let $Z:\,\Phi\to\mathcal{P}(\Phi)$ be a
set-valued function on $\Phi$ with a closed graph and the property
that, for all $\phi\in\Phi,$ $Z(\phi)$ is non-empty and convex.
Then $Z$ has a fixed point. 
\end{thm}
The mapping $\tilde{T}^{S\circ F}$ is a set-valued function on $\mathbb{V},$
which is a non-empty, compact, and convex subset of $\mathbb{R}^{N}.$
$\tilde{T}^{S\circ F}$ also has a closed graph, and the set of its
outputs is non-empty and convex. Therefore, $\tilde{T}^{S\circ F}$
has a fixed-point, which is precisely the definition of a GNE. Hence,
we have Theorem \ref{thm:exist}.
\begin{thm}
\label{thm:exist}(GNE Existence) Let the utility functions in the
signaling game satisfy Assumptions A1-A5. Then a GNE exists. \end{thm}
\begin{IEEEproof}
The proof has been constructed from Lemmas \ref{lem:regimes}-\ref{lem:sigGameClosed}
and Theorem \ref{thm:kakutani}.
\end{IEEEproof}
\begin{algorithm}
\caption{Adaptive defense algorithm for iSTRICT}
\label{algorithm} 
\begin{enumerate}
\item Initialize parameters $\alpha_{\mathcal{A}}^{i}$, $\alpha_{\mathcal{D}}^{i}$,
$p_{\mathcal{A}}^{i}$, $p_{\mathcal{D}}^{i}$, $\forall i\in\mathbb{S}$,
in each \texttt{FlipIt} game, and \textcolor{black}{$\sigma_{\mathcal{A}}^{{i}}$
and $\sigma_{\mathcal{D}}^{{i}}$, $\forall i\in\mathbb{S}$, $\sigma_{\mathcal{R}}$}
in the signaling game \\
 \textbf{Signaling game:} 
\item Solve optimization problems in Eq. \eqref{sigma_a} and Eq. \eqref{sigma_d},
respectively, and obtain $\sigma_{\mathcal{A}}^{{i}*}$ and $\sigma_{\mathcal{D}}^{{i}*}$,
$\forall i\in\mathbb{S}$ \label{sig} 
\item Update belief $\mu^{i}(\theta_{\mathcal{A}}\,|\,m^{i})$ based on
Eq. \eqref{belief_a}, and $\mu^{i}(\theta_{\mathcal{D}}\,|\,m^{i})=1-\mu^{i}(\theta_{\mathcal{A}}\,|\,m^{i})$,
$\forall i\in\mathbb{S}$ \label{belief}
\item Solve receiver's problem in Eq. \eqref{sigma_r} and obtain $\sigma_{\mathcal{R}}^{*}$
\label{receiver}
\item If $\sigma_{\mathcal{A}}^{{i}*},\sigma_{\mathcal{D}}^{{i}*},\sigma_{\mathcal{R}}^{*}$
do not change, go to step \ref{nex}; otherwise, go back to step \ref{sig}\label{endSig} 
\item Obtain $v_{\mathcal{A}}^{i}$ and $v_{\mathcal{D}}^{i}$, $\forall i\in\mathbb{S}$,
from Eq. \eqref{va} and Eq. \eqref{vd}, respectively \label{nex}
\\
 \texttt{FlipIt}\textbf{ game:} 
\item Solve defenders' and attackers' problems in Eq. \eqref{defender_p}
and Eq. \eqref{attacker_p} jointly, and obtain $f_{\mathcal{A}}^{i*}$
and $f_{\mathcal{D}}^{i*}$, $\forall i\in\mathbb{S}$\label{flip} 
\item Map the frequency pair $(f_{\mathcal{A}}^{i*},f_{\mathcal{D}}^{i*})$
to the probability pair $(p_{\mathcal{A}}^{i*},p_{\mathcal{D}}^{i*})$
through Eq. \eqref{eq:pA}, $\forall i\in\mathbb{S}$ 
\item If $(p_{\mathcal{D}}^{i*},p_{\mathcal{A}}^{i*})$, $\forall i\in\mathbb{S}$,
do not change, go to step \ref{last}; otherwise, go back to step
\ref{sig}\label{exitCond} 
\item \textbf{Return} $p_{\mathcal{A}}^{\dagger}\coloneqq p_{\mathcal{A}}^{*}$, \textcolor{black}{
$\sigma_{\mathcal{A}}^{{i}\dagger}\coloneqq\sigma_{\mathcal{A}}^{{i}*}$,
$\sigma_{\mathcal{D}}^{{i}\dagger}\coloneqq\sigma_{\mathcal{D}}^{{i}*}$,}
$\forall i\in\mathbb{S},$ and $\sigma_{\mathcal{R}}^{\dagger}\coloneqq\sigma_{\mathcal{R}}^{*}$
\label{last}\end{enumerate}
\end{algorithm}

\subsection{Adaptive Algorithm\label{sub:adaptiveAlg}}

Numerical simulations suggest that Assumptions A1-A5 often hold. If this is not the case, however, Algorithm \ref{algorithm} can be used to compute the GNE. The main idea of the adaptive algorithm is to update the strategic decision-making of different entities in iSTRICT iteratively. 

Given the probability vector $p_{\mathcal{A}},$ Lines \ref{sig}-\ref{endSig}
of Algorithm \ref{algorithm} compute a PBNE for the signaling game
which consists of the strategy profile $\left(\sigma_{\mathcal{R}}^{*};\sigma_{\mathcal{A}}^{1*},\ldots\sigma_{\mathcal{A}}^{N*};\sigma_{\mathcal{D}}^{1*},\ldots\sigma_{\mathcal{D}}^{N*}\right)$
and belief vector $\mu\left(\theta\,|\,m\right).$ The algorithm computes
the PBNE iteratively using best response. The vectors of equilibrium
utilities $\left(v_{\mathcal{A}}^{*},v_{\mathcal{D}}^{*}\right)$
are given by Eq. \eqref{va} and Eq. \eqref{vd}. Using $\left(v_{\mathcal{A}}^{*},v_{\mathcal{D}}^{*}\right),$
Line \ref{flip} of Algorithm \ref{algorithm} updates the equilibrium
strategies of the \texttt{FlipIt} games and arrives at a new prior
probability pair $(p_{\mathcal{A}}^{i*},p_{\mathcal{D}}^{i*}),$ $\forall i\in\mathbb{S},$
through the mapping in Eq. \eqref{eq:pA}. This initializes the next
round of the signaling game with the new $(p_{\mathcal{A}}^{i*},p_{\mathcal{D}}^{i*}).$
The algorithm terminates when the probabilities remain unchanged between
rounds. 

\begin{figure}
\begin{centering}
\subfloat[\label{fig:deviceBelief}Device belief]{\begin{centering}
\includegraphics[width=0.75\columnwidth]{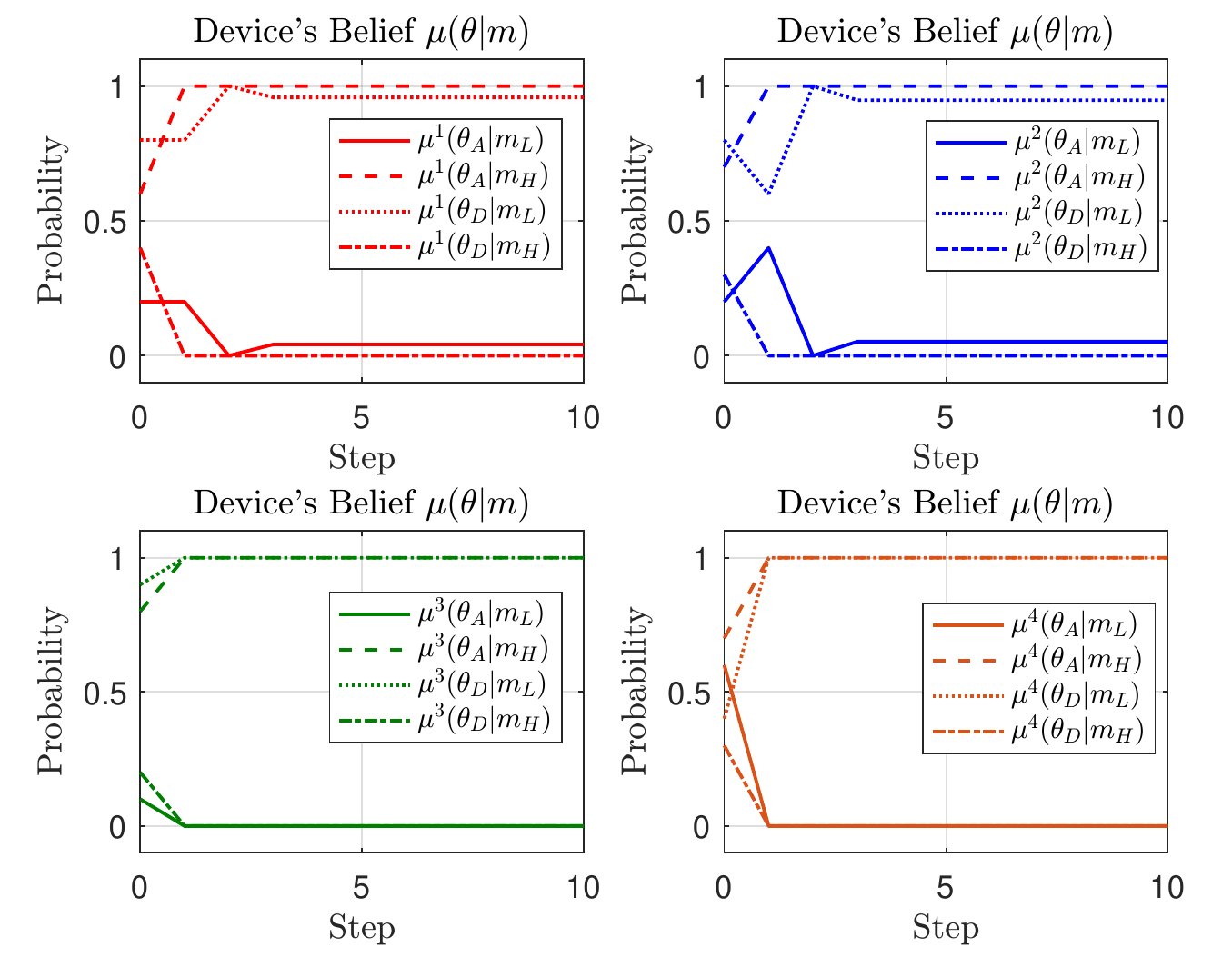}
\par\end{centering}
}
\par\end{centering}

\begin{centering}
\medskip{}
\subfloat[\label{fig:deviceAction}Device action]{\begin{centering}
\includegraphics[width=0.75\columnwidth]{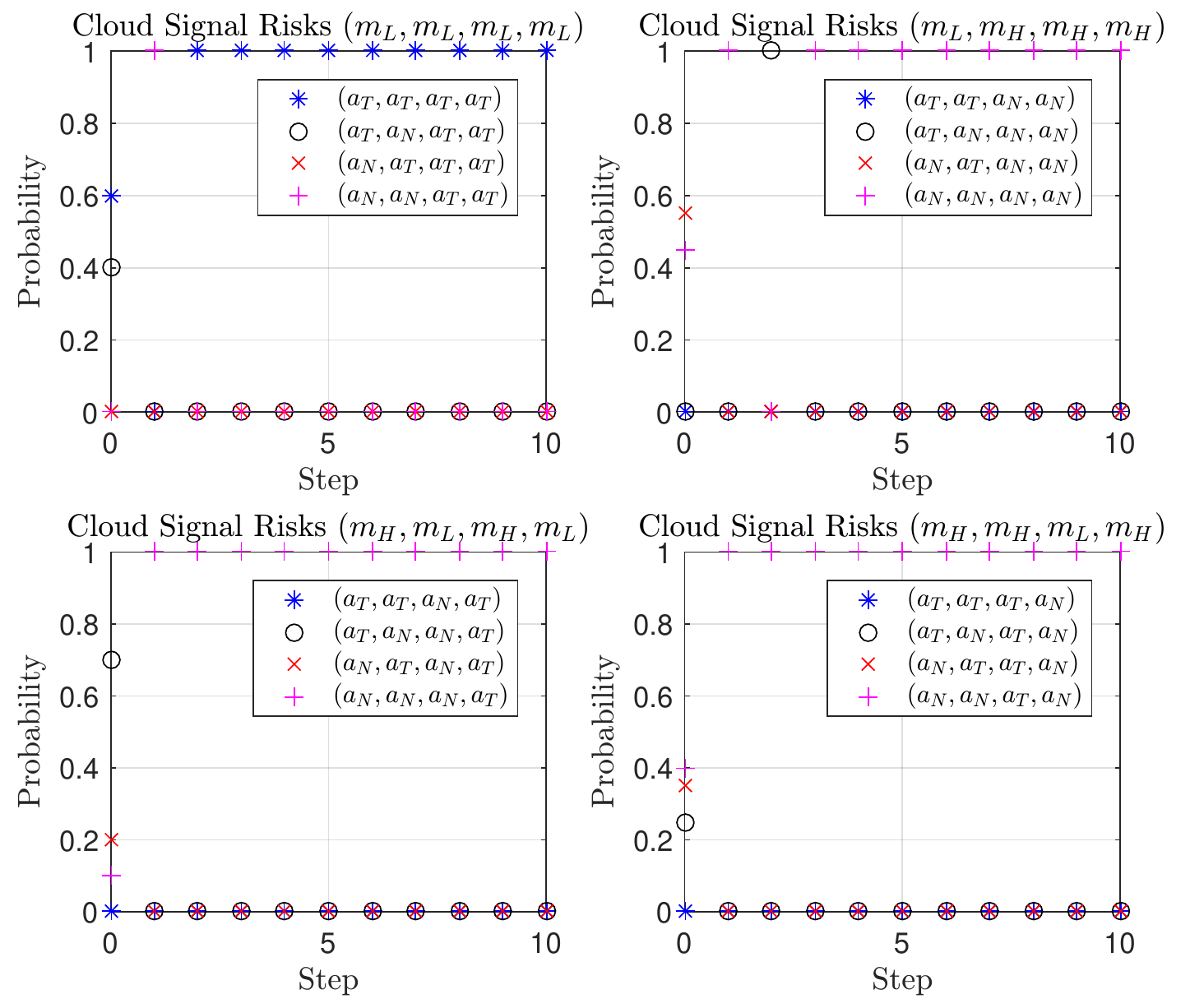}
\par\end{centering}

}
\par\end{centering}

\begin{centering}
\medskip{}
\subfloat[\label{fig:GNE-flip}\texttt{FlipIt} game]{\begin{centering}
\includegraphics[width=0.6\columnwidth]{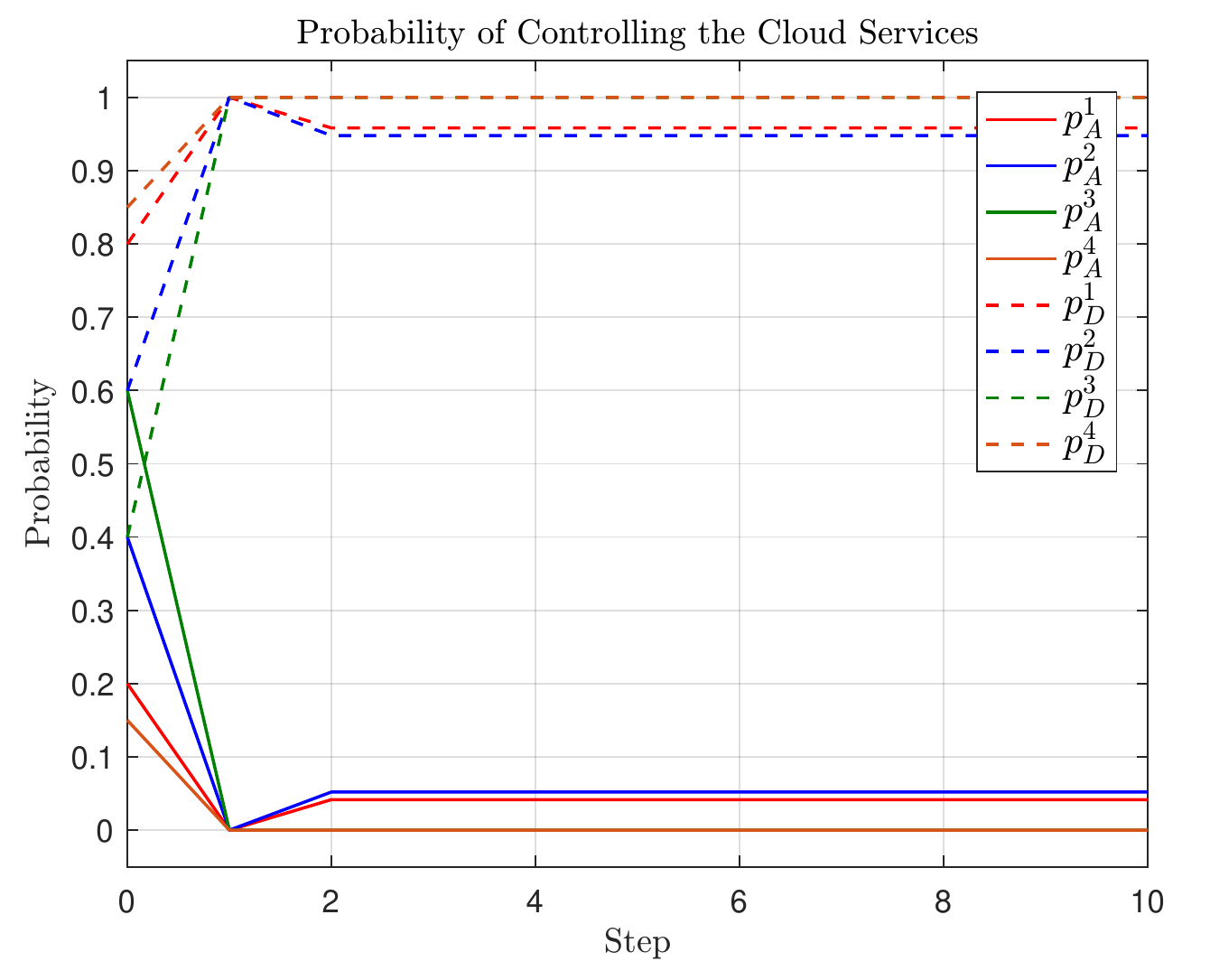}
\par\end{centering}

}
\par\end{centering}

\caption{\label{fig:allAlgPlots} \textcolor{black}{Adaptive Algorithm with four cloud services. (a) and (b) depict each device's belief and action, respectively. (c) shows the result of cloud security. The algorithm converges to a GNE in four steps, where one step represents a round of updates including the \texttt{FlipIt} game and signaling game.}}
\end{figure}

\textcolor{black}{To illustrate Algorithm \ref{algorithm}, we next present an example including $N=4$ cloud services. The detailed physical meaning of each service will be presented in Section \ref{sec:app}. Specifically, the costs of renewing control of cloud services are $\alpha_{\mathcal{A}}^{1}=\$2$k, $\alpha_{\mathcal{A}}^{2}=\$0.8$k, $\alpha_{\mathcal{A}}^{3}=\$10$k, $\alpha_{\mathcal{A}}^{4}=\$12$k, and  $\alpha_{\mathcal{D}}^{1}=\$0.2$k, $\alpha_{\mathcal{D}}^{1}=\$0.1$k, $\alpha_{\mathcal{D}}^{1}=\$0.05$k, $\alpha_{\mathcal{D}}^{1}=\$0.03$k, for the attackers and defenders, respectively. The initial proportions of time of each attacker and defender controlling the cloud services are $p_{\mathcal{A}}^{1}=0.2$, $p_{\mathcal{A}}^{2}=0.4$, $p_{\mathcal{A}}^{3}=0.6$,
$p_{\mathcal{A}}^{4}=0.15$, and $p_{\mathcal{D}}^{1}=0.8$, $p_{\mathcal{D}}^{2}=0.6$, $p_{\mathcal{D}}^{3}=0.4$, $p_{\mathcal{D}}^{4}=0.85$, respectively. For the signaling game at the communication layer, the initial probabilities that attacker sends low-risk message at each cloud service are equal to 0.2, 0.3, 0.1, and 0.4, respectively. Similarly, the defender's initial probabilities of sending low-risk message are equal to 0.9, 0.8, 0.95, and 0.97, respectively.
Figure \ref{fig:allAlgPlots} presents the results of Algorithm
\ref{algorithm} on this example system. 
The result in Fig. \ref{fig:deviceBelief} shows that the cloud services 1 and 2 can be compromised by the attacker. Figure. \ref{fig:deviceBelief} shows the device's belief on the received information. At the GNE, the attacker also sends low-risk message to deceive the receiver and gain utility when controlling the cloud service. Four representative devices' actions are shown in Fig. \ref{fig:deviceAction}, where the devices strategically reject low-risk message in some cases due to the couplings between layers in iSTRICT.  Because of the large attack and defense cost ratios and the crucial impact on physical system performance of services 3 and 4, $p_{\mathcal{D}}^3=p_{\mathcal{D}}^4=1$, insuring a secure information provision. In addition, the defense strategies at the
cloud layer and the communication layer are adjusted adaptively according
to the attackers' behaviors. Within each layer, all players are required
to best respond to the strategies of the other players. This cross-layer
approach enables a defense-in-depth mechanism for the devices in iSTRICT.}

\section{Application to Autonomous Vehicle Control\label{sec:app}}

In this section, \textcolor{black}{we apply iSTRICT to a cloud-enabled autonomous vehicle network in which the framework of vehicular cloud computing is similar to the one in \cite{cordeschi2015distributed}.} \textcolor{black}{Two
autonomous vehicles use an observer to estimate their positions and
velocities based on measurements from six sources, four of which 
may be compromised}. They also implement optimal feedback control
based on the estimated state.

\subsection{Autonomous Vehicle Security}

Autonomous vehicle technology will make a powerful impact on several
industries. In the automotive industry, traditional car companies
as well as technology firms such as Google \cite{guizzo2011google}
are racing to develop autonomous vehicle technology. Maritime shipping
is also an attractive application of autonomous vehicles. Autonomous
ships are expected to be safer, higher-capacity, and more resistant
to piracy attacks \cite{levander2017forget}. Finally, unmanned aerial
vehicles (UAVs) have the potential to reshape fields such as mining,
disaster relief, and precision agriculture \cite{zhang2012application}.

Nevertheless, autonomous vehicles pose clear safety risks. In ground
transportation, in March of 2018, an Uber self-driving automobile
struck and killed a pedestrian \cite{wakabayashi2018ubser}. On the
sea, multiple crashes of ships in the Unites States Navy during 2017
\cite{shane2017fitzgerald} have prompted concerns about too much
reliance on automation. In the air, cloud-enabled UAVs could be subject
to data integrity or availability attacks \cite{xu2015secure}. In
general, autonomous vehicles rely on many remote sources (\emph{e.g.},
other vehicles, GPS signals, location-based services) for information.
In the most basic case, these sources are subject to errors that must
be handled robustly. In addition, the sources could also be selfish
and strategic. For instance, an autonomous ship could transmit its
own coordinates dishonestly in order to clear its own shipping path
of other vessels. In the worst case, the sources could be malicious.
An attacker could use a spoofed GPS signal in order to destroy a UAV
or to use the UAV to attack another target. In all of these cases,
autonomous vehicles must decide whether to trust the remote sources
of information.

\subsection{Physical-Layer Implementation}

\begin{figure}[t]
\begin{centering}
\includegraphics[width=0.55\columnwidth]{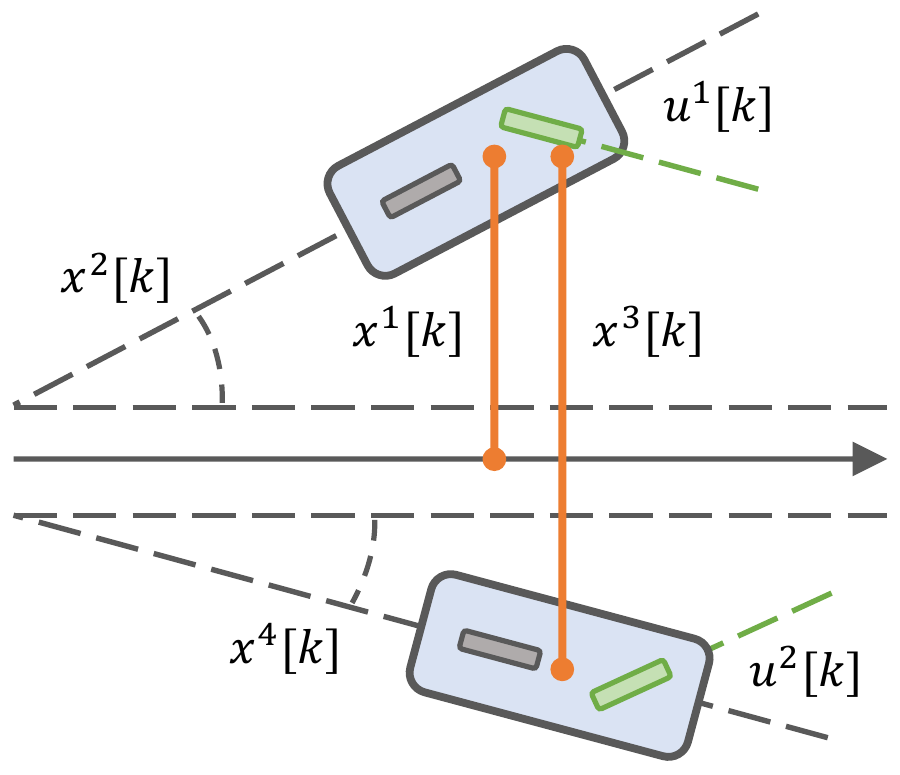}
\par\end{centering}

\caption{\label{fig:vehStates}\textcolor{black}{We use bicycle steering models
from \cite{astrom2010feedback} to conceptually capture the vehicle
dynamics. The vehicle states are given by $x^{1}[k]:$ first vehicle
position, $x^{2}[k]:$ first vehicle angle; $x^{3}[k]:$ offset between
vehicles; $x^{4}[k]:$ second vehicle angle. Controls $u^{1}[k]$
and $u^{2}[k]$ represent the steering angles of the first and second
vehicles, respectively.}}
\end{figure}

\textcolor{black}{We consider an interaction between nine agents. Two
autonomous vehicles implement} observer-based optimal feedback control
according to the iSTRICT framework. Each vehicle has two states: position
and angle. \textcolor{black}{Thus, the combined system has the state vector $x[k]\in\mathbb{R}^{4}$
described in Fig. \ref{fig:vehStates}.} The states evolve over finite
horizon $k\in\{0,1,\ldots,T\}.$

\begin{figure}[t]
\begin{centering}
\includegraphics[width=0.55\columnwidth]{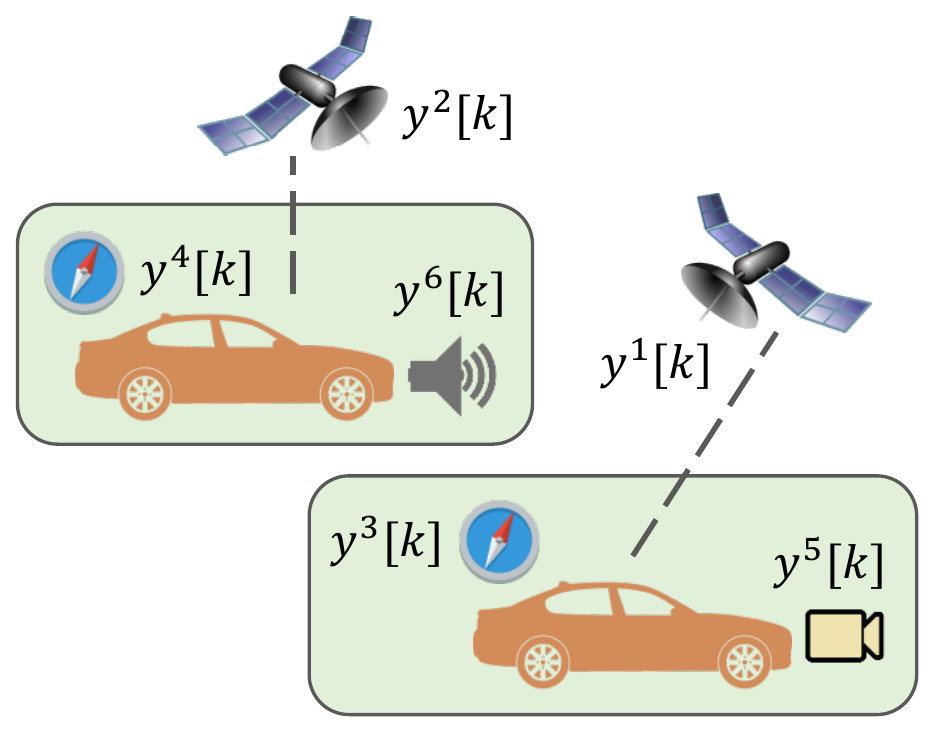}
\par\end{centering}

\caption{\label{fig:vehMsmts}\textcolor{black}{Local sensors include a localization camera on
vehicle 1 and a range finding device on vehicle 2. Remote sensors
include magnetic compass sensors and GPS receivers on both vehicles. }}

\end{figure}

These states are observed through both remote and local measurements.
Figure \ref{fig:vehMsmts} describes these measurements. \textcolor{black}{The
local measurements $y^{5}[k]$ and $y^{6}[k]$ originate from sensors
on the autonomous vehicle, so these are secure. Hence, the autonomous
vehicle always trusts $y^{5}[k]$ and $y^{6}[k].$ In addition, while
the magnetic compass sensors are subject to electromagnetic attack,
this involves high attack costs $\alpha_{\mathcal{A}}^{3}$ and $\alpha_{\mathcal{A}}^{4}$. The defense algorithm 
yields that $\mathcal{R}$ always trusts $y^{3}[k]$ and $y^{4}[k]$ at the GNE.}
The remote measurements $\tilde{y}^{1}[k]$ and $\tilde{y}^{2}[k]$
are received from cloud services that may be controlled by defenders
$\mathcal{D}^{1}$ and $\mathcal{D}^{2},$ or that may be compromised
by attackers $\mathcal{A}^{1}$ and $\mathcal{A}^{2}.$ 

In the signaling game, attackers $\mathcal{A}^{1}$ and $\mathcal{A}^{2}$
may add bias terms $\Delta_{\mathcal{A}}^{1}[k]$ or $\Delta_{\mathcal{A}}^{2}[k]$
if $\theta^{1}=\theta_{\mathcal{A}}$ or $\theta^{2}=\theta_{\mathcal{A}},$
respectively. Therefore, the autonomous vehicles must strategically
decide whether to trust these measurements. Each $\tilde{y}^{i}[k],$
$i\in\{1,2\},$ is classified as a low-risk ($m^{i}=m_{L})$ or high-risk
($m^{i}=m_{H})$ message according to an innovation filter. $\mathcal{R}$
decides whether to trust each message according to the action vector
$a=[\begin{array}{cc}
a^{1} & a^{2}\end{array}]',$ where $a^{1},a^{2}\in\{a_{N},a_{T}\}.$ We seek an equilibrium of
the signaling game that satisfies Definition \ref{def:PBNEsig}.

\subsection{Signaling Game Results}

\begin{figure*}
\subfloat[\label{fig:High-unfilteredInnov}Innovation]{\begin{centering}
\includegraphics[width=0.3\textwidth]{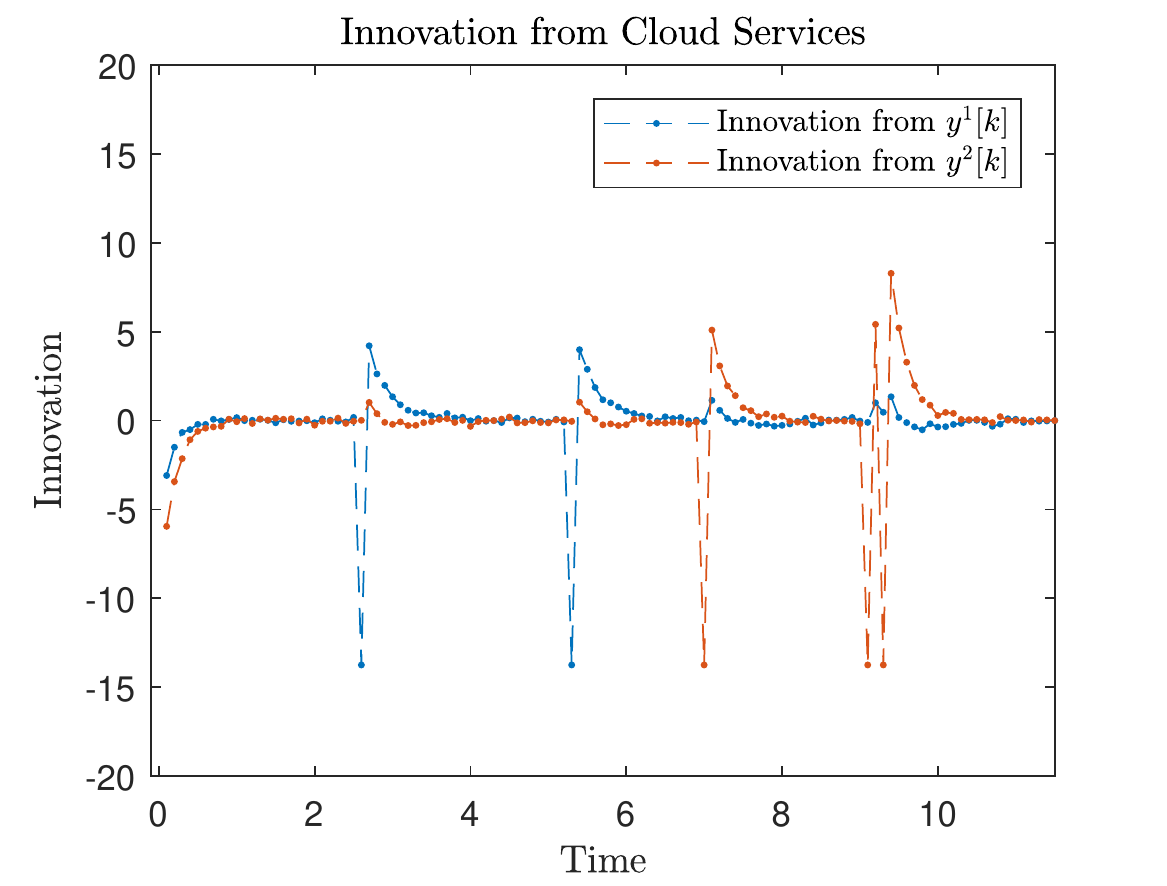}
\par\end{centering}

}\hfill{}\subfloat[\label{fig:High-filteredInnov}Innovation]{\begin{centering}
\includegraphics[width=0.3\textwidth]{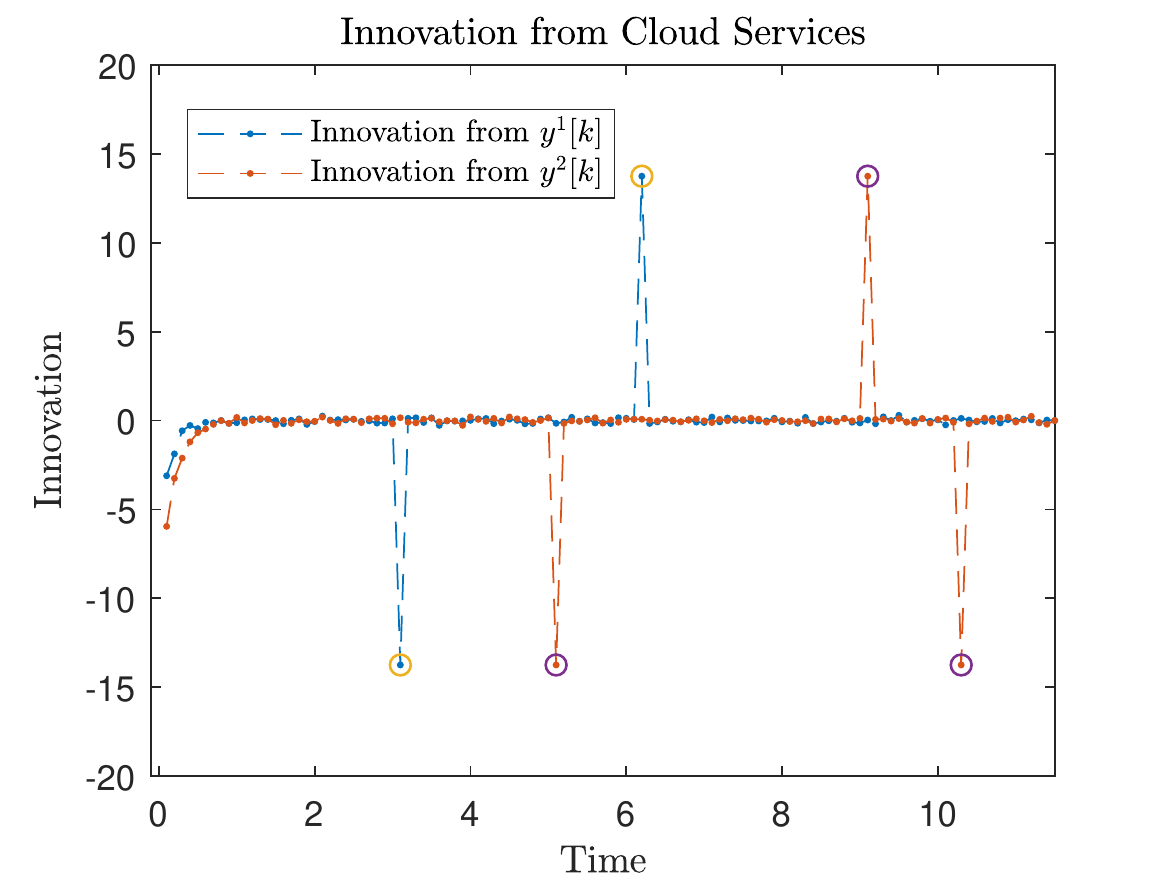}
\par\end{centering}

}\hfill{}\subfloat[\label{fig:lowInnov}Innovation]{\begin{centering}
\includegraphics[width=0.3\textwidth]{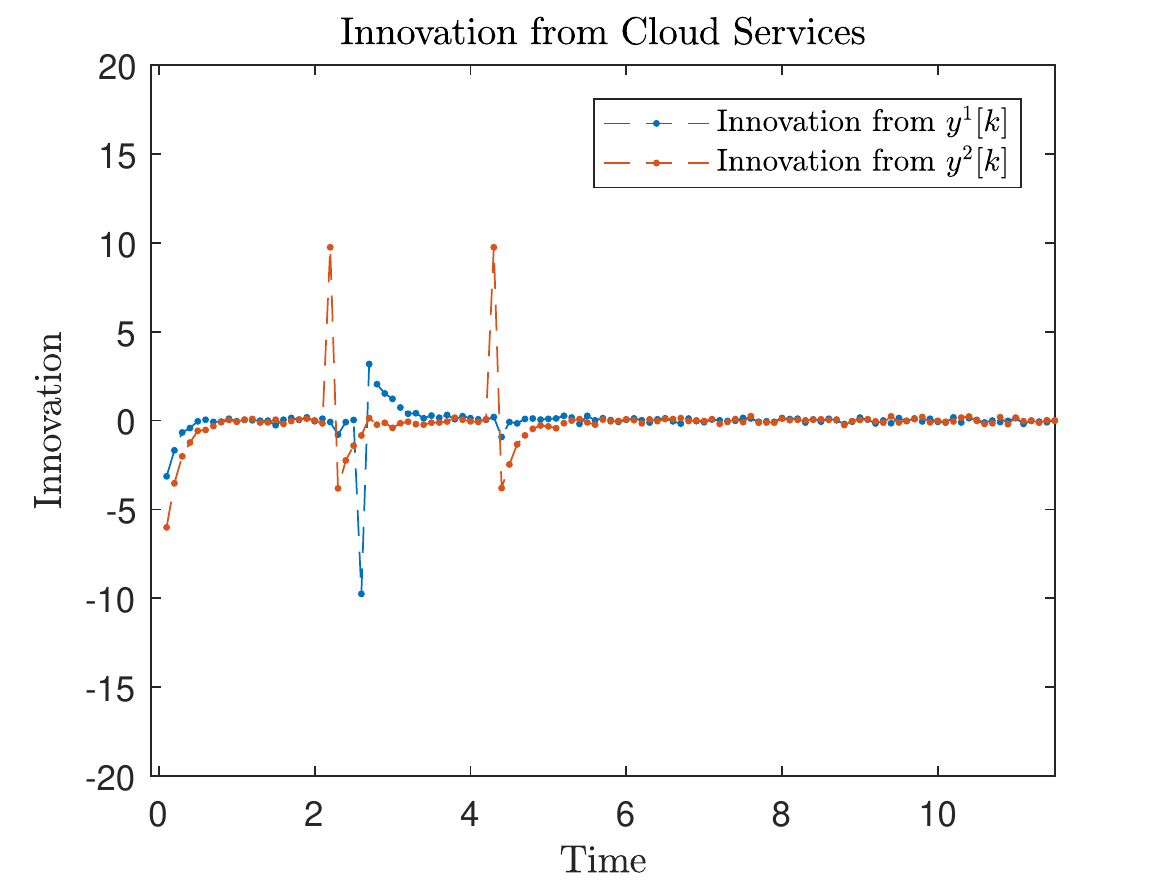}
\par\end{centering}

}

\medskip{}

\subfloat[\label{fig:High-unfilteredState}State trajectories]{\begin{centering}
\includegraphics[width=0.3\textwidth]{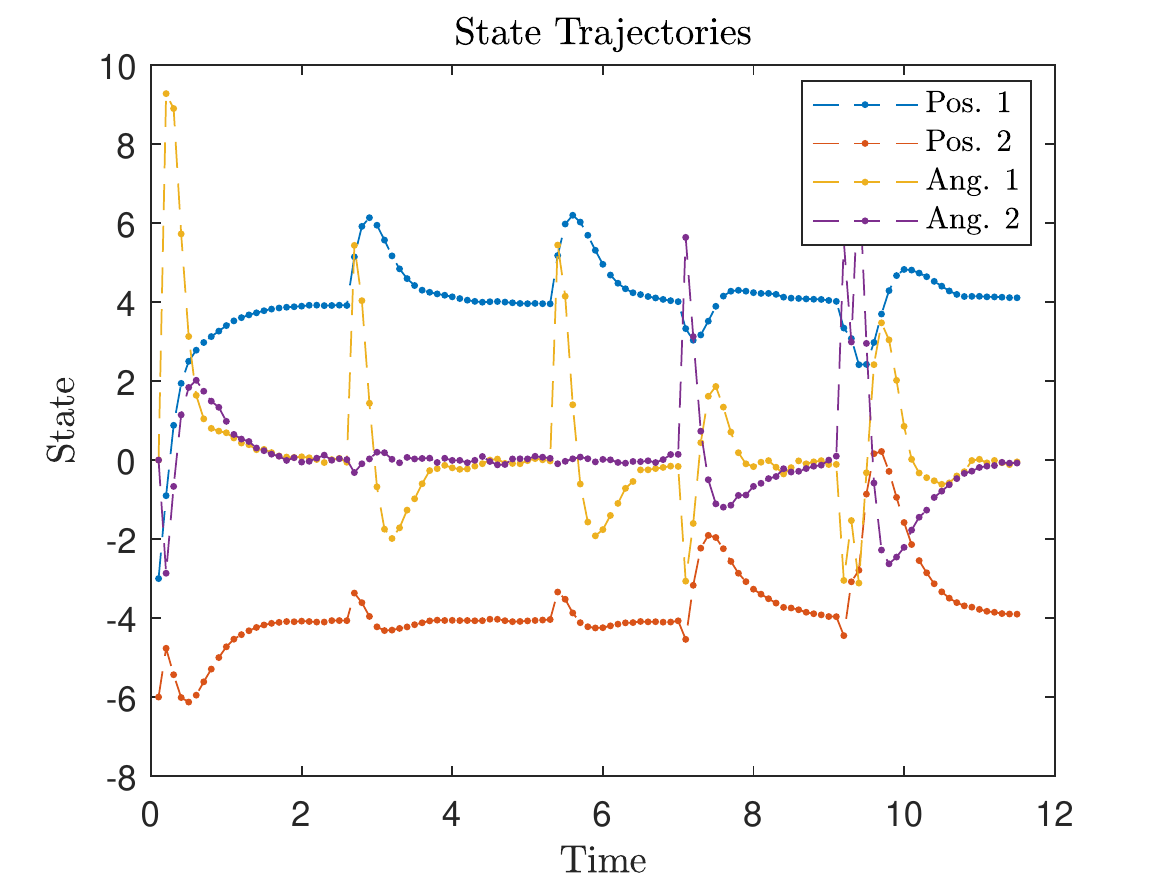}
\par\end{centering}

}\hfill{}\subfloat[\label{fig:High-filteredState}State trajectories]{\begin{centering}
\includegraphics[width=0.3\textwidth]{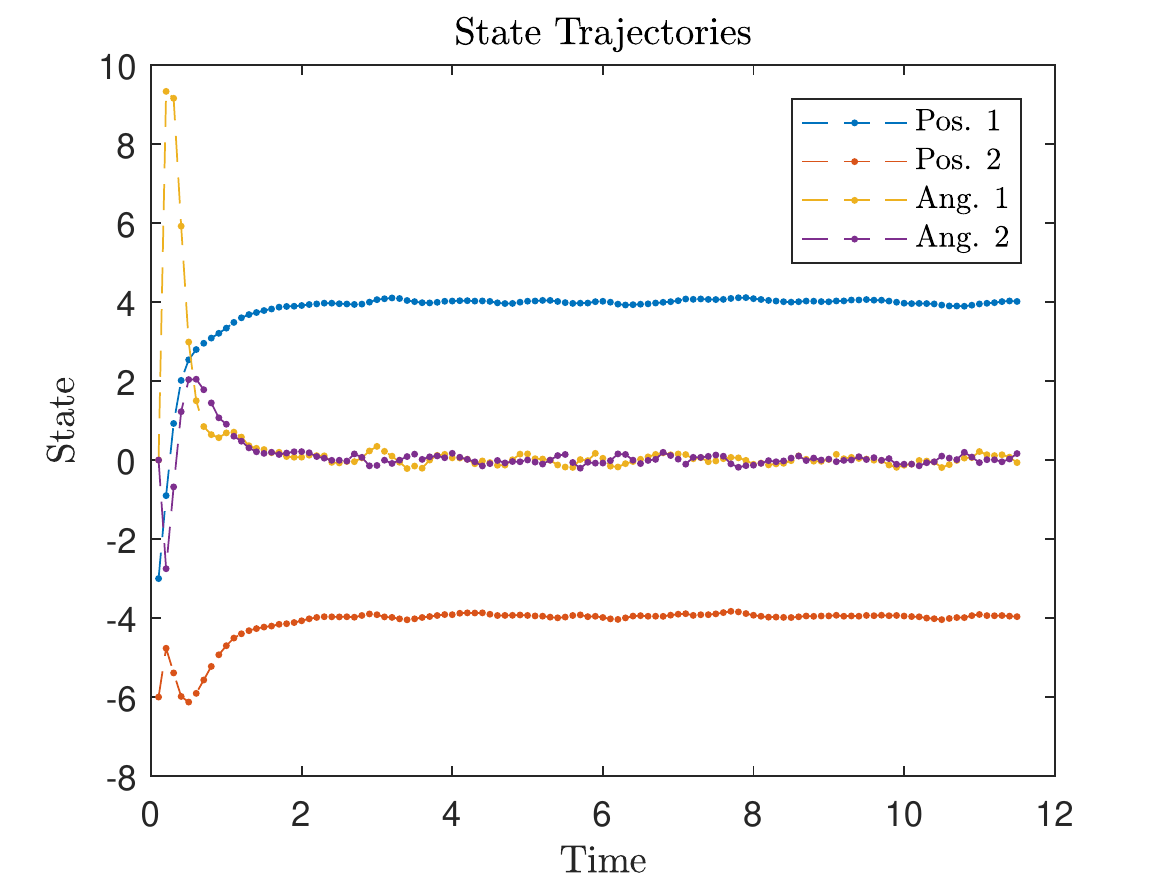}
\par\end{centering}

}\hfill{}\subfloat[\label{fig:lowState}State trajectories]{\begin{centering}
\includegraphics[width=0.3\textwidth]{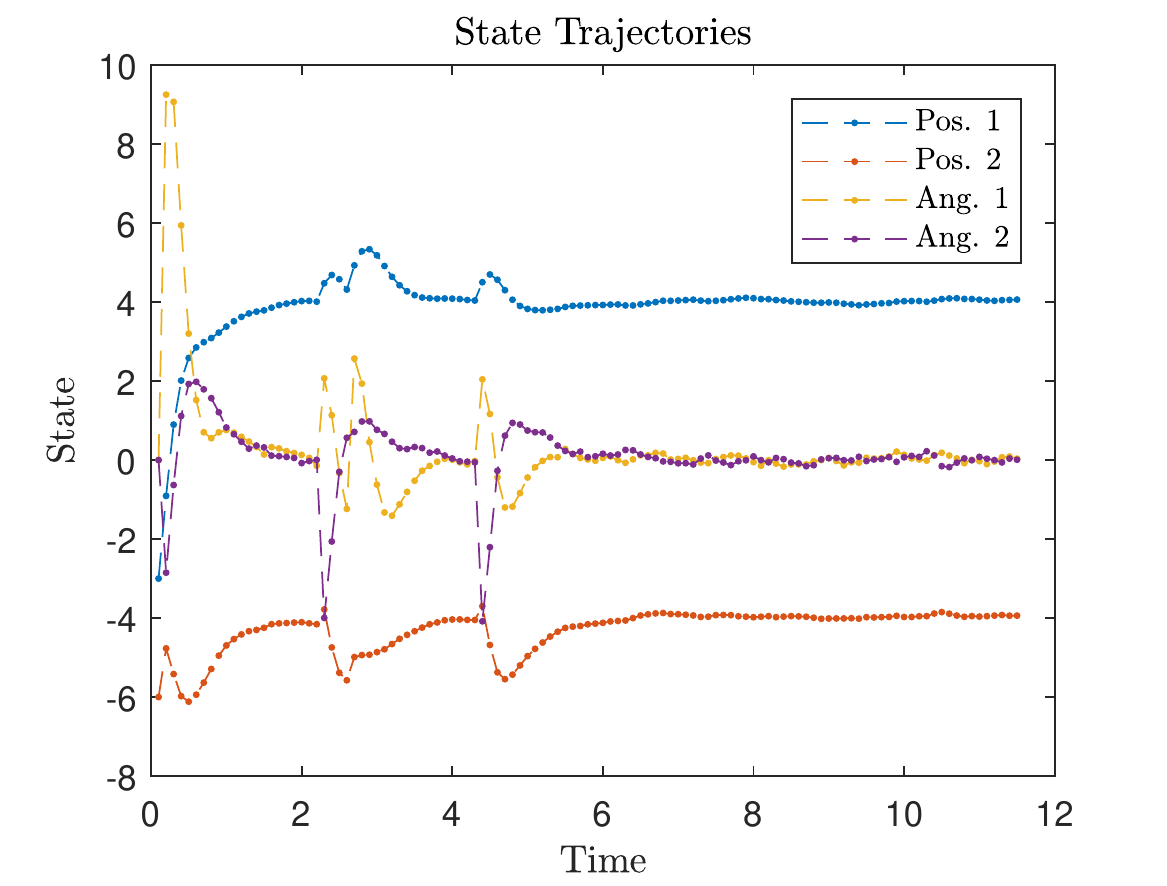}
\par\end{centering}

}

\caption{\label{fig:allSimResults}Column 1: $\mathcal{A}^{1}$ and $\mathcal{A}^{2}$
send $m_{H}$ and $\mathcal{R}$ plays $[\protect\begin{array}{cc}
a_{T} & a_{T}\protect\end{array}]',$ Column 2: $\mathcal{A}^{1}$ and $\mathcal{A}^{2}$ send $m_{H}$
and $\mathcal{R}$ plays $[\protect\begin{array}{cc}
a_{N} & a_{N}\protect\end{array}]',$ Column 3: $\mathcal{A}^{1}$ and $\mathcal{A}^{2}$ send $m_{L}$
and $\mathcal{R}$ plays $[\protect\begin{array}{cc}
a_{T} & a_{T}\protect\end{array}]'.$ Row 1: Innovation, Row 2: State trajectories.}
\end{figure*}

\textcolor{black}{Figure \ref{fig:allSimResults} depicts the results of three different signaling-game strategy profiles for the attackers, defenders, and device, using the software MATLAB \cite{MATLAB}. The observer and controller are linear, so the computation is rapid. Each iteration of the computational elements of the control loop depicted in Fig. \ref{control_system} takes less than 0.0002s on a Lenovo ThinkPad L560 laptop with 2.30 GHz processor and 8.0 GB of installed RAM.}

\textcolor{black}{In all of the scenarios, we set the position of the first vehicle to a track a reference trajectory of $x^1[k]=4,$ the offset between the second vehicle and the first vehicle to track a reference trajectory of $x^3[k]=8,$ and both angles to target $x^2[k]=x^4[k]=0.$} Column 1 depicts a scenario in which $\mathcal{A}^{1}$ and
$\mathcal{A}^{2}$ send $m_{H}$ and $\mathcal{R}$ plays $a^{1}=a^{2}=a_{T}.$
\textcolor{black}{The spikes in the innovation represent the bias terms added by the attacker when he controls the cloud. The spikes in Fig.
\ref{fig:allSimResults}(a) are large because these the attacker adds bias terms corresponding to high risk messages.} These bias terms cause large deviations
in the position and angle from their desired values (Fig.
\ref{fig:allSimResults}(d)). \textcolor{black}{For instance, at time $10,$ the two vehicles come within approximately $4$ units of each other.}

Column 2 depicts the best response of $\mathcal{R}$ to this strategy.
The vehicle uses an innovation filter (here, at $\epsilon^{1}=\epsilon^{2}=10$)
which categorizes the biased innovations as $m_{H}.$ The best response
is to choose 
\[
\sigma_{\mathcal{R}}\left(\left[\begin{array}{c}
a_{T}\\
a_{T}
\end{array}\right]|\left[\begin{array}{c}
m_{L}\\
m_{L}
\end{array}\right]\right)=\sigma_{\mathcal{R}}\left(\left[\begin{array}{c}
a_{T}\\
a_{N}
\end{array}\right]|\left[\begin{array}{c}
m_{L}\\
m_{H}
\end{array}\right]\right)=1,
\]
\[
\sigma_{\mathcal{R}}\left(\left[\begin{array}{c}
a_{N}\\
a_{T}
\end{array}\right]|\left[\begin{array}{c}
m_{H}\\
m_{L}
\end{array}\right]\right)=\sigma_{\mathcal{R}}\left(\left[\begin{array}{c}
a_{N}\\
a_{N}
\end{array}\right]|\left[\begin{array}{c}
m_{H}\\
m_{H}
\end{array}\right]\right)=1,
\]
\emph{i.e.}, to trust only low-risk messages. The circled data points
in Fig. \ref{fig:allSimResults}(b) denote high-risk innovations from
the attacker that are rejected. Figure \ref{fig:allSimResults}(e)
shows that this produces very good results \textcolor{black}{in which the positions of the first and second vehicle converge to their desired values of $4$ and $-4,$ respectively, and the angles converge to $0.$}

But iSTRICT assumes that the attackers are also strategic. $\mathcal{A}^{1}$
and $\mathcal{A}^{2}$ realize that high-risk messages will be rejected,
so they add smaller bias terms $\Delta_{\mathcal{A}}^{1}[k]$ and
$\Delta_{\mathcal{A}}^{2}[k]$ which are classified as $m_{L}.$ This
is depicted by Fig. \ref{fig:allSimResults}(c). It is not optimal
for the autonomous vehicle to reject all low-risk messages, because
most such messages come from a cloud controlled by the defender. Therefore,
the device must play $a^{1}=a^{2}=a_{T}.$ Nevertheless, Fig. \ref{fig:allSimResults}(f)
shows that these low-risk messages create significantly less disturbance
than the disturbances from high-risk messages in Fig. \ref{fig:allSimResults}(d).
In summary, the signaling-game equilibrium is for $\mathcal{A}^{1},$
$\mathcal{A}^{2},$ $\mathcal{D}^{1},$ and $\mathcal{D}^{2}$ to
transmit low-risk messages and for $\mathcal{R}$ to trust low-risk
messages while rejecting high-risk messages off the equlibrium path.

\subsection{Results of the \texttt{FlipIt} Games}

Meanwhile, $\mathcal{A}^{1}$ and $\mathcal{D}^{1}$ play a \texttt{FlipIt}
game for control of Cloud Service 1, and $\mathcal{A}^{2}$ and $\mathcal{D}^{2}$
play a \texttt{FlipIt} game for control of Cloud Service 2. Based
on the equilibrium of the signaling game, all players realize that
the winners of the \texttt{FlipIt} games will be able to send trusted
low-risk messages, but not trusted high-risk messages. Based on Assumption
A2, low-risk messages are more beneficial to the defenders than to
the attackers. Hence, the incentives to control the cloud are larger
for defenders than for attackers. This results in a low $p_{\mathcal{A}}^{1}$ and $p_{\mathcal{A}}^{2}$ from the \texttt{FlipIt} game. If the equilibrium from the previous subsection holds for these prior probabilities, then the overall five-player interaction is at a GNE as described in Definition
\ref{def:GNE} and Theorem \ref{thm:exist}.

\textcolor{black}{Table \ref{tab:ctrlCosts} is useful for benchmarking the performance of iSTRICT. The table lists the empirical value of the control criterion given by Eq. (\ref{eq:criterion}). The first three columns quantify the performance depicted in Fig. \ref{fig:allSimResults}. Column $1$ is the benchmark case, in which $\mathcal{A}^1$ and $\mathcal{A}^2$ add high-risk noise, and the noise is mitigated somewhat by a Kalman filter, but the bias is not handled optimally. Column 2 shows the improvement provided by iSTRICT against a  nonstrategic attacker, and Column 3 shows the improvement provided by iSTRICT against a strategic attacker. The improvement is largest in Column 2, but it is significant against a strategic attacker as well.}

\begin{table*}
\caption{\label{tab:ctrlCosts}\textcolor{black}{Control Costs and Benchmarks for the Simulations
depicted in Fig. 11-12}}

\centering{}%
\begin{tabular}{|c|c|c|c|c|c|c|}
\hline 
 & Ungated $m_{H}$ & Gated $m_{H}$ & Trusted $m_{L}$ & Trusted, Frequent $m_{L}$ & Untrusted, Frequent $m_{L}$ & Mixed Trust with $m_{L}$ \tabularnewline
\hline 
\hline 
Trial 1 & 274,690 & 42,088 & 116,940  & 185,000  & 128,060  & 146,490 \tabularnewline
\hline 
Trial 2 & 425,520  & 42,517 & 123,610  & 211,700  & 121,910  & 143,720 \tabularnewline
\hline 
Trial 3 & 119,970  & 42,444 & 125,480  & 213,500  & 144,090  & 130,460 \tabularnewline
\hline 
Trial 4 & 196,100  & 42,910 & 89,980  & 239,400  & 138,350  & 135,930 \tabularnewline
\hline 
Trial 5 & 229,870  & 42,733 & 66,440  & 94,400  & 135,160  & 139,680 \tabularnewline
\hline 
Trial 6 & 139,880  & 42,412 & 69,510  & 2,581,500  & 119,700  & 125,270 \tabularnewline
\hline 
Trial 7 & 129,980  & 42,642 & 116,560  & 254,000  & 138,160  & 122,790 \tabularnewline
\hline 
Trial 8 & 97,460  & 42,468 & 96,520  & 1,020,000  & 130,260  & 146,370 \tabularnewline
\hline 
Trial 9 & 125,490  & 42,633 & 50,740  & 250,900  & 138,960  & 151,470 \tabularnewline
\hline 
Trial 10 & 175,670  & 42,466 & 78,700  & 4,182,600  & 135,780  & 126,550 \tabularnewline
\hline 
Average & 191,463  & 42,531 & 93,448  & 923,300  & 133,043  & 136,873 \tabularnewline
\hline 
\end{tabular}
\end{table*}

\subsection{GNE for Different Parameters}

\begin{figure*}
\subfloat[\label{fig:innov_aTtwo}Innovation]{\begin{centering}
\includegraphics[width=0.3\textwidth]{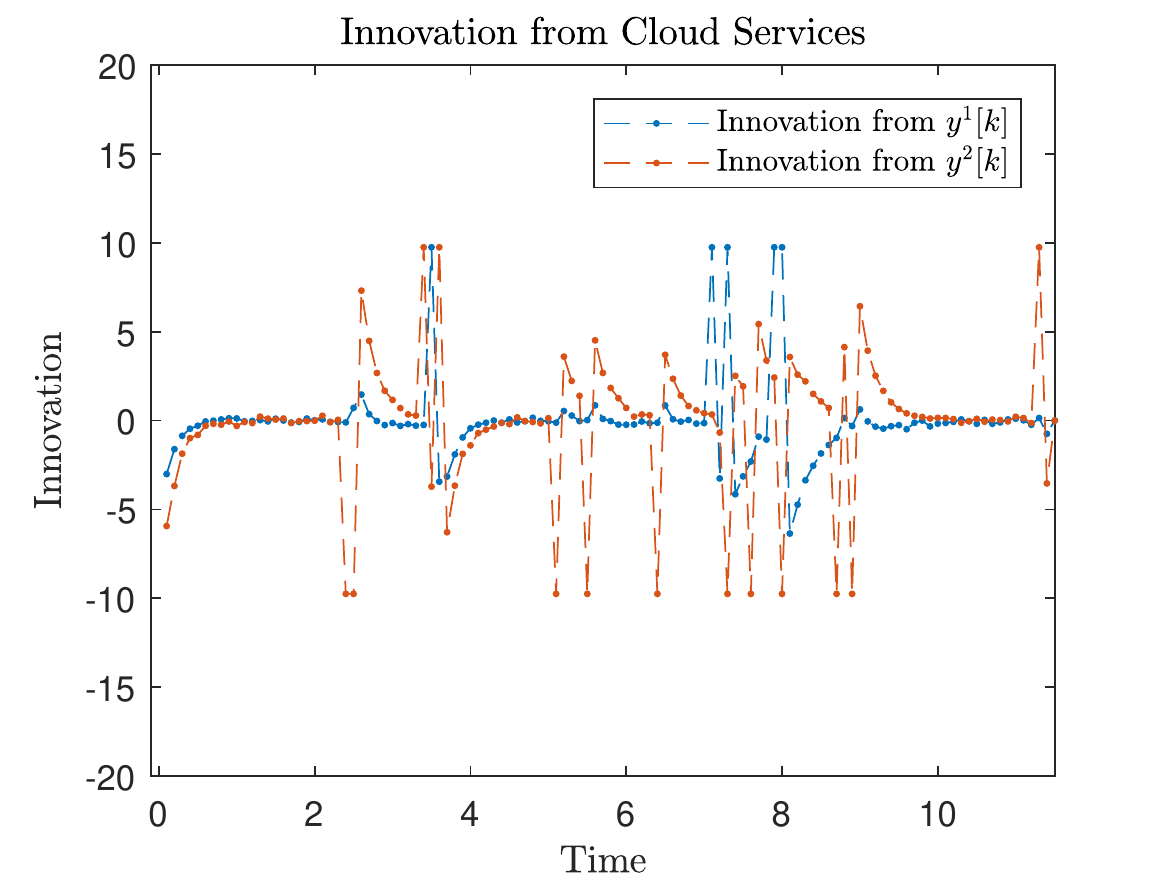}
\par\end{centering}

}\hfill{}\subfloat[\label{fig:innov_aNtwo}Innovation]{\begin{centering}
\includegraphics[width=0.3\textwidth]{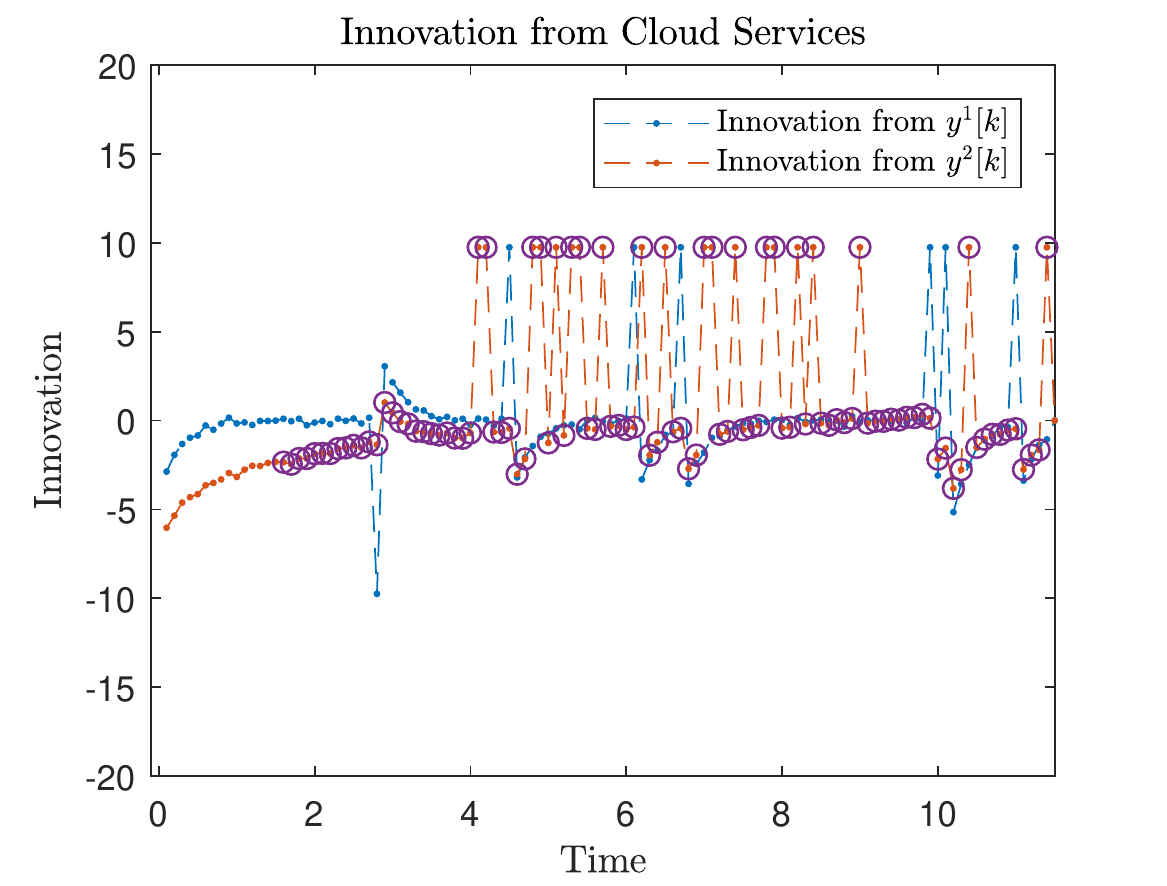}
\par\end{centering}

}\hfill{}\subfloat[\label{fig:innovMixTwo}Innovation]{\begin{centering}
\includegraphics[width=0.3\textwidth]{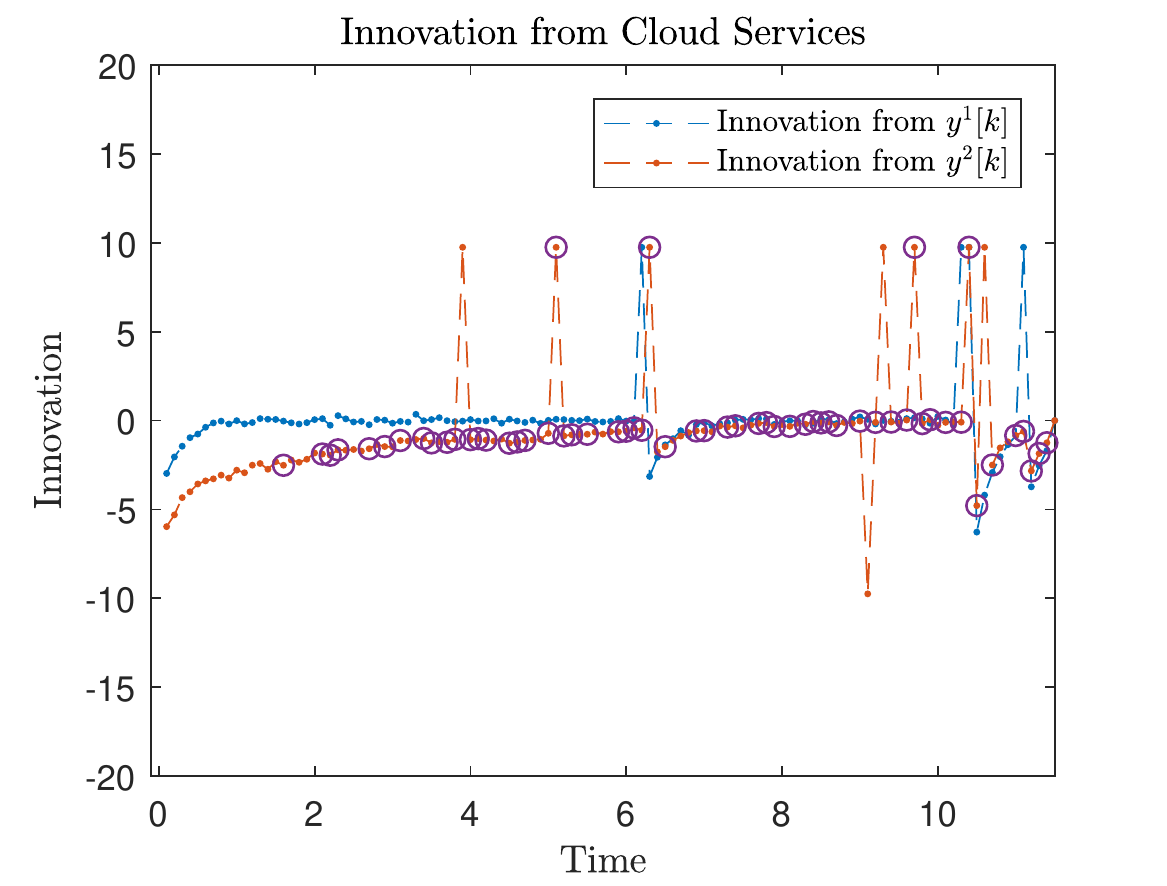}
\par\end{centering}

}

\medskip{}

\subfloat[\label{fig:states_aTtwo}State trajectories]{\begin{centering}
\includegraphics[width=0.3\textwidth]{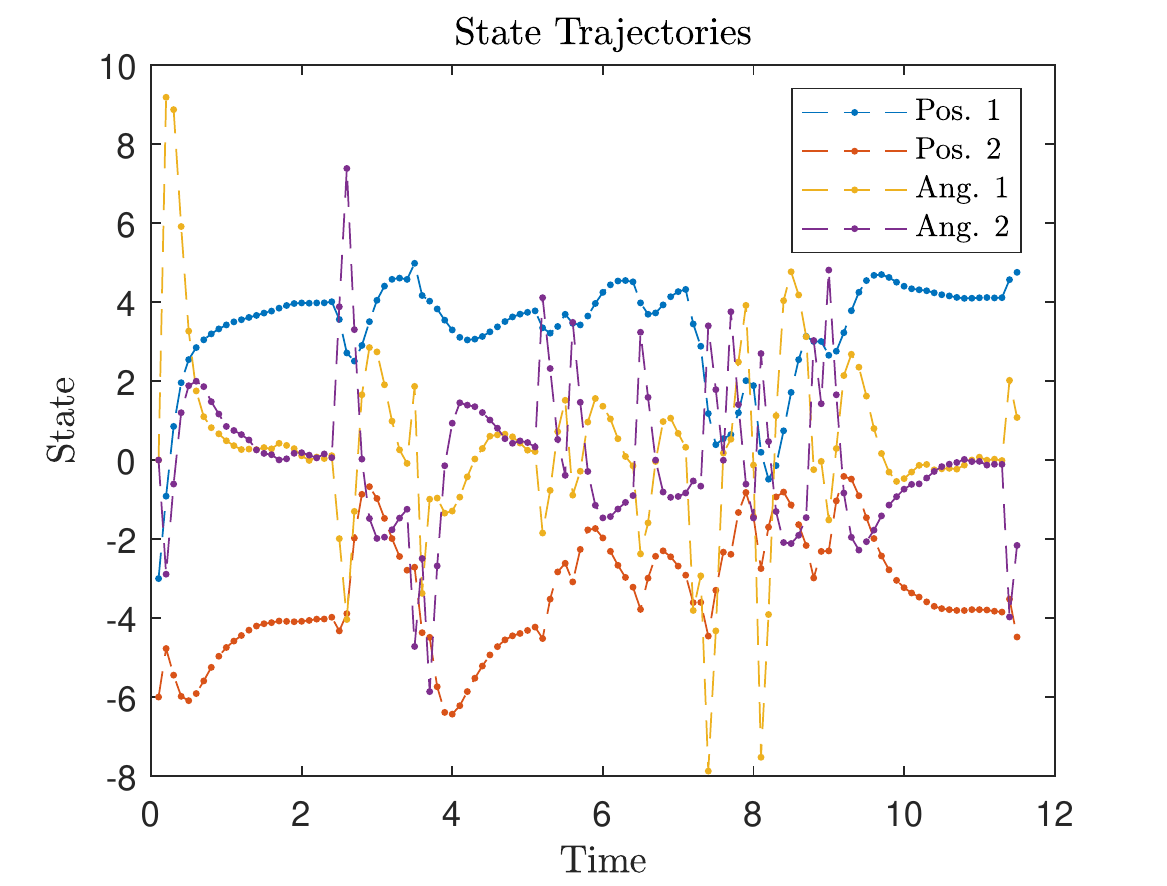}
\par\end{centering}

}\hfill{}\subfloat[\label{fig:states_aNtwo}State trajectories]{\begin{centering}
\includegraphics[width=0.3\textwidth]{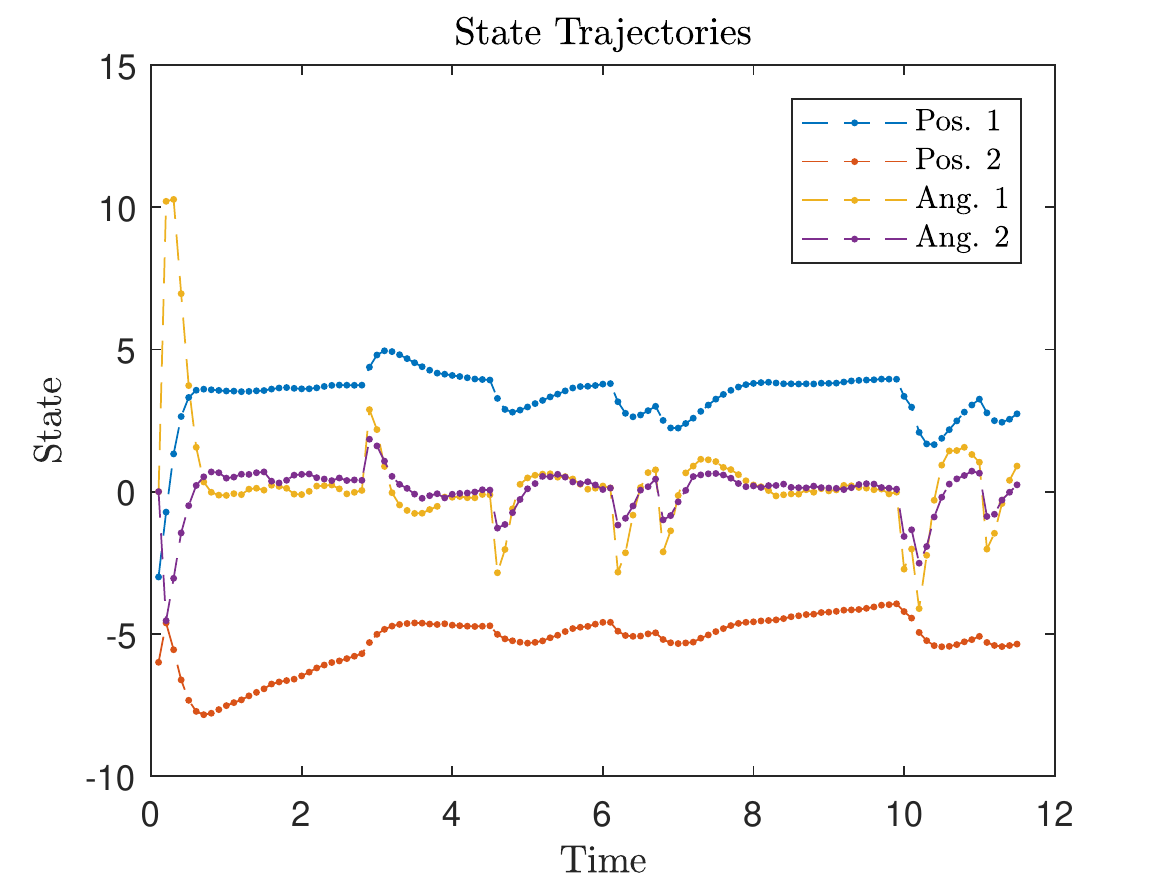}
\par\end{centering}

}\hfill{}\subfloat[\label{fig:states_MixTwo}\texttt{FlipIt} game]{\begin{centering}
\includegraphics[width=0.3\textwidth]{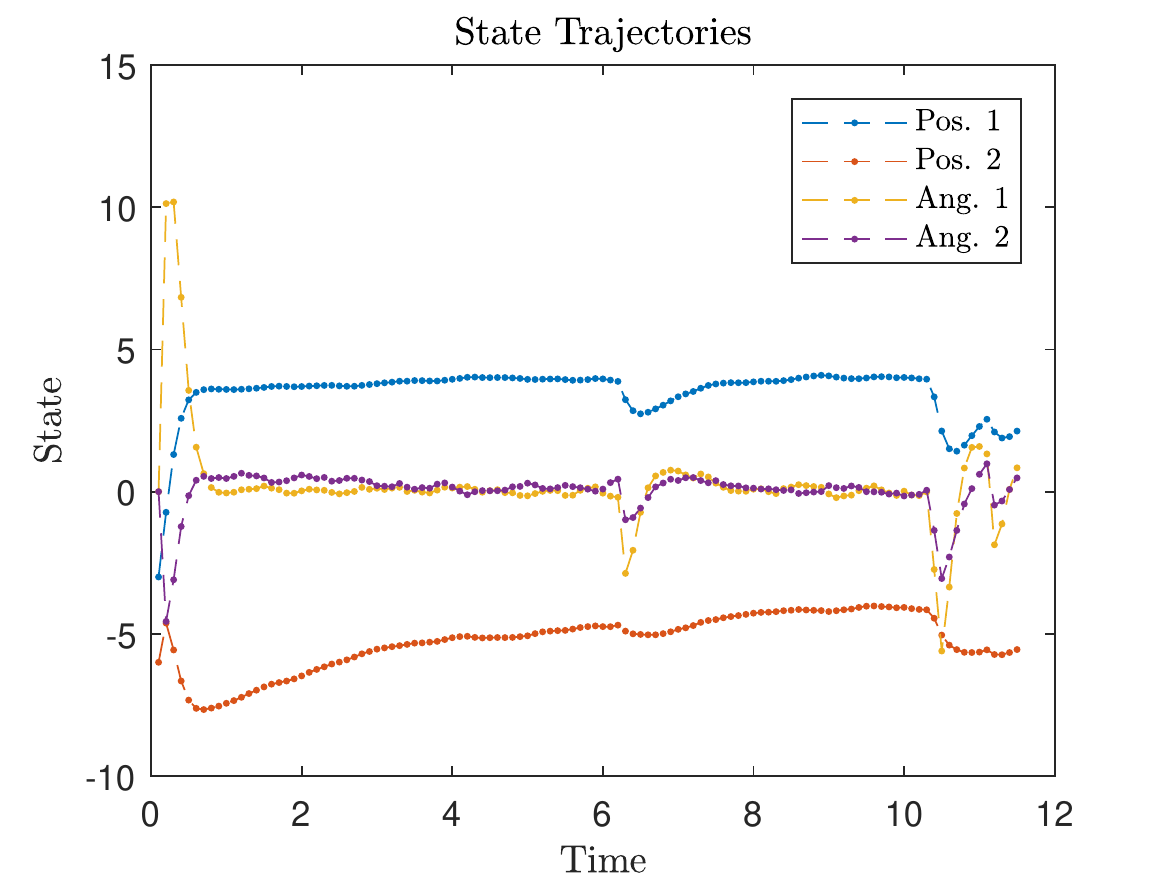}
\par\end{centering}

}

\caption{\label{fig:experimentsGNE}Column 1: Innovation and state trajectories
for $p_{\mathcal{A}}^{2}=0.10$ and $\mathcal{R}$ plays $[\protect\begin{array}{cc}
a_{T} & a_{T}\protect\end{array}]',$ Column 2: Innovation and state trajectories for $p_{\mathcal{A}}^{2}=0.10$
and $\mathcal{R}$ plays $[\protect\begin{array}{cc}
a_{T} & a_{N}\protect\end{array}]',$ Column 3: Innovation and state trajectories in which $\mathcal{R}$ mixes strategies between $[\protect\begin{array}{cc}
a_{T} & a_{T}\protect\end{array}]'$ and $[\protect\begin{array}{cc}
a_{T} & a_{N}\protect\end{array}]'.$}
\end{figure*}

Now consider a parameter change in which $\mathcal{A}^{2}$ develops
new malware to compromise the GPS position signal $\tilde{y}^2[k]$ at a much lower cost $\alpha_{\mathcal{A}}^{2}.$
(See Subsection \ref{sec:cyber}). In equilibrium, this increases
$p_{\mathcal{A}}^{2}$ from $0.03$ to $0.10.$ A higher number of perturbed innovations
are visible in Fig. \ref{fig:experimentsGNE}(a). This leads to the
poor state trajectories of Fig. \ref{fig:experimentsGNE}(d). The
control cost from Eq. (\ref{eq:criterion}) increases,  \textcolor{black}{and the two vehicles nearly collide at time $8.$ The large changes in angles show that the vehicles turn rapidly in different directions.}

In this case, $\mathcal{R}$'s best response is 
\[
\sigma_{\mathcal{R}}\left(\left[\begin{array}{c}
a_{T}\\
a_{N}
\end{array}\right]|\left[\begin{array}{c}
m_{L}\\
m_{L}
\end{array}\right]\right)=\sigma_{\mathcal{R}}\left(\left[\begin{array}{c}
a_{T}\\
a_{N}
\end{array}\right]|\left[\begin{array}{c}
m_{L}\\
m_{H}
\end{array}\right]\right)=1,
\]
\[
\sigma_{\mathcal{R}}\left(\left[\begin{array}{c}
a_{N}\\
a_{N}
\end{array}\right]|\left[\begin{array}{c}
m_{H}\\
m_{L}
\end{array}\right]\right)=\sigma_{\mathcal{R}}\left(\left[\begin{array}{c}
a_{N}\\
a_{N}
\end{array}\right]|\left[\begin{array}{c}
m_{H}\\
m_{H}
\end{array}\right]\right)=1,
\]
\textcolor{black}{\emph{i.e.}, to not trust even the low-risk messages from the remote GPS signal.} The circles on $\nu^{2}[k]$ for all $k$ in Fig. \ref{fig:experimentsGNE}(b)
represent not trusting. The performance improvement can be seen in
Fig. \ref{fig:experimentsGNE}(e). 

Interestingly, though, Remark \ref{rem:notUseless} states that this
cannot be an equilibrium. In the \texttt{FlipIt} game, $\mathcal{A}^{2}$
would have no incentive to capture Cloud Service 2, since $\mathcal{R}$
never trusts that cloud service. This would lead to $p_{\mathcal{A}}^{2}=0.$
Moving forward, $\mathcal{R}$ would trust Cloud Service 2 in the
next signaling game, and $\mathcal{A}^{2}$ would renew his attacks.
iSTRICT predicts that this pattern of compromising, not trusting,
trusting, and compromising would repeat in a limit cycle, and not
converge to equilibrium.

A mixed-strategy equilibrium, however, does exist. $\mathcal{R}$
chooses a mixed strategy in which he trusts low-risk messages on Cloud
Service 2 with some probability. This probability incentivizes $\mathcal{A}^{2}$
to attack the cloud with a frequency between those that best respond
to either of $\mathcal{R}$'s pure strategies. \textcolor{black}{At the GNE, the attack
frequency of $\mathcal{A}^{2}$ produces $0<p_{\mathcal{A}}^{2}<0.10$
in the \texttt{FlipIt} game.} In fact, this is the worst case $p_{\mathcal{A}}^{2}=p_{\mathcal{A}}^{2\diamond}$
from Remark \ref{rem:worst-case}. In essence, $\mathcal{R}$'s mixed-strategy
serves as a last-resort countermeasure to the parameter change due
the new malware obtained by $\mathcal{A}^{2}.$

Figure \ref{fig:experimentsGNE}(c) depicts the innovation with a
mixed strategy in which $\mathcal{R}$ sometimes trusts Cloud Service
2. Figure \ref{fig:experimentsGNE}(f) shows the impact on state trajectories. At this mixed-strategy equilibrium, $\mathcal{A}^{1},$
$\mathcal{A}^{2},$ $\mathcal{D}^{1},$ and $\mathcal{D}^{2}$ choose
optimal attack/recapture frequencies in the cloud-layer and send optimal
messages in the communications layer, and $\mathcal{R}$ optimally
chooses which messages to trust in the communication layer based on
an innovation filter and observer-based optimal control in the physical
layer. No players have incentives to deviate from their strategies
at the GNE.

\textcolor{black}{Columns 4-6 of Table \ref{tab:ctrlCosts} quantify the improvements provided by iSTRICT in these cases. Column 4 is the benchmark case, in which an innovation gate forces $\mathcal{A}^2$ to add low-risk noise, but his frequent attacks still cause large damages. Column 5 gives the performance of iSTRICT against a strategic attacker, and Column 6 gives the performance of iSTRICT against a nonstrategic attacker. In both cases, the cost criterion decreases by a factor of at least six.}

\section{Conclusion and Future Work}

iSTRICT attains robustness through a combination of multiple, interdependent
layers of defense. At the lowest physical layer, a Kalman filter handles
sensor noise. The Kalman filter, however, is not designed for the
large bias terms that can be injected into sensor measurements by
attackers. We use an innovation gate in order to reject these large
bias terms. But even measurements within the innovation gate should
be rejected if there is a sufficiently high risk that a cloud service
is compromised. We determine this threshold risk level strategically,
using a signaling game. Now, it may not be possible to estimate these
risk levels using past data. Instead, iSTRICT estimates the risk proactively
using \texttt{FlipIt} games. The equilibria of the \texttt{FlipIt}
games depend on the incentives of the attackers and defenders to capture
or reclaim the cloud. These incentives result from the outcome of
the signaling game, which means that the equilibrium of the overall
interaction consists of a fixed point between mappings that characterize
the \texttt{FlipIt} games and the signaling game. This equilibrium
is a GNE.

We have proved the existence of GNE under a set of natural assumptions,
and provided an algorithm to iteratively compute the GNE. We have
shown that a device can use iSTRICT to guarantee a worst-case compromise
probability, even without fully rejecting measurements from any of
the cloud services. \textcolor{black}{Through an application to autonomous vehicle networks,} we have shown the performance gains achieved by iSTRICT over naive
strategies. Because of the modularity of the GNE concept, the solutions
to each layer do not need to be completely recomputed when devices
enter or leave the IoCT. 

Future work can extend the framework to a fourth layer composed of
a cloud radio access network and a fifth layer of resource management
for economic and policy issues of the IoCT. Another promising extension
is to incorporate intelligent control designs that further mitigate
the performance loss due to cyber threats by addressing them at the
physical layer. These future directions would further contribute to
policies for strategic trust management in the dynamic and heterogeneous
IoCT.

\bibliographystyle{IEEEtran}
\bibliography{IEEEabrv,refs}

\begin{IEEEbiography}[{\includegraphics[width=1in,height=1.25in,clip,keepaspectratio]{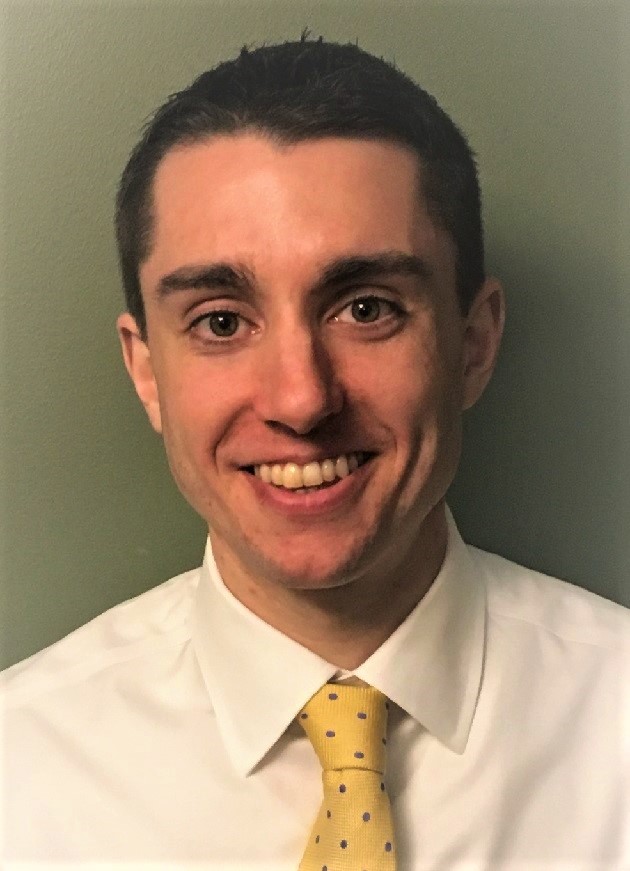}}]{Jeffrey Pawlick}

received the B.S. degree 
from Rensselaer Polytechnic Institute in Troy, NY in 2013, and the Ph.D. 
degree from New York University (NYU) in 2018, both in Electrical Engineering. At NYU he studied in the Laboratory for Agile and Resilient Complex Systems within the Tandon School of Engineering. His research interests include game theory, privacy, security, trust 
management, and deception in cyber-physical systems and the Internet
of things.\end{IEEEbiography}

\begin{IEEEbiography}[{\includegraphics[width=1in,height=1.25in,clip,keepaspectratio]{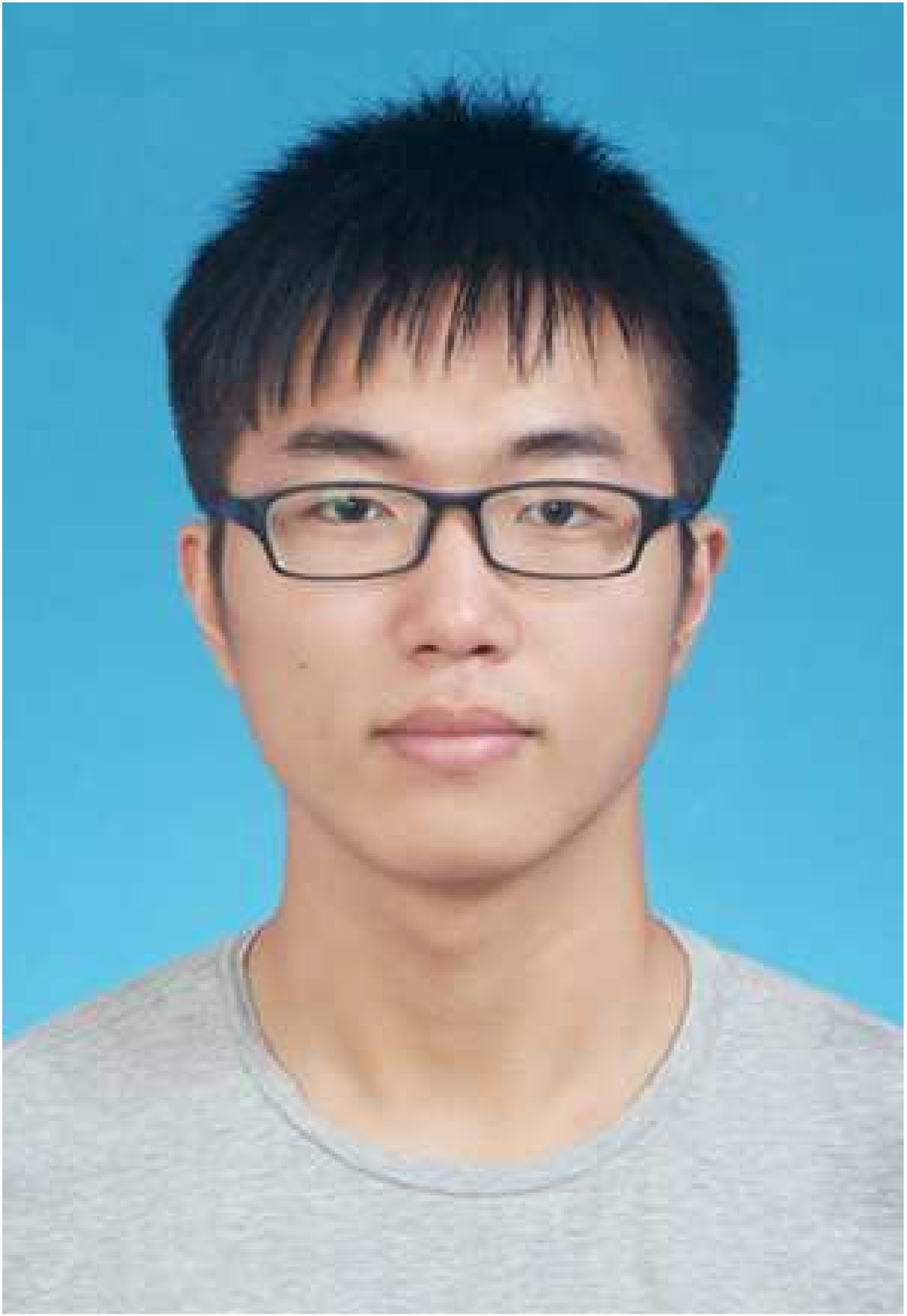}}]{Juntao Chen}
(S'15) received the B.Eng. degree in Electrical Engineering and Automation from Central South University, Changsha, China, in 2014. He is currently pursuing the Ph.D. degree in the Laboratory for Agile and Resilient Complex Systems, Tandon School of Engineering, New York University, NY, USA. 
His research interests include mechanism design, game theory, and cyber-physical systems security.
\end{IEEEbiography}

\begin{IEEEbiography}[{\includegraphics[width=1in,height=1.25in,clip,keepaspectratio]{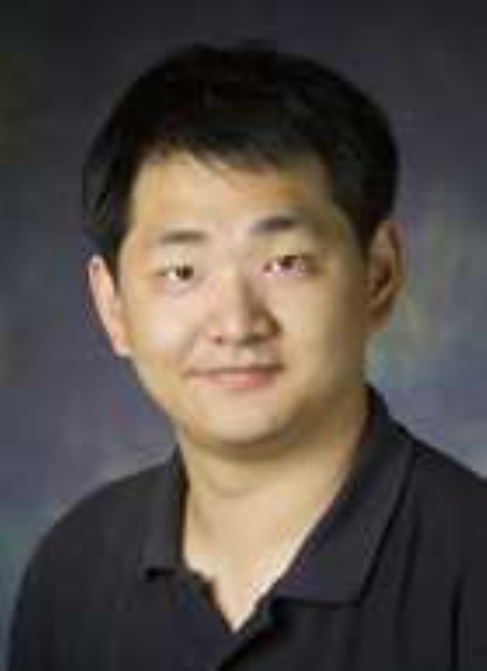}}]{Quanyan Zhu}
 (M\textquoteright 14) received the Ph.D. degree from the
University of Illinois at Urbana-Champaign (UIUC) in 2013. After a
short stint at Princeton University, he joined the Department of Electrical
and Computer Engineering at New York University (NYU) as an assistant
professor in 2014. He spearheaded and chaired INFOCOM Workshop on Communications and Control on Smart Energy Systems (CCSES), Midwest Workshop on Control and Game Theory (WCGT), and 7th Game and Decision Theory for Cyber Security (GameSec). His research interests include game theory, smart
grid, network security and privacy, resilient critical infrastructures,
cyber-physical systems and cyber 
deception. \end{IEEEbiography}


\end{document}